\documentclass[
    reprint, twocolumn,
    aps,prx,10pt,amsmath,amssymb,
    longbibliography,superscriptaddress,
    showpacs,preprintnumbers,
    groupedaddress]{revtex4-1}

\usepackage{xr}
\makeatletter
\newcommand*{\addFileDependency}[1]{
  \typeout{(#1)}
  \@addtofilelist{#1}
  \IfFileExists{#1}{}{\typeout{No file #1.}}
}
\makeatother

\newcommand*{\myexternaldocument}[1]{
    \externaldocument{#1}
    \addFileDependency{#1.tex}
    \addFileDependency{#1.aux}
}

\myexternaldocument{supp}

\usepackage[colorlinks=true,linkcolor=blue,citecolor=blue,urlcolor=blue]{hyperref}
\usepackage{oubraces}
\usepackage{bm,bbm,mathbbol}
\usepackage{physics}
\usepackage{mathtools}
\usepackage{amsmath, amsthm, amssymb, amsfonts}
\usepackage{bbold}
\usepackage{esvect}

\usepackage{epsfig}
\usepackage{array}
\usepackage{multirow}
\usepackage{graphicx}
\usepackage[normalem]{ulem}

\usepackage{float}
\usepackage[section]{placeins}
\usepackage{afterpage}

\usepackage{lipsum}

\usepackage{xifthen}

\makeatletter
\newcommand\footnoteref[1]{\protected@xdef\@thefnmark{\ref{#1}}\@footnotemark}
\makeatother

\newcommand{\nnnl}{\nonumber \\}

\newcommand{\myeqref}[1]{Eq.~\eqref{#1}}

\newcommand{\mb}[1]{{\boldsymbol{\mathbf{#1}}}}
\newcommand{\nk}[0]{{n\mathbf{k}}}
\newcommand{\mk}[0]{{m\mathbf{k}}}

\newcommand{\nkq}[0]{{n\mathbf{k}+\mathbf{q}}}
\newcommand{\mkq}[0]{{m\mathbf{k}+\mathbf{q}}}

\newcommand{\pkq}[0]{{p\mathbf{k}+\mathbf{q}}}
\newcommand{\pk}[0]{{p\mathbf{k}}}
\newcommand{\qnu}[0]{{\mathbf{q}\nu}}
\newcommand{\primesum}[1]{\sideset{}{'}\sum_{{#1}}}

\newcommand{\veps}[0]{\varepsilon}

\newcommand{\vepszero}[0]{\varepsilon^{\rm(0)}}

\newcommand{\one}[0]{\mathbb{1}}

\newcommand{\opH}[0]{\hat{H}}
\newcommand{\opV}[0]{\hat{V}}

\newcommand{\NWan}[0]{{N_\mathrm{W}}}

\newcommand{\ord}[1]{{(\mathrm{#1})}}
\newcommand{\oord}[1]{^{(\mathrm{#1})}}
\newcommand{\mcO}[0]{{\mathcal{O}}}
\newcommand{\mcW}[0]{{\mathcal{W}}}
\newcommand{\WP}[0]{{\mathcal{W}_\mathrm{P}}}
\newcommand{\WF}[0]{{\mathcal{W}_\mathrm{F}}}
\newcommand{\WD}[0]{{\mathcal{W}_\mathrm{D}}}
\newcommand{\expixy}[2]{{e^{i{#1}\cdot{#2}}}}
\newcommand{\expmixy}[2]{{e^{-i{#1}\cdot{#2}}}}
\newcommand{\iR}[0]{{i\mathbf{R}}}
\newcommand{\jR}[0]{{j\mathbf{R}}}

\newcommand{\ipRp}[0]{{i'\mathbf{R'}}}

\newcommand{\mkp}[0]{{m\mathbf{k'}}}
\newcommand{\nkp}[0]{{n\mathbf{k'}}}
\newcommand{\npkp}[0]{{n'\mathbf{k'}}}
\newcommand{\mpkp}[0]{{m'\mathbf{k'}}}

\newcommand{\opP}[0]{{\hat{P}}}
\newcommand{\opQ}[0]{{\hat{Q}}}

\newcommand{\opPP}[0]{{\hat{P}^{(0)}_{\mathrm{P}}}}
\newcommand{\opPD}[0]{{\hat{P}^{(0)}_{\mathrm{D}}}}
\newcommand{\opQD}[0]{{\hat{Q}^{(0)}_{\mathrm{D}}}}
\newcommand{\opPW}[0]{{\hat{P}^{(0)}_{\mathrm{W}}}}

\newcommand{\NPk}[0]{{N_{\mathrm{P},\mb{k}}}}
\newcommand{\NFk}[0]{{N_{\mathrm{F},\mb{k}}}}
\newcommand{\NDk}[0]{{N_{\mathrm{D},\mb{k}}}}

\newcommand{\psizero}[0]{\psi^{(0)}}

\usepackage{tikz}
\newcommand*\circled[1]{\tikz[baseline=(char.base)]{
\node[circle,draw,scale=1,inner sep=1pt] (char) {#1};}}
\begin{document}

\title{Wannier Function Perturbation Theory:\\ Localized Representation and Interpolation of Wavefunction Perturbation}

\author{Jae-Mo Lihm}
\email{jaemo.lihm@gmail.com}
\author{Cheol-Hwan Park}
\email{cheolhwan@snu.ac.kr}
\affiliation{Center for Correlated Electron Systems, Institute for Basic Science, Seoul 08826, Korea}
\affiliation{Department of Physics and Astronomy, Seoul National University, Seoul 08826, Korea}
\affiliation{Center for Theoretical Physics, Seoul National University, Seoul 08826, Korea}

\date{\today}

\footnotetext[1]{See Supplemental Material, which includes Refs.~\cite{2021PonceMobility, 1996AdolphVelocity, 2012LopezOrbitalMag}, at [URL will be inserted by publisher] for
the details of the calculation of the velocity and related matrix elements
and the convergence study.
}
\newcommand{\citeSupp}[0]{Note1}
 
\begin{abstract}
Thanks to the nearsightedness principle, the low-energy electronic structure of solids can be represented by localized states such as the Wannier functions.
Wannier functions are actively being applied to a wide range of phenomena in condensed matter systems.
However, the Wannier-function-based representation is limited to a small number of bands and thus cannot describe the change of wavefunctions due to various kinds of perturbations, which require sums over an infinite number of bands.
Here, we introduce the concept of the Wannier function perturbation, which provides a localized representation of wavefunction perturbations.
Wannier function perturbation theory allows efficient calculation of numerous quantities involving wavefunction perturbation, among which we provide three applications.
First, we calculate the temperature-dependent indirect optical absorption spectra of silicon near the absorption edge nonadiabatically, i.e., differentiating phonon-absorption and phonon-emission processes, and without arbitrary temperature-dependent shifts in energy.
Second, we establish a theory to calculate the shift spin conductivity without any band-truncation error.
Unlike the shift charge conductivity, an exact calculation of the shift spin conductivity is not possible within the conventional Wannier function methods because it cannot be obtained from geometric quantities for low-energy bands.
We apply the theory to monolayer WTe$_2$.
Third, we calculate the spin Hall conductivity of the same material again without any band-truncation error.
Wannier function perturbation theory is a versatile method that can be readily applied to calculate a wide range of quantities related to various kinds of perturbations.
\end{abstract}

\maketitle

\section{Introduction}
Most physical and chemical systems are ``nearsighted''~\cite{1996KohnNearsighted,2005ProdanNearsighted}: Their properties can be studied by inspecting one local region at a time and the long-range effects can be easily approximated.
A localized representation of a system is highly desirable because it enables one to exploit the nearsightedness to its full extent.
Wannier functions (WFs) provide a localized representation of wavefunctions~\cite{1937Wannier,1959Kohn}.
The invention of the maximally localized Wannier function (MLWF) method~\cite{1997Marzari,2001Souza} enabled the calculation of the localized WFs for electronic structures computed using density functional theory (DFT).
Using the MLWFs, the low-energy eigenvalues and eigenstates can be interpolated very efficiently.

Applications of MLWFs are stretched over all fields of condensed matter physics and materials science~\cite{2012MarzariRMP}.
MLWFs are used to efficiently compute electron-phonon coupling (EPC) at a dense grid of electron and phonon wavevectors~\cite{2007GiustinoEPW}.
Wannier interpolation of EPC is actively being applied to study transport~\cite{2016ZhouTransport,2018PonceTransport}, phonon-assisted optical absorption~\cite{2012NoffsingerIndabs}, superconductivity~\cite{2013MargineSC}, and ultrafast carrier dynamics~\cite{2017JhalaniUltrafast}.
Various linear and nonlinear responses of solids to dc and ac electric fields are also routinely calculated using Wannier interpolation~\cite{2006WangAHC,2007YatesWannier,2018IbanezAzpirozShift,2018TsirkinGyro,2018QiaoSHC,2019RyooSHC}.

However, one of the most severe limitations of WF interpolation is the limited number of bands being interpolated.
Consequently, WFs cannot represent the wavefunction perturbation, which is the sum of contributions from an infinite number of bands.
Obtaining MLWFs for the high-energy, plane-wave-like bands is a highly nontrivial problem.
Even assuming that the MLWFs are obtained, calculations in the MLWF basis will be much more costly, negating the efficiency of Wannier interpolation.
In DFT calculations, the wavefunction perturbation can be calculated without referring to the high-energy states by solving a linear equation known as the Sternheimer equation~\cite{1954Sternheimer}.
This method forms the basis of the calculation of the linear response properties of solids using density functional perturbation theory (DFPT)~\cite{2001BaroniRMP}.
However, the Sternheimer equation is not directly applicable to WFs since the WFs do not accurately represent the wavefunction perturbation.
Therefore, WF methods cannot deal with properties that require the calculation of wavefunction perturbations.
Ge \textit{et al.}~\cite{2015GeLocalized} developed a local representation of wavefunction perturbation for the dielectric response of insulators.
However, this formalism cannot be applied to WFs for entangled bands~\cite{2001Souza}, while such generalization is necessary for many applications.

A representative example where the interpolation of wavefunction perturbations is needed is the real part of the phonon-induced electron self-energy.
This self-energy causes the temperature dependence and zero-point renormalization of electron band structures~\cite{2010GiustinoPRL,2011GonzeAnnPhys}.
The state-of-the-art method for the perturbative calculation of the real part of the self-energy is to solve the Sternheimer equation using a complete basis set such as the plane waves~\cite{2011GonzeAnnPhys}.
However, such a calculation is too costly to be done on a large number of $k$ points.
The high computational cost prohibits the consideration of phonon-induced renormalization at a dense $k$-point mesh in the study of optical and transport properties of materials.
Also, band disentanglement~\cite{2001Souza} is necessary even for the valence band of insulators because the wavefunctions and energy eigenvalues of the low-lying conduction bands are required.

Another example of a quantity involving the wavefunction perturbation is the shift current~\cite{2000SipeBPVE}.
For the shift charge current, the wavefunction perturbation can alternatively be expressed as a derivative with respect to the crystal momentum $\mb{k}$, allowing an exact WF-based calculation~\cite{2018IbanezAzpirozShift}.
However, such an alternative expression cannot be made for the shift spin current in spin-orbit coupled systems~\cite{2021LihmSpin}.
Previous calculations of shift spin current in spin-orbit coupled systems truncated the infinite sum over bands~\cite{2020XuSpinPhotocurrent,2020MuSpinPhotocurrent} or used the tight-binding approximation~\cite{2021LihmSpin}.
Similarly, conventional Wannier interpolation schemes for the spin Hall conductivity~\cite{2018QiaoSHC,2019RyooSHC} resorts to the band truncation.
To our knowledge, the effect of band truncation on the shift spin or spin Hall conductivity has not been investigated.

In this work, we develop Wannier function perturbation theory (WFPT), which enables a localized representation of wavefunction perturbation, alleviating the conceptual and practical limitations of the WF method.
We define the Wannier function perturbation (WFP) as the change of the WF due to a perturbation.
The WFPs are spatially localized and thus can be used to efficiently interpolate the perturbation of eigenstate wavefunctions and the related matrix elements.
We provide three applications of WFPT.
The first application is the temperature-dependent indirect optical absorption of silicon.
We calculate the phonon renormalization of the electron band structure using WFPT.
Using the renormalized band structures, we perform a predictive calculation of the temperature dependence of the phonon-assisted optical absorption spectra near the absorption edge.
Our calculation takes into account nonadiabatic effects and does not require arbitrary shifts in energy depending on the temperature.
The second and third applications are the shift spin and spin Hall conductivity of monolayer WTe$_2$, respectively.
In these applications, we show that WFPT is necessary to use an ultradense $k$-point grid and eliminate the band-truncation error at the same time.

\section{Formalism of Wannier function perturbation theory}
Before discussing the formalism of WFPT in detail, let us briefly summarize the key ideas.
First, we Wannierize both the unperturbed and perturbed systems.
We call the corresponding WFs the unperturbed WFs and the perturbed WFs, respectively.
To study the perturbative properties of the WFs, the perturbed WFs must converge smoothly to the unperturbed WFs in the limit of a small perturbation.
To achieve this property, we use the unperturbed WFs as the initial guesses and construct projection-only WFs~\cite{2001Souza} of the perturbed system.
The resulting first-order derivatives of the WFs with respect to the perturbation strength are the WFPs.
Next, we show that the perturbation of the eigenstate wavefunction can be efficiently interpolated using WFPs.
Finally, we study the interpolation of matrix elements that involve wavefunction perturbations.

\subsection{Wannier function perturbation}
In this subsection, we first introduce the setting for our theory.
Then, we derive the expression for the WFs of the perturbed system and the WFPs.
Finally, we argue that the WFPs are spatially localized.

\subsubsection{Setting}
The WFs of a perfect crystal can be written as
\begin{equation} \label{eq:wfpt_WF0_def}
    \ket{w\oord{0}_\iR} = \frac{1}{\sqrt{N_k}} \sum_{n,\mb{k}} \ket{\psi\oord{0}_\nk} U\oord{0}_{ni;\mb{k}} \expmixy{\mb{k}}{\mb{R}}.
\end{equation}
Here, $\ket{\psi\oord{0}_\nk}$ is the unperturbed electron eigenstate for band $n$ and crystal momentum $\mb{k}$, $i=1,\cdots,\NWan$ the WF index, $\mb{R}$ the unit cell position, and
$N_k$ the size of the Born-von Karman supercell.
The semi-unitary gauge transformation matrix $U\oord{0}_{ni;\mb{k}}$ is chosen to make $\ket{w\oord{0}_\iR}$ spatially localized.
The unperturbed eigenstates satisfy
\begin{equation}
    \braket{\psi\oord{0}_\mk}{\psi\oord{0}_\nkp}
    = \delta_{mn} \delta_{\mb{k}\mb{k'}}.
\end{equation}

We apply a perturbation $\lambda \opV$, where $\lambda$ is a small perturbation parameter.
The perturbation $\opV$ may or may not break the periodicity of the lattice but should respect the Born-von Karman periodic boundary condition.
The only additional condition we impose on the perturbation $\opV$ is that $\mel{\mb{r}}{\opV}{\mb{r'}}$ should be small if $\abs{\mb{r} - \mb{r'}}$ is large.
This condition is necessary for the perturbed system to be nearsighted.
Any local potentials, phonon perturbation potentials that contain contributions from nonlocal pseudopotentials, velocity operator $\hat{v}^a = [\hat{r}^a,\opH^{(0)}]/i\hbar$, where $\opH^{(0)}$ is the unperturbed Hamiltonian, and spin-velocity operator $\hat{j}^{s;a} = \frac{1}{2} \acomm{\hat{S}_s}{\hat{v}^a}$ all fulfill this condition.
Here, $\hat{S}_s$ is the spin operator for spin polarization along $s$.

Eigenstates of the perturbed system can be expanded as
\begin{equation} \label{eq:wfpt_perturb}
    \ket{\psi_\nk} = \ket{\psi\oord{0}_\nk} + \lambda\ket{\psi\oord{1}_\nk} + \mcO(\lambda^2),
\end{equation}
where the wavefunction perturbation is
\begin{equation} \label{eq:wfpt_wf_series}
    \ket{\psi\oord{1}_\nk} = \primesum{\npkp} \ket{\psizero_\npkp} \frac{\mel{\psizero_\npkp}{\opV}{\psizero_\nk}} {\veps^\ord{0}_\nk - \veps^\ord{0}_\npkp}.
\end{equation}
Hereafter, the primed sum denotes that the terms making the denominator zero are excluded.
For degenerate states, the true perturbed eigenstates in the $\lambda \rightarrow 0$ limit will, in general, converge to the unperturbed eigenstates that diagonalize $\opV$ in the degenerate subspace, not to $\ket{\psizero_\nk}$.
In this case, we linearly combine the perturbed eigenstates so that they are no longer eigenstates of the perturbed Hamiltonian but satisfy Eqs.~(\ref{eq:wfpt_perturb}, \ref{eq:wfpt_wf_series}).
This operation is allowed because our goal is to construct WFs of the perturbed system, and there is gauge freedom in constructing the WFs that allows such a linear combination.
We note that this treatment is needed only to make the derivation more concise, while, in practice, one does not have any issue with degeneracy because of the careful use of energy windows that will be described in the following.

\subsubsection{WFs of the perturbed system and the WFPs}
\begin{figure*}[htbp]
\centering
\includegraphics[width=0.8\textwidth]{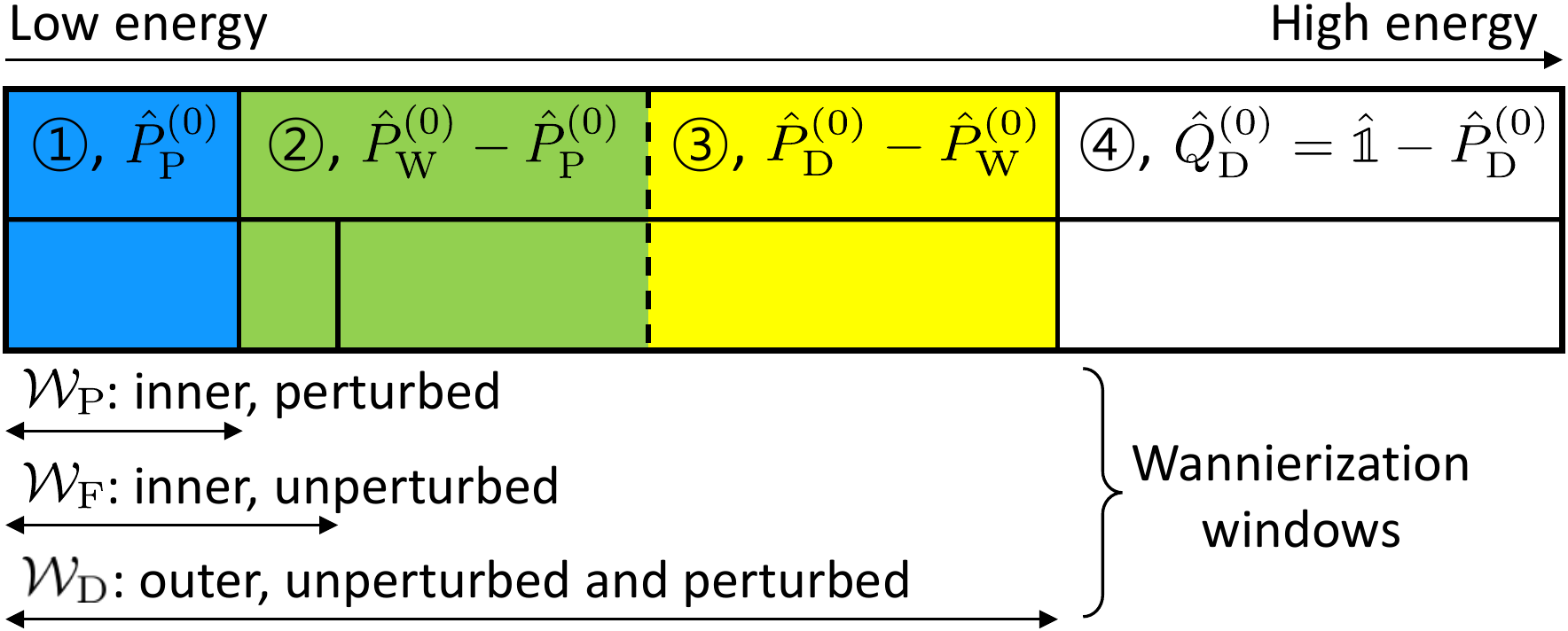}
\caption{
Schematic illustration of the subspaces and energy windows.
The vertical lines indicate the separation of the full Hilbert space into subspaces.
Solid vertical lines separating different regions indicate the separation by energy eigenvalues.
The dashed vertical line separates the subspace spanned by the unperturbed WFs from its complement and does not indicate an energetic separation.
The ``high-energy'' states may also include deep semicore states which are excluded in the Wannierization process.
}
\label{fig:subspaces}
\end{figure*}

We aim to accurately describe the wavefunction perturbation of the low-energy states whose eigenvalues are inside the ``perturbation window'' $\WP$.
We denote the inner (frozen) and outer (disentanglement) windows for the unperturbed system by $\WF$ and $\WD$, respectively~\cite{2001Souza}.
The following inclusion relation should hold:
\begin{equation} \label{eq:wfpt_window_inclusion}
    \WP \subseteq \WF \subseteq \WD
\end{equation}
This relation is illustrated in Fig.~\ref{fig:subspaces}.
In principle, $\WP$ can be chosen to be identical to $\WF$.
Using $\WP$ slightly narrower than $\WF$ is useful when the perturbation mixes the states right above and below the boundary of the inner window for the unperturbed system.
For later use, we define the projection operators onto the subspaces spanned by the unperturbed eigenstates inside $\WP$ and $\WD$ as $\opPP$ and $\opPD$, respectively, and we also define $\opQD = \hat{\one} - \opPD$.
Additionally, $N_X$ and $N_{X,\mb{k}}$ denote the number of unperturbed states inside $\mcW_X$, and those with crystal momentum $\mb{k}$, respectively, for $X=\mathrm{P}, \mathrm{F}, \mathrm{D}$.
We only consider small $\lambda$ such that no eigenvalues cross the boundary of the windows as $\lambda$ increases from 0.
Another important subspace is the unperturbed Wannier subspace, of which the projection operator is
\begin{equation}
    \opPW = \sum_{\mb{R}} \sum_{i=1}^{\NWan} \ket{w_\iR\oord{0}} \bra{w_\iR\oord{0}}.
\end{equation}
By construction, $\opPP \subseteq \opPW \subseteq \opPD$ holds, as illustrated in Fig.~\ref{fig:subspaces}.
We define ``subspaces \circled{1}, \circled{2}, \circled{3}, and \circled{4}'' as shown in Fig.~\ref{fig:subspaces}.

In general, the perturbed WFs need not make any reference to the unperturbed WFs.
However, to study the perturbative property of the WFs, the WFs must change smoothly around $\lambda=0$.
Our key idea to make the WFs smooth functions of $\lambda$ is to use the unperturbed WFs as the initial guesses for the projection-only WFs of the perturbed system.
For Wannierization of the perturbed system, the inner and outer windows are set to $\mcW_{\rm P}$ and $\mcW_{\rm D}$, respectively.
Using the perturbation window as the inner window is natural because only the states inside the perturbation window need to be accurately represented.
The outer window is set to make the conditions for Wannierization of the perturbed system as similar as possible to that of the unperturbed system.
This condition guarantees that the perturbed WFs are close to the unperturbed WFs; hence, it makes the WFPs well localized.
See Appendix~\ref{sec:app_cases} for the numerical verification that this choice is indeed beneficial.

Now, we explicitly construct the perturbed WFs.
First, we select the subspace to be Wannierized following the initialization recipe described in Sec.~III.G of Ref.~\cite{2001Souza}.
We take the perturbed eigenstates inside the inner window $\WP$ directly.
For the states not included in $\WP$, we project $\opPW$, the projection operator for the unperturbed Wannier subspace, onto the subspace $\WD - \WP$ and diagonalize it.
Since we are Wannierizing a supercell, we need $N_\mb{k}\NWan$ WFs.
Hence, we select $N_\mb{k}\NWan - N_{\rm P}$ states inside $\WD - \WP$ with the largest eigenvalues~\cite{2001Souza} together with $N_{\rm P}$ states inside $\WP$ and use them as the basis states for the perturbed WFs.

After we obtain the disentangled subspace, we compute the perturbed WFs.
Following the scheme of Ref.~\cite{1997Marzari}, we project the initial guess functions, the unperturbed WFs, onto the disentangled subspace and perform L\"owdin orthonormalization.
The resulting WFs of the perturbed system are (see Appendix~\ref{sec:app_wfp_derivation} for the detailed derivation)
\begin{align} \label{eq:wfpt_WF_result}
    \ket{w_\iR (\lambda)}
    = \ket{w^{(0)}_\iR}
    + \lambda \ket{w^{(1)}_\iR}
    + \mcO(\lambda^2),
\end{align}
where
\begin{align} \label{eq:wfpt_WFP}
    \ket{w^{(1)}_\iR}
    =& \frac{1}{\sqrt{N_k}}
    \sum_{m,\mb{k}}
    \opQD \ket{\psi^\ord{1}_\mk}
    U\oord{0}_{mi,\mb{k}} e^{-i\mb{k}\cdot\mb{R}} \\
    +& \frac{1}{\sqrt{N_k}} \sum_{\substack{m,\mb{k}\\ \mk \in \WP}}
    \left( \opPD - \opPW \right) \ket{\psi^\ord{1}_\mk}
    U\oord{0}_{mi;\mb{k}} e^{-i\mb{k}\cdot\mb{R}}. \nonumber
\end{align}
We call $\ket{w^{(1)}_\iR}$, the first-order derivative of a WF at $\lambda = 0$, the ``Wannier function perturbation.''
The concept of WFP and its explicit form [\myeqref{eq:wfpt_WFP}] is the first main theoretical result of this paper.

The WFPs can be computed without explicitly referring to the states outside the outer window.
The first term of \myeqref{eq:wfpt_WFP} can be evaluated by solving the Sternheimer equation to obtain the wavefunction perturbation for states $\mk$ inside $\WD$.
The second term of \myeqref{eq:wfpt_WFP} can be rewritten using
\begin{align} \label{eq:wfpt_wfp_second}
    \sum_{\substack{m,\mb{k}\\ \mk \in \WP}} &\left( \opPD - \opPW \right) \ket{\psi^\ord{1}_\mk}
    = \sum_{\substack{m,\mb{k}\\ \mk \in \WP}} \sum_{m'} \primesum{n',\mb{k'}} \ket{\psizero_\mpkp} \nnnl
    &\times \left( P\oord{0}_{\rm D} - P\oord{0}_{\rm W} \right)_{m'n';\mb{k'}}
    \frac{\mel{\psizero_\npkp}{\opV}{\psizero_\mk}}{\veps^\ord{0}_\mk - \veps^\ord{0}_\npkp}.
\end{align}
Evaluating this expression requires only the eigenenergies and matrix elements of $\opV$ and $\opPW$ for states inside $\WD$.

\begin{figure*}[htbp]
\centering
\includegraphics[width=0.8\textwidth]{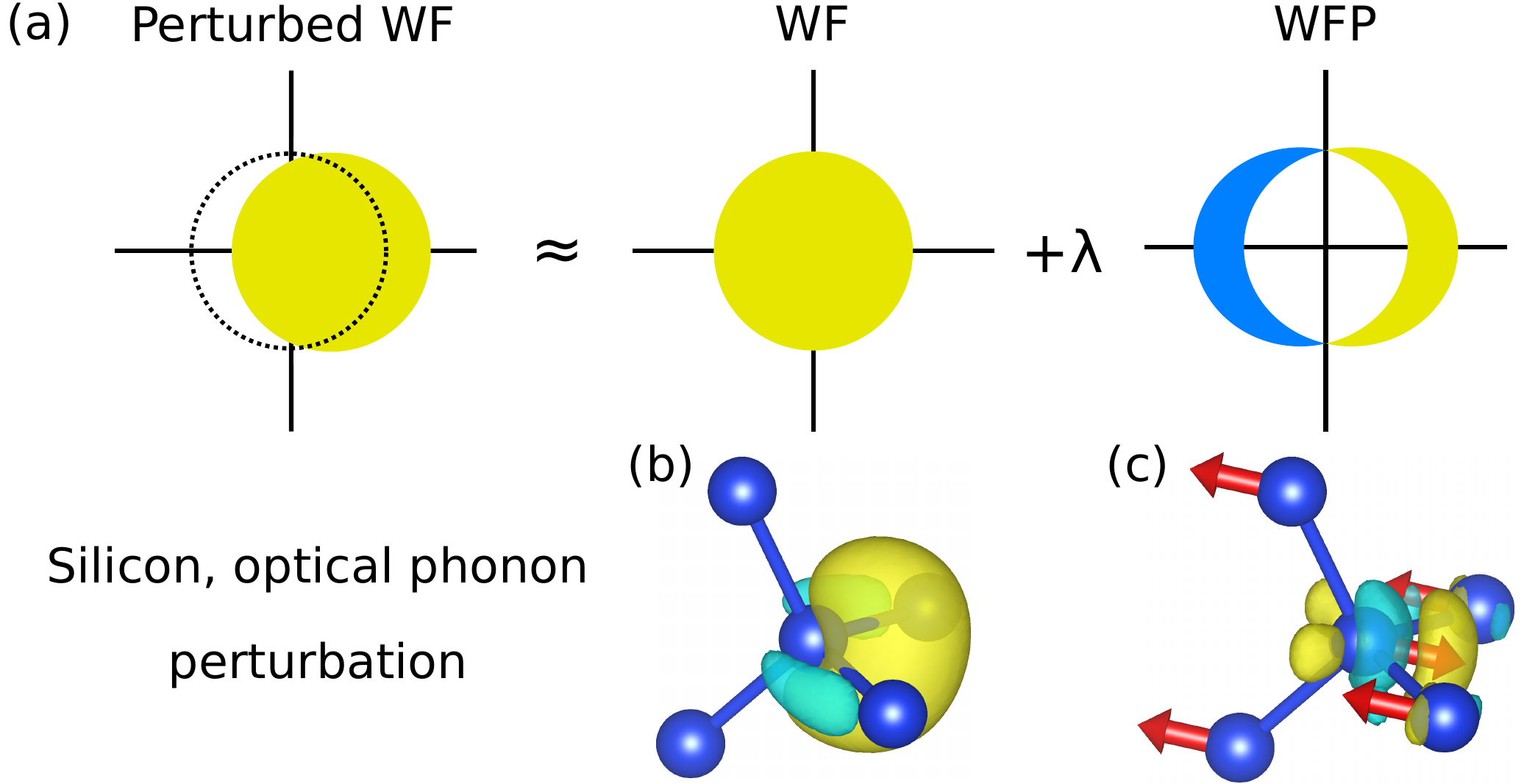}
\caption{
(a) Schematic illustration of the definition of a WFP.
(b) Isosurface of a WF of silicon. Isovalues are $+0.21$ and $-0.21$ times the maximum absolute value for the yellow and cyan isosurfaces, respectively.
(c) Isosurface of a WFP for an optical phonon perturbation whose vibration mode is indicated by the arrows. 
Isovalues are $+0.13$ and $-0.13$ times the maximum absolute value for the yellow and cyan isosurfaces, respectively.
}
\label{fig:schematic}
\end{figure*}

Figure \ref{fig:schematic}(a) schematically illustrates the definition of the WFP as the change of the WF due to a perturbation.
Figures \ref{fig:schematic}(b) and \ref{fig:schematic}(c) show the actual WF of silicon and the corresponding WFP for an optical phonon perturbation, respectively.
They visually demonstrate that the WFP is the change of the WF due to the perturbation and that the WFP is spatially localized.

The formula for the WFPs [\myeqref{eq:wfpt_WFP}] can be intuitively understood from the partition of subspaces shown in Fig.~\ref{fig:subspaces}.
First, consider subspaces \circled{1} and \circled{2}, whose union is the unperturbed Wannier subspace. The wavefunction perturbation in this subspace is not included in the WFPs because it can be reconstructed from the perturbation theory of the WF-based tight-binding Hamiltonian.
Second, consider subspace \circled{3}, which includes the states inside the outer window but not selected for Wannierization. To accurately describe the wavefunction perturbation, these states should be included in the WFPs.
However, the contribution of these states to the perturbation of eigenstates in subspace \circled{2} might diverge because the separation of subspaces \circled{2} and \circled{3} is done by Wannierization not by the energy eigenvalues.
In \myeqref{eq:wfpt_WFP}, this problem is absent as the contribution of subspace \circled{3} is included only if $\mk \in \WP$.
From \myeqref{eq:wfpt_wfp_second}, one can find that this condition ensures that the denominator $\veps^\ord{0}_\mk - \veps^\ord{0}_\npkp$ is not zero because states $\mk$ and $\npkp$ are inside subspaces \circled{1} and \circled{3}, respectively.
The last subspace, subspace \circled{4}, is energetically separated from other states by the outer window $\WD$.
Its contribution to the wavefunction perturbation is included for all states as shown in the first term of \myeqref{eq:wfpt_WFP}.
The separation of energy ranges by the windows is the key idea in WFPT which makes the WFPs spatially localized.

\subsubsection{Spatial localization of the WFPs}
The most important property of the WFPs is their spatial localization.
We provide two arguments for the localization of the WFPs.
First, the localization of WFPs follows from the localization of the perturbed WFs.
There exists much empirical evidence that the WFs are well localized in real space if a proper initial guess is used~\cite{2001HeWannier,2007YatesWannier,2012MarzariRMP}.
The initial guesses we used for the WFs of the perturbed system are the MLWFs of the unperturbed system, which is an ideal guess in the $\lambda \rightarrow 0$ limit.
Thus, it is very likely that the perturbed WFs are also localized, at least for small $\lambda$.
Since the WFPs are the derivative of the WFs at $\lambda=0$,
the localization of the WFs for a finite range of $\lambda$ including $\lambda=0$ implies the localization of the WFPs.

The second argument is that the summands of \myeqref{eq:wfpt_WFP} are almost smooth functions of $\mb{k}$.
The summand being smooth in reciprocal space implies localization in the real space.
We emphasize that the condition $\mk \in \WP$ in the second term of \myeqref{eq:wfpt_WFP} plays a crucial role.
If this condition is absent, the denominator $\veps^\ord{0}_\mk - \veps^\ord{0}_\npkp$ in \myeqref{eq:wfpt_wfp_second} can be zero, leading to a divergence in the summand and rendering the WFPs delocalized.
In our construction, the denominator never becomes zero because $\veps^\ord{0}_\mk$ is forced to be inside $\WP$ (subspace \circled{1}), while $\veps^\ord{0}_\npkp$ is forced to be in subspace \circled{3}.

Although these two arguments are not rigorous in a mathematical sense, they provide a solid justification that the WFPs are spatially localized.
We note that the localization of disentangled MLWFs is also an empirical fact without a mathematical proof~\cite{2012MarzariRMP,2019CorneanWannier}.
Later, we numerically demonstrate that the WFPs and the matrix elements involving WFPs are localized.

One source of singularity in \myeqref{eq:wfpt_WFP} is the discontinuity that occurs when $\veps^\ord{0}_\mk$ crosses $\WP$. In this case, the summand may change discontinuously because of the condition $\mk \in \WP$, although it does not diverge.
This discontinuity can be removed by smoothing the inner window, as described in Appendix~\ref{sec:app_smooth_window}.
However, for the systems studied in this work, we find that the WFPs are already sufficiently localized without this additional step.
Hence, we use the WFPs as defined in \myeqref{eq:wfpt_WFP}.

We also note that if two bands lie close to the boundary of the outer window from inside and outside, respectively, at some $k$ point, the first term of \myeqref{eq:wfpt_WFP} can give a large contribution rapidly changing with $\mb{k}$.
In such a rare case, one can locally, i.e., at that particular $k$ point, extend the outer window to include the band outside the original outer window to make the summand change slowly.
The local extension of the outer window can be done just for the construction of WFPs without affecting the unperturbed WFs.
In the applications studied in this paper, we did not perform this modification as the WFPs were already sufficiently localized.

\subsection{Interpolation of wavefunction perturbation}
Next, we study how to interpolate the wavefunction perturbation of Bloch eigenstates using the WFPs.
In this subsection, we first construct the perturbed tight-binding model using the matrix elements of the perturbed Hamiltonian in the perturbed WF basis.
Then, we use first-order perturbation theory to obtain the eigenvector of this tight-binding model.
Finally, we calculate the wavefunction perturbation.

\subsubsection{Perturbed tight-binding Hamiltonian}
Let us calculate the Hamiltonian matrix elements of the perturbed system in the perturbed WF basis.
The Hamiltonian matrix element is
\begin{align} \label{eq:int_hop}
    H_{\ipRp,\iR}
    =& \mel{w_\ipRp}{(\opH\oord{0} + \lambda \opV)}{w_\iR} \nnnl
    =& H^\ord{0}_{\ipRp,\iR}
    + \lambda H^\ord{1}_{\ipRp,\iR}
    + \mcO(\lambda^2),
\end{align}
where
\begin{equation} \label{eq:int_hop_0}
    H^\ord{0}_{\ipRp,\iR}
    = \mel{w^\ord{0}_\ipRp}{\opH^\ord{0}}{w^\ord{0}_{\iR}}
\end{equation}
is the unperturbed Hamiltonian matrix element and
\begin{align} \label{eq:int_hop_1}
    H^\ord{1}_{\ipRp,\iR}
    =& \mel{w^\ord{0}_\ipRp}{\opV}{w^\ord{0}_{\iR}} \\
    +& \mel{w^\ord{1}_\ipRp}{\opH^\ord{0}}{w^\ord{0}_{\iR}}
    + \mel{w^\ord{0}_\ipRp}{\opH^\ord{0}}{w^\ord{1}_{\iR}} \nonumber
\end{align}
is the perturbative correction.
Note that $H\oord{1}_{\ipRp,\iR}$ contains a direct contribution from $\opV$ as well as indirect contributions from the change of the WFs.

We write the perturbation $\opV$ as a sum of monochromatic perturbations
\begin{equation} \label{eq:int_Vq_expansion}
    \opV = \sum_\mb{q} \opV_\mb{q},
\end{equation}
where each monochromatic perturbation satisfies
\begin{equation} \label{eq:int_Vq_condition}
    \mel{\mb{r+R}}{\opV_\mb{q}}{\mb{r'+R}}
    = \expixy{\mb{q}}{\mb{R}} \mel{\mb{r}}{\opV_\mb{q}}{\mb{r'}}
\end{equation}
for every lattice vector $\mb{R}$.
Correspondingly, the WFPs can be written as
\begin{equation}
    \ket{w\oord{1}_\iR} = \sum_\mb{q} \ket{w\oord{1}_{\iR;\mb{q}}},
\end{equation}
where
\begin{equation} \label{eq:rs_WFP_trans}
    \braket{\mb{r+R'}}{w\oord{1}_{i\mb{R+R'};\mb{q}}}
    = \expixy{\mb{q}}{\mb{R'}} \braket{\mb{r}}{w\oord{1}_{i\mb{R};\mb{q}}}.
\end{equation}
We also define real-space matrix elements
\begin{equation} \label{eq:int_g_def}
    g_{ij\mb{R};\mb{q}}
    \equiv \mel{w\oord{0}_{i\mb{0}}}{\opV_\mb{q}}{w\oord{0}_{j\mb{R}}},
\end{equation}
\begin{equation} \label{eq:int_deltag_def}
    \delta g_{ij\mb{R};\mb{q}}
    \equiv
    \mel{w^\ord{0}_{i\mb{0}}}{\opH^\ord{0}}{w^\ord{1}_{\jR;\mb{q}}}
    + \mel{w^\ord{1}_{i\mb{0};\mb{-q}}}{\opH^\ord{0}}{w^\ord{0}_\jR},
\end{equation}
and
\begin{equation} \label{eq:int_gtilde_def}
    \widetilde{g}_{ij\mb{R};\mb{q}}
    \equiv g_{ij\mb{R};\mb{q}} + \delta g_{ij\mb{R};\mb{q}}.
\end{equation}
Here, $g_{\mb{R};\mb{q}}$ is the usual WF-based matrix element of $\opV_\mb{q}$, and $\delta g_{\mb{R};\mb{q}}$ is a correction originating from the change of the WFs.
This correction term is not needed for calculating the scattering matrix elements, but it is necessary when calculating the perturbed wavefunctions.
We define the Fourier transformation of $g_{\mb{R};\mb{q}}$ as
\begin{equation}
    g^\mathrm{(W)}_{ij\mb{k};\mb{q}}
    = \sum_{\mb{R}} e^{i\mb{k}\cdot\mb{R}} g_{ij\mb{R};\mb{q}},
\end{equation}
and analogously, we define $\widetilde{g}^\mathrm{(W)}_{\mb{k};\mb{q}}$.

\subsubsection{Perturbed tight-binding eigenvector}
The Hamiltonian of the perturbed system in the Bloch representation is
\begin{equation} \label{eq:int_HW0_def}
    H\oord{\rm W}_{i'\mb{k'},i\mb{k}}
    \equiv \frac{1}{N_k} \sum_{\mb{R}, \mb{R'}} H_{\ipRp,\iR} \expmixy{\mb{k'}}{\mb{R'}} \expixy{\mb{k}}{\mb{R}}.
\end{equation}
Since the unperturbed system has lattice translation symmetry, the zeroth-order part of $H\oord{\rm W}$ is block diagonal:
\begin{equation} \label{eq:int_HW0_FT}
    H\oord{\rm W0}_{i'\mb{k'},i\mb{k}}
    = \delta_{\mb{k},\mb{k'}} H\oord{\rm W0}_{i'i;\mb{k}}
    = \delta_{\mb{k},\mb{k'}} \sum_\mb{R} H\oord{0}_{i'\mb{0},i\mb{R}} e^{i\mb{k}\cdot\mb{R}}.
\end{equation}
The unperturbed Hamiltonian $H\oord{\rm W0}_{\mb{k}}$ can be diagonalized as
\begin{equation} \label{eq:int_HW0_diag}
    (V^{(0)\dagger}_{\mb{k}} H\oord{\rm W0}_\mb{k} V\oord{0}_\mb{k})_{mn}
    = \veps\oord{\rm H0}_\mk \delta_{m,n},
\end{equation}
where $V\oord{0}_\mb{k}$ is an $\NWan \times \NWan$ unitary matrix.
In the Born-von Karman supercell, the unperturbed eigenvector is
\begin{equation}
    V\oord{0}_{\iR,\nk} = \frac{1}{\sqrt{N_k}} e^{i\mb{k}\cdot\mb{R}} V\oord{0}_{in;\mb{k}}.
\end{equation}

Next, we consider the perturbative correction.
Using Eqs.~(\ref{eq:int_Vq_condition}, \ref{eq:rs_WFP_trans}), one can show that the Fourier transformation of $H\oord{1}_{\ipRp,\iR}$ is
\begin{align} \label{eq:int_HW1}
    H^\ord{\rm W1}_{i'\mb{k'},i\mb{k}}
    =& \widetilde{g}^\mathrm{(W)}_{i'i\mb{k};\mb{k'-k}}
\end{align}
(see Appendix~\ref{sec:app_wf_interpol} for a proof).
The matrix element of $H\oord{\rm W1}$ in the unperturbed eigenstate basis is
\begin{equation} \label{eq:int_H_H1}
    H^\ord{H1}_{\npkp,\nk}
    = \left( V^{\ord{0}\dagger}_{\mb{k'}} \widetilde{g}^\mathrm{(W)}_{\mb{k};\mb{k'-k}} V^\ord{0}_{\mb{k}} \right)_{n'n}
    \equiv \widetilde{g}^\ord{H}_{n'n\mb{k};\mb{k'-k}}.
\end{equation}
We also define $g^\ord{H}$ analogously.

We now apply first-order perturbation theory to the tight-binding Hamiltonian.
The perturbed eigenvector in the WF basis is
\begin{align}
    &V_{\iR,\nk}
    = \frac{1}{\sqrt{N_k}} e^{i\mb{k}\cdot\mb{R}} V^\ord{0}_{in;\mb{k}} \\
    &+ \lambda \frac{1}{\sqrt{N_k}} \sum_\mb{k'} \primesum{n'=1}^{\NWan}
    e^{i\mb{k'}\cdot\mb{R}} V^\ord{0}_{in';\mb{k'}} \frac{\widetilde{g}^\ord{H}_{n'n\mb{k};\mb{k'-k}}}{\veps^\ord{H0}_\nk - \veps^\ord{H0}_\npkp}
    + \mcO(\lambda^2). \nonumber
\end{align}

\subsubsection{Calculation of wavefunction perturbation}
We can now calculate the wavefunction perturbation as we know the basis functions (the perturbed WF $\ket{w_\iR}$) as well as the coefficients (the perturbed eigenvector $V_{\iR,\nk}$) for the perturbed wavefunction.
The perturbed wavefunction reads
\begin{align} \label{eq:int_wf}
    \ket{\psi^\ord{\rm H}_\nk}
    =& \sum_{i,\mb{R}} \ket{w_\iR} V_{\iR,\nk} \nnnl
    =& \ket{\psi^\ord{\rm H0}_\nk}
    + \lambda \sum_\mb{q} \ket{\psi^\ord{\rm H1}_{\nk;\mb{q}}}
    + \mcO(\lambda^2),
\end{align}
where
\begin{equation} \label{eq:int_wf_H0}
    \ket{\psi^\mathrm{(H0)}_\nk}
    = \frac{1}{\sqrt{N_k}} \sum_{i,\mb{R}} \ket{w^\mathrm{(0)}_\iR} V^{(0)}_{in,\mb{k}} e^{i\mb{k}\cdot\mb{R}}
\end{equation}
and
\begin{align} \label{eq:int_wf_H1}
    \ket{\psi^\ord{\rm H1}_{\nk;\mb{q}}}
    =& \frac{1}{\sqrt{N_k}} \sum_{i,\mb{R}} e^{i\mb{k}\cdot\mb{R}}
    \ket{w^\ord{1}_{\iR;\mb{q}}} V^\ord{0}_{in;\mb{k}} \nnnl
    +& \primesum{m=1}^{\NWan} \ket{\psi^\ord{H0}_{m\mb{k+q}}} \frac{\widetilde{g}^\ord{H}_{mn\mb{k};\mb{q}}}{\veps^\ord{H0}_\nk - \veps^\ord{H0}_{m\mb{k+q}}}.
\end{align}
Equation \eqref{eq:int_wf_H1} contains contributions from both the WFP and the perturbation of the tight-binding eigenvector $V^\ord{0}_{in;\mb{k}}$.

Equation \eqref{eq:int_wf_H1} is the interpolation equation for the first-order wavefunction perturbation.
This equation is the second main theoretical result of this paper.
The contribution of the states outside $\opPW$, the unperturbed Wannier subspace, to the wavefunction perturbation is taken into account by the WFPs and is the first term of \myeqref{eq:int_wf_H1}.
The contribution of the states inside $\opPW$ is the second term, which is derived by the perturbation theory in the tight-binding model.
Since the WFPs are spatially localized, the wavefunction perturbation converges quickly with the number of lattice vectors used in the sum over $\mb{R}$.

\subsection{Interpolation of matrix elements}

Finally, we demonstrate how the matrix elements that involve wavefunction perturbations can be interpolated.
We consider interpolation of two different types of quantities, $S_{mn\mb{k}}$ [\myeqref{eq:int_s_def}] and $K_{mn\mb{k}}$ [\myeqref{eq:kubo_k_def}].

\begin{table*}[]
\caption{Key quantities of WFPT.}
\begin{tabular}{|c|c|c|}
\hline
Description &
    Real space, Wannier basis &
    Reciprocal space, eigenstate basis \\ \hline
Wavefunction perturbation &
    $\ket{w\oord{1}_{i\mb{R}}}$, \myeqref{eq:wfpt_WFP} &
    $\ket{\psi\oord{H1}_{\nk;\mb{q}}}$, \myeqref{eq:int_wf_H1} \\ \hline
\begin{tabular}[c]{@{}c@{}}Perturbation to the\\ Hamiltonian matrix elements\end{tabular} &
\multicolumn{1}{c|}{\begin{tabular}[c]{@{}c@{}}
    WF contribution: $g_{ij\mb{R};\mb{q}}$, \myeqref{eq:int_g_def} \\
    WFP contribution: $\delta g_{ij\mb{R};\mb{q}}$, \myeqref{eq:int_deltag_def} \\
    Total: $\widetilde{g}_{ij\mb{R};\mb{q}}$, \myeqref{eq:int_gtilde_def} \end{tabular}} &
\multicolumn{1}{c|}{$H\oord{H1}_{\npkp,\nk}$, \myeqref{eq:int_H_H1}} \\ \hline
Sternheimer matrix elements &
    $s_{ij\mb{R};\mb{q}}$, \myeqref{eq:int_s_R_def} &
    $S_{mn\mb{k}}$, Eqs.~(\ref{eq:int_s_def}, \ref{eq:int_s_wfp_derivation}) \\ \hline
Kubo matrix elements &
    $k_{ij\mb{R};\mb{q}}$, \myeqref{eq:kubo_k_R_def} &
    $K_{mn\mb{k}}$, Eqs.~(\ref{eq:kubo_k_def}, \ref{eq:kubo_k_wfp}) \\ \hline
\end{tabular}
\label{table:key_quantities}
\end{table*}

\subsubsection{Interpolation of $S_{mn\mathbf{k}}$}
First, we consider quantities of the form
\begin{align} \label{eq:int_s_def}
    S_{mn\mb{k}}
    =& \mel{\psi^\ord{1}_{\mk;\mb{q}}}{\opV'_\mb{q}}{\psi^\ord{0}_\nk} \nnnl
    =& \primesum{p}
    \frac{
        \mel{\psi^\ord{0}_\mk}{\opV_\mb{q}^\dagger}{\psi^\ord{0}_{p\mb{k+q}}}
        \mel{\psi^\ord{0}_{p\mb{k+q}}}{\opV'_\mb{q}}{\psi^\ord{0}_\nk}
    }{\veps^\ord{0}_\mk - \veps^\ord{0}_{p\mb{k+q}}},
\end{align}
where $\opV'_\mb{q}$ is a monochromatic perturbation that may or may not equal $\opV_\mb{q}$.
If $\opV_\mb{q}$ is a phonon potential, interpolation of $S_{\mb{k}}$ enables an efficient calculation of the phonon-induced renormalization of the electron band structure.
For $\opV_\mb{0}$ being the spin-velocity operator and $\opV_{\mb{q}\neq\mb{0}}=0$, $S_\mb{k}$ is relevant to the shift spin current.

For $\mk,\nk \in \WP$, one can use the Wannier interpolation of the unperturbed wavefunction and the wavefunction perturbation [Eqs.~(\ref{eq:int_wf_H0}, \ref{eq:int_wf_H1})] to find
\begin{widetext}
\begin{align} \label{eq:int_s_wfp_derivation}
    S_{mn\mb{k}}
    =& \mel{\psi^\ord{H1}_{\mk;\mb{q}}}{\opV'_\mb{q}}{\psi^\ord{H0}_{\nk}} \nnnl
    =& \frac{1}{N_k} \sum_{i,i',\mb{R},\mb{R'}}
    V^{(0)\dagger}_{mi',\mb{k}} \mel{w^\ord{1}_{\ipRp;\mb{q}}}{\opV'_\mb{q}}{w^\ord{0}_\iR} V^\ord{0}_{in;\mb{k}}
    e^{i\mb{k}\cdot(\mb{R-R'})}
    + \primesum{p=1}^{\NWan} \frac{\left( \widetilde{g}^\ord{H}_{pm\mb{k};\mb{q}} \right)^* \mel{\psi^\ord{H0}_{p\mb{k+q}}}{\opV'_\mb{q}}{\psi^\ord{H0}_\nk}}{\veps^\ord{H0}_\mk - \veps^\ord{H0}_{p\mb{k+q}}} \nnnl
    =& \sum_{i,i',\mb{R}}
    V^{(0)\dagger}_{mi',\mb{k}} \mel{w^\ord{1}_{i'\mb{0};\mb{q}}}{\opV'_\mb{q}}{w^\ord{0}_{\iR}} V^\ord{0}_{in;\mb{k}}
    e^{i\mb{k}\cdot\mb{R}}
    + \primesum{p=1}^{\NWan} \frac{\left( \widetilde{g}^\ord{H}_{pm\mb{k};\mb{q}} \right)^* {g'}^\ord{H}_{pn\mb{k};\mb{q}}}{\veps^\ord{H0}_\mk - \veps^\ord{H0}_{p\mb{k+q}}}
\end{align}
\end{widetext}
In the third equality, we use the discrete translation properties of $\opV'_\mb{q}$ and the WFPs [Eq.~(\ref{eq:int_Vq_condition}, \ref{eq:rs_WFP_trans})].

We define the real-space matrix element
\begin{equation} \label{eq:int_s_R_def}
    s_{ij\mb{R};\mb{q}}
    \equiv \mel{w^\ord{1}_{i\mb{0};\mb{q}}}{\opV'_\mb{q}}{w^\ord{0}_{\jR}}.
\end{equation}
Since the WFs and the WFPs are localized, $s_{ij\mb{R};\mb{q}}$ rapidly decays as a function of $\abs{\mb{R}}$.
Defining
\begin{equation} \label{eq:int_s_RtoW}
    s^\ord{W}_{ij\mb{k};\mb{q}}
    \equiv \frac{1}{N_k} \sum_{\mb{R}} s_{ij\mb{R};\mb{q}} e^{i\mb{k}\cdot\mb{R}}
\end{equation}
and
\begin{equation} \label{eq:int_s_WtoH}
    s^\ord{H}_{mn\mb{k};\mb{q}}
    \equiv \left( V^{(0)\dagger}_{\mb{k}} s^\ord{W}_{\mb{k};\mb{q}} V^{(0)}_{\mb{k}} \right)_{mn},
\end{equation}
one can rewrite \myeqref{eq:int_s_wfp_derivation} as
\begin{align} \label{eq:int_s_wfp}
    S_{mn\mb{k}}
    = s^\ord{H}_{mn\mb{k};\mb{q}}
    + \primesum{p=1}^{\NWan} \frac{\left( \widetilde{g}^\ord{H}_{pm\mb{k};\mb{q}} \right)^* {g'}^\ord{H}_{pn\mb{k};\mb{q}}}{\veps^\ord{H0}_\mk - \veps^\ord{H0}_{p\mb{k+q}}}.
\end{align}
Since all quantities appearing in \myeqref{eq:int_s_wfp} can be calculated using a WF- or WFP-based interpolation, $S_{mn\mb{k}}$ can also be interpolated.

\subsubsection{Interpolation of $K_{mn\mathbf{k}}$}
Next, we show that the matrix element
\begin{align} \label{eq:kubo_k_def}
    K_{mn\mb{k}}
    =& \braket{\psi^\ord{1}_{\mk;\mb{q}}}{\psi'^{(1)}_{\nk;\mb{q}}} \\
    =& \primesum{p}
    \frac{
        \mel{\psi^\ord{0}_\mk}{\opV_\mb{q}^\dagger}{\psi^\ord{0}_{p\mb{k+q}}}
        \mel{\psi^\ord{0}_{p\mb{k+q}}}{\opV'_\mb{q}}{\psi^\ord{0}_\nk}
    }{(\veps^\ord{0}_\mk - \veps^\ord{0}_{p\mb{k+q}})(\veps^\ord{0}_\nk - \veps^\ord{0}_{p\mb{k+q}})}. \nonumber
\end{align}
can similarly be interpolated using WFPT.
This matrix element is relevant to the evaluation of the Kubo formula for noninteracting quasiparticles.
When $\mb{q}=0$ and $\opV_\mb{0}$ and $\opV'_\mb{0}$ are the velocity and spin-velocity operators, respectively, $K_{nn\mb{k}}$ is directly related to the spin Berry curvature and the spin Hall conductivity.

Defining the real-space matrix element
\begin{equation} \label{eq:kubo_k_R_def}
    k_{ij\mb{R};\mb{q}} \equiv \braket{w^{(1)}_{i\mb{0};\mb{q}}} {w'^{(1)}_{\jR;\mb{q}}},
\end{equation}
we show in Appendix~\ref{sec:app_interp_kubo_formula} that $K_{mn\mb{k}}$ can be interpolated as
\begin{align} \label{eq:kubo_k_wfp}
    K_{mn\mb{k}}
    = k^\ord{H}_{mn\mb{k};\mb{q}}
    + \primesum{p=1}^{\NWan}\frac{\left( \widetilde{g}^{\rm (H)}_{pm\mb{k};\mb{q}} \right)^* \widetilde{g}'^{\rm (H)}_{pn\mb{k};\mb{q}}}{(\veps^\ord{H0}_\mk - \veps^\ord{H0}_\pkq)(\veps^\ord{H0}_\nk - \veps^\ord{H0}_\pkq)}.
\end{align}
Here, we define
\begin{equation} \label{eq:kubo_k_WtoH}
    k^\ord{H}_{mn\mb{k};\mb{q}}
    \equiv \left( V^{(0)\dagger}_{\mb{k}} k^\ord{W}_{\mb{k};\mb{q}} V^{(0)}_{\mb{k}} \right)_{mn}
\end{equation}
and
\begin{equation} \label{eq:kubo_k_RtoW}
    k^\ord{W}_{ij\mb{k};\mb{q}}
    \equiv \frac{1}{N_k} \sum_{\mb{R}} k_{ij\mb{R};\mb{q}} e^{i\mb{k}\cdot\mb{R}}\,.
\end{equation}
In the actual evaluation of response quantities, positive infinitesimal $\eta$ that ensures the causality of the response is included in the denominator of the low-energy-band contribution, as in \myeqref{eq:s_lofan_def} and \myeqref{eq:shc_berry}.
Numerically, the value of $\eta$ is decreased until the calculated quantity converges.

In Appendix~\ref{sec:app_real_space}, we detail how to calculate the real-space matrix elements $g_{ij\mb{R};\mb{q}}$, $\widetilde{g}_{ij\mb{R};\mb{q}}$, $s_{ij\mb{R};\mb{q}}$, and $k_{ij\mb{R};\mb{q}}$ from the Bloch eigenstates at the coarse $k$ grid.
Table~\ref{table:key_quantities} summarizes the key quantities of WFPT.

\subsection{Proof-of-principles example: Berry curvature} \label{sec:theory_example}

\begin{figure}[htbp]
\includegraphics[width=1.0\columnwidth]{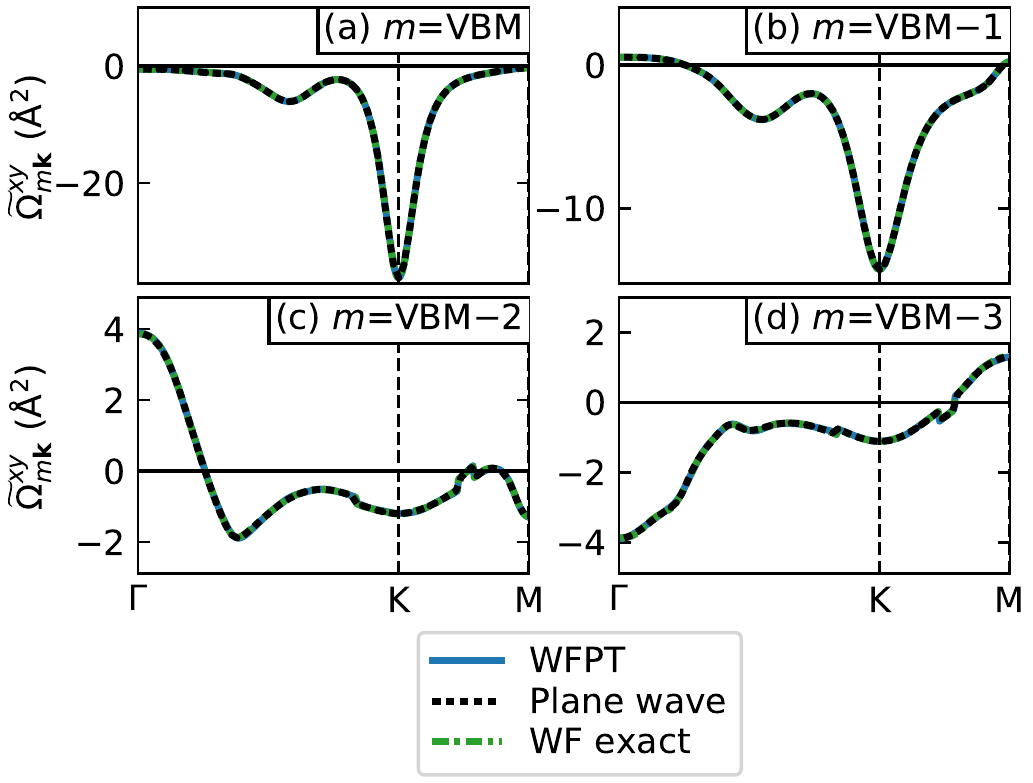}
\caption{
Band-resolved Berry curvature of monolayer WTe$_2$. VBM stands for the highest-energy valence band, or valence band maximum.
The plane-wave calculation is done by solving the Sternheimer equation, \myeqref{eq:kubo_dpsi_dk_sternheimer}, to directly evaluate \myeqref{eq:kubo_omega_tilde_def}.
}
\label{fig:wte2_bcurv}
\end{figure}

Now, let us present a proof-of-principles example of WFPT interpolation by interpolating the band-resolved Berry curvature.
Since the Berry curvature strongly oscillates in momentum space due to the contribution from nearby bands, we exclude the intra-valence-band contribution as follows:
\begin{equation} \label{eq:kubo_omega_tilde_def}
    \widetilde{\Omega}_\mk^{ab}
    = -2 \Im \mel{\frac{\partial u_\mk}{\partial k^a}}{\opQ_{\mb{k},{\rm valence}}}{\frac{\partial u_\mk}{\partial k^b}}\,.
\end{equation}
Here, $\ket{u_\mk} = e^{-i\mb{k}\cdot\hat{\mb{r}}} \ket{\psi_\mk}$ and $\opQ_{\mb{k},{\rm valence}} = \hat{\one} - \opP_{\mb{k},{\rm valence}}$, with $\opP_{\mb{k},{\rm valence}} = \sum_{n \in {\rm valence}} \ket{u_\nk} \bra{u_\nk}$ the projection operator to the valence bands at $\mb{k}$.
By ``valence bands,'' we mean only the occupied bands inside the outer window for Wannierization, $\WD$: The low-lying bands that are outside $\WD$ are not included in $\opP_{\mb{k},{\rm valence}}$.

The Berry curvature can be calculated using WFPT interpolation.
Since the velocity operator is the derivative of the Hamiltonian with respect to $\mb{k}$, the derivative of the wavefunction is equivalent to the wavefunction perturbation due to a velocity operator perturbation.
Hence, by choosing the velocity operators as the perturbations for the WFPs, $\hat{V}_\mb{q} = \hat{v}^a \delta_{\mb{q}, \mb{0}}$ and $\hat{V}'_\mb{q} = \hat{v}^b \delta_{\mb{q}, \mb{0}}$, we have
\begin{equation}
    \ket{\psi^{(1)}_\mk} = e^{i\mb{k}\cdot\mb{r}} \ket{\frac{\partial u_\mk}{\partial k^a}}
    \text{ and }
    \ket{\psi^{'(1)}_\mk} = e^{i\mb{k}\cdot\mb{r}} \ket{\frac{\partial u_\mk}{\partial k^b}}.
\end{equation}
Using \myeqref{eq:kubo_k_def} and \myeqref{eq:kubo_k_wfp}, we find
\begin{widetext}
\begin{align} \label{eq:kubo_Omega_tilde_k}
    \widetilde{\Omega}_\mk^{ab}
    =& -2 \Im \left( \braket{\frac{\partial u_\mk}{\partial k^a}}{\frac{\partial u_\mk}{\partial k^b}}
    - \sum_{n \in {\rm valence}} \braket{\frac{\partial u_\mk}{\partial k^a}}{u_\nk}
    \braket{u_\nk}{\frac{\partial u_\mk}{\partial k^b}} \right) \nnnl
    =& -2 \Im \left( \braket{\psi^{(1)}_\mk}{\psi^{'(1)}_\mk} - \sum_{n \in {\rm valence}} \braket{\psi^{(1)}_\mk}{\psi^{(0)}_\nk} \braket{\psi^{(0)}_\nk}{\psi^{'(1)}_\mk} \right) \nnnl
    =& -2\Im \left( K_{mm\mb{k}} - \primesum{{\substack{n=1 \\ n \in {\rm valence}}}}^{\NWan}\frac{\left( g^{\rm (H)}_{nm\mb{k}} \right)^* g'^{{\rm (H)}}_{nm\mb{k}}}{(\veps^\ord{H0}_\mk - \veps^\ord{H0}_\nk)^2} \right) \nnnl
    =& -2 \Im \left( k^{\mathrm{(H)}}_{mm\mb{k}}
    + \primesum{{\substack{n=1 \\ n \notin {\rm valence}}}}^{\NWan}\frac{\left( \widetilde{g}^{{\rm (H)}}_{nm\mb{k}} \right)^* \widetilde{g}'^{{\rm (H)}}_{nm\mb{k}}}{(\veps^\ord{H0}_\mk - \veps^\ord{H0}_\nk)^2} \right).
\end{align}
\end{widetext}
This equation is the WFPT interpolation formula for $\widetilde{\Omega}_\mk^{ab}$.
In the last equality, we used the fact that in the limit of exact interpolation, $g^{\rm (H)}_{nm\mb{k}} = \widetilde{g}^{\rm (H)}_{nm\mb{k}}$ holds for $\mk, \nk \in \WP$, which can easily be shown to follow from the $P^\ord{0}_\mathrm{D} - P^\ord{0}_\mathrm{W}$ terms in \myeqref{eq:k2r_deltag}.
Because of the interpolation error, $\delta g^\ord{H}_{nm\mb{k}}=0$ holds exactly only if $\mb{k}$ belongs to the coarse $k$-point grid for Wannier interpolation.
But, the deviation at a generic $\mb{k}$ is small and can be diminished by using a denser coarse $k$-point grid for Wannierization, as with all other Wannier-interpolated quantities.

We also calculate the Berry curvature directly in the plane-wave basis by solving the following Sternheimer equation to calculate $\hat{Q}_{\mb{k},\rm{valence}} \ket{\frac{\partial u_\mk}{\partial k^a}}$:
\begin{align} \label{eq:kubo_dpsi_dk_sternheimer}
    &\left( \opH(\mb{k}) - \veps_\mk \right) \hat{Q}_{\mb{k},\rm{valence}} \ket{\frac{\partial u_\mk}{\partial k^a}} \nnnl
    =& -\left( \sum_{\substack{n=1 \\ n \notin {\rm valence}}}^{\infty} \ket{u_{n\mb{k}}}\bra{u_{n\mb{k}}} \right) \frac{\partial \opH(\mb{k})}{\partial k^a} \ket{u_\mk},
\end{align}
where $\opH(\mb{k}) = e^{-i\mb{k}\cdot\hat{\mb{r}}} \opH e^{i\mb{k}\cdot\hat{\mb{r}}}$.
The conventional Wannier interpolation method is also exact for the calculation of Berry curvature~\cite{2006WangAHC}.
There, $\widetilde{\Omega}_\mk^{ab}$ is calculated by first calculating the full Berry curvature and then subtracting the intra-valence-band contribution.

Figure~\ref{fig:wte2_bcurv} shows the Berry curvature of four valence bands of monolayer WTe$_2$ calculated using WFPT interpolation, direct plane-wave calculations, and exact conventional Wannier interpolation.
The WFPT result agrees perfectly with the plane-wave and Wannier interpolation results, demonstrating the validity of the WFPT interpolation method.

\section{Results}
\subsection{Application 1: Temperature-dependent electron band structure and indirect absorption}

\begin{figure*}[htbp]
\includegraphics[width=0.9\textwidth]{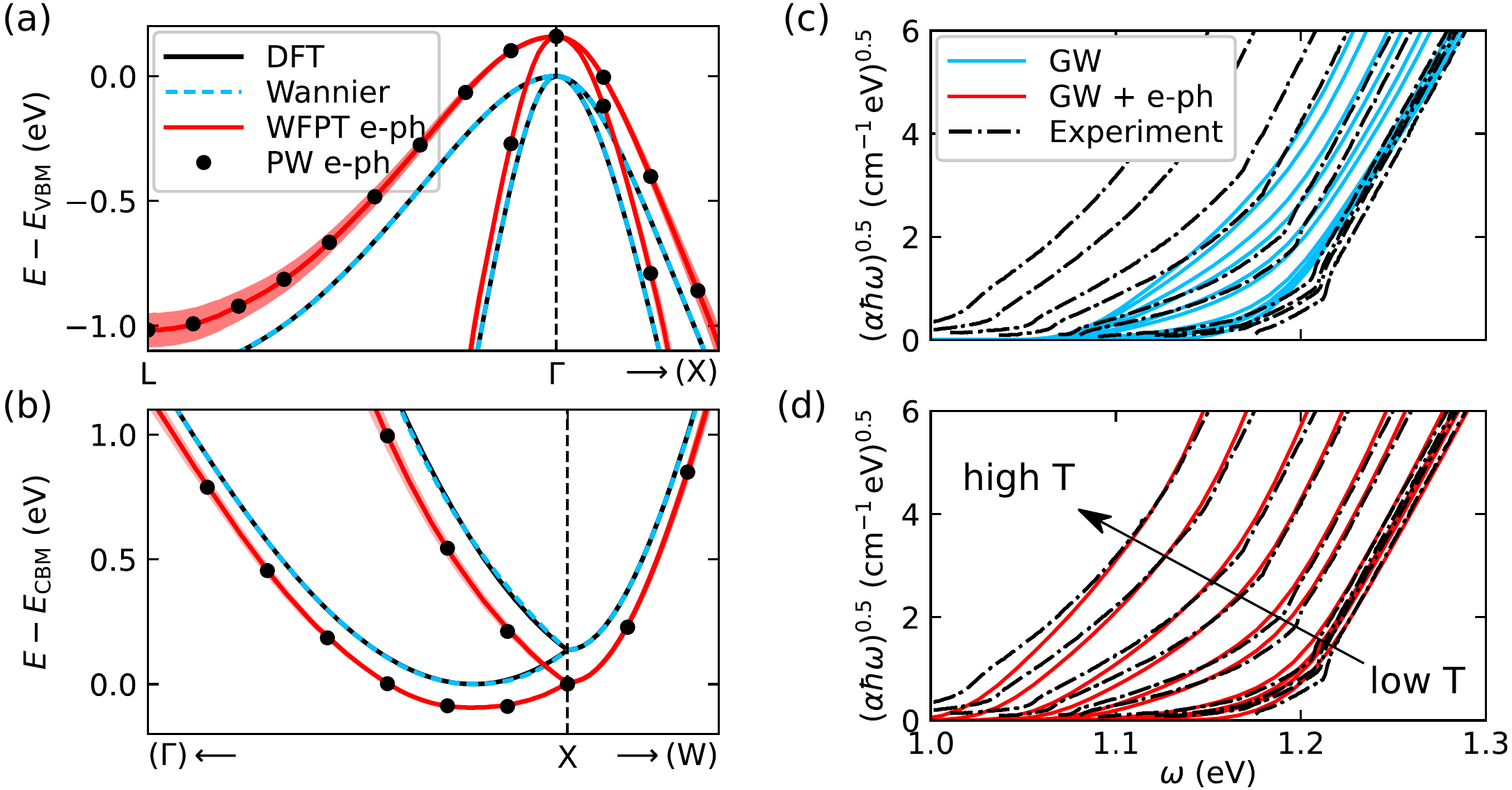}
\caption{
(a, b) Band structure of silicon around the (a) VBM and (b) conduction band minimum (CBM) computed without and with the EPC at $T$=1000~K. The red shades around the red curves indicate the phonon-induced broadening, with the width the same as the imaginary part of the self-energy.
(c, d) Indirect absorption spectra of silicon computed (c) without and (d) with the EPC-induced band renormalization, and the experimental data from Ref.~\cite{1958MacfarlaneExpZPR}. The curves correspond to temperatures 20, 77, 90, 112, 170, 195, 249, 291, 363, and 415~K, from right to left.}
\label{fig:si_indabs}
\end{figure*}

\begin{figure}[htbp]
\includegraphics[width=1.0\columnwidth]{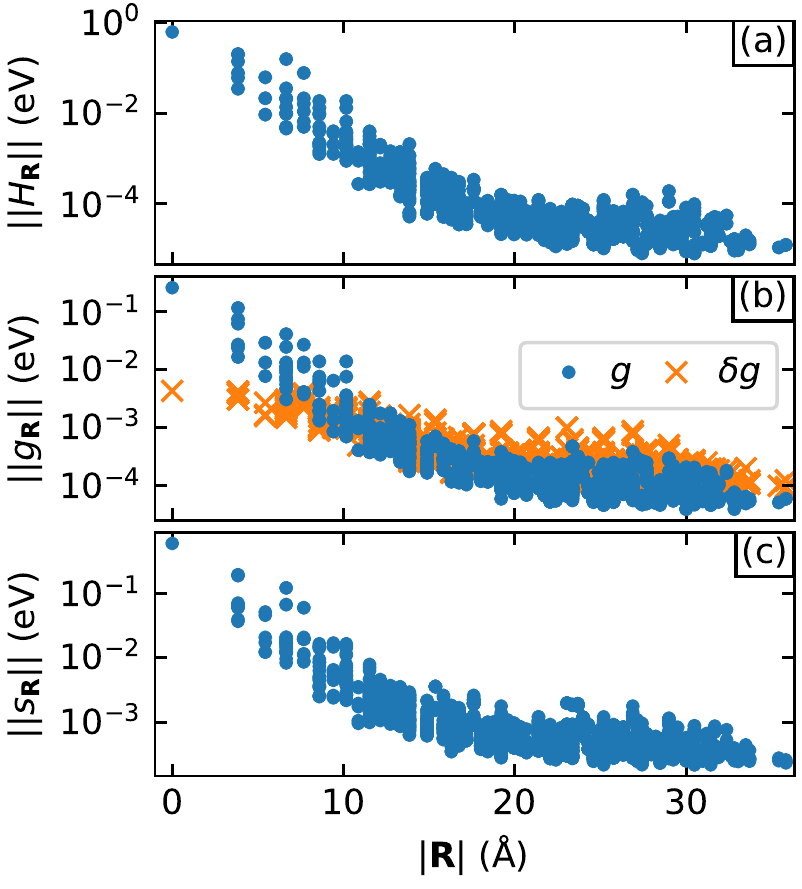}
\caption{
Spatial decay of (a) Hamiltonian, (b) EPC, and (c) WFP matrix elements.
Each data point corresponds to the maximum absolute value over all WF pairs, phonon momentum, and atomic displacements for each lattice vector $\mb{R}$. The $y$ axes are on a logarithmic scale.
}
\label{fig:si_decay}
\end{figure}

\begin{table}[htbp]
\centering
\caption{Time required to calculate the phonon-induced electron self-energy for 379 $k$ points using direct plane-wave calculations and WFPT interpolation.}
\begin{tabular}{c|c|c} 
    \hline
    & \multicolumn{2}{c}{Time (cpu$\cdot$h)} \\ \cline{2-3}
    & Plane wave & WFPT \\ \hline
    SCF, DFPT (coarse $q$) & 269 & 269 \\ 
    NSCF + Sternheimer (coarse $k$, coarse $q$) & - & 424  \\
    Wannierization & - & 171 \\
    WFPT interpolation (fine $k$, coarse $q$) & - & 23 \\
    NSCF + Sternheimer (fine $k$, coarse $q$) & 1,241$^*$ & - \\
    Lower Fan (fine $k$, fine $q$) & 4,450,591$^*$ & 2,846 \\
    \hline
    Total & 4,452,101$^*$ & 3,733 \\
    \hline
\end{tabular} \\
\footnotesize{$^*$ Estimated from a calculation with smaller numbers of $k$ and $q$ points.}
\label{table:si_time}
\end{table}

As the first application of WFPT, we calculate the temperature-dependent electronic structure and indirect optical absorption spectra of silicon.
The electron self-energy in the dynamical Allen-Heine-Cardona formalism~\cite{1976Allen,1981Allen,1983Allen} is the sum of the Fan and the Debye-Waller (DW) terms~\cite{2017GiustinoRMP}:
\begin{equation} \label{eq:s_sigma_def}
    \Sigma_{nn'\mb{k}}(\omega) = \frac{1}{N_q} \sum_{\mb{q},\nu} \left[ \Sigma^{\rm Fan}_{nn'\mb{k};\qnu} (\omega) + \Sigma^{\rm DW}_{nn'\mb{k};\qnu} \right],
\end{equation}
\begin{align} \label{eq:s_fan_def}
    \Sigma^{\rm Fan}_{nn'\mb{k};\qnu} (\omega)
    &= \sum_{m} g_{mn\mb{k};\mb{q}\nu}^* \, g_{mn'\mb{k};\mb{q}\nu} \nnnl
    &\times \sum_{\pm} \frac{n_\qnu + [1 \pm (2f_\mkq - 1)]/2}{\omega - \veps_\mkq \pm \omega_\qnu + i\eta},
\end{align}
\begin{align} \label{eq:s_dw_def}
    \Sigma^{\rm DW}_{nn'\mb{k};\qnu}
    =& \frac{n_\qnu +\frac{1}{2}}{2\omega_{\qnu}} \sum_{\substack{m,\kappa,a\\\kappa',a'}} \mathcal{D}_{nn'}^{\kappa a\kappa' a'}(\mb{k},\mb{q})
    U_{\kappa a,\nu}^*(\mb{q}) U_{\kappa' a',\nu}(\mb{q}).
\end{align}
Here, $U_{\kappa a,\nu}(\mb{q})$ is the displacement pattern for the phonon eigenmode with branch index $\nu$ and wavevector $\mb{q}$ in units of inverse square root of mass, $n_\qnu$ the Bose-Einstein occupation factor, $f_\mkq$ the Fermi-Dirac occupation factor, and $\mathcal{D}_{nn'}^{\kappa a\kappa' a'}(\mb{k},\mb{q})$ the DW matrix element [see Eq.~(5) of Ref.~\cite{2020LihmAHC} for the definition].
In practice, the DW matrix element is calculated within the rigid-ion approximation~\cite{1976Allen,2014PoncePRB}.

Under the on-the-mass-shell approximation~\cite{2018NeryPRB,2020MiglioAHC}, the diagonal part of the self-energy of the state $\nk$ is $\Sigma_\nk = \Sigma_{nn\mb{k}}(\veps_\nk)$.
In this work, we only consider the diagonal part of the self-energy because the off-diagonal part has little effect on silicon, which has a band gap much larger than the band energy renormalization~\cite{2020LihmAHC}.

In practical calculations, the Fan term is separated into the lower Fan and the upper Fan terms, which are the contributions from the low-energy and high-energy bands, respectively.
We write the set of low-energy states as $\mcW_\mathrm{Lower}$.
The upper bound of the energy window for $\mcW_\mathrm{Lower}$ is chosen to be a few eV above $\veps^\ord{0}_\nk$.
We set $\mcW_\mathrm{Lower}$ to be inside the perturbation window, which is necessary because the lower Fan term [\myeqref{eq:s_lofan_def}] requires an explicit Wannier interpolation of the EPC matrix elements for bands inside $\mcW_\mathrm{Lower}$.

In the upper Fan term, one neglects the phonon frequency $\omega_{\mb{q}\nu}$ and $\eta$ in the denominator, which is a good approximation because $\abs{\veps_\nk - \veps_\mkq}$ is always much larger than $\omega_{\mb{q}\nu}$ and $\eta$ as one is only interested in the self-energy of low-energy states whose energy is much lower than the upper energy bound for $\mcW_\mathrm{Lower}$.
The lower and upper Fan terms read
\begin{align} \label{eq:s_lofan_def}
    \Sigma^{\rm lower\ Fan}_{n\mb{k};\qnu}(\eta)
    &= \sum_{\substack{m \\ m\mb{k+q} \in \mcW_\mathrm{Lower}}} \abs{g_{mn\mb{k};\mb{q}\nu}}^2 \nnnl
    &\times \sum_{\pm} \frac{n_\qnu + [1 \pm (2f_\mkq - 1)]/2}{\veps_\nk - \veps_\mkq \pm \omega_\qnu + i\eta}\,,
\end{align}
and
\begin{align} \label{eq:s_upfan_def}
    &\Sigma^{\rm upper\ Fan}_{n\mb{k};\qnu}
    = (2n_\qnu + 1) \sum_{\substack{m \\ m\mb{k+q} \notin \mcW_\mathrm{Lower}}}
    \frac{\abs{g_{mn\mb{k};\mb{q}\nu}}^2}{\veps_\nk - \veps_\mkq}\,,
\end{align}
respectively.

The DW self-energy under the rigid-ion approximation~\cite{1976Allen,2014PoncePRB} can be easily Wannier interpolated without any infinite sum by exploiting the recently developed momentum-operator-based representation of the DW matrix elements~\cite{2020LihmAHC}.
The lower Fan term [\myeqref{eq:s_lofan_def}] can also be calculated using the conventional Wannier interpolation method.

Calculation of the upper Fan term [\myeqref{eq:s_upfan_def}] is where the WF interpolation fails and WFPT becomes necessary.
The reason is that the upper Fan term involves a sum over an infinite number of bands.
Typically, a few hundred bands are needed to reach convergence~\cite{2010GiustinoPRL,2011GonzeAnnPhys}.
Wannier interpolation of such a large number of bands is impractical, if not impossible.

Using WFPT, the upper Fan term can be easily interpolated without explicitly performing the sum over bands.
By choosing $\opV_\mb{q}$ and $\opV'_\mb{q}$ as the perturbation potential for the same phonon mode $\nu$ in the definition of $S_{mn\mb{k}}$ [\myeqref{eq:int_s_def}] and using the WFPT interpolation formula [\myeqref{eq:int_s_wfp}], we find
\begin{align} \label{eq:eph_interpolation}
    &\sum_{\substack{m \\ \mkq \notin \mcW_\mathrm{Lower}}} \frac{\abs{g_{mn\mb{k};\nu\mb{q}}}^2}{\veps^\ord{0}_\nk - \veps^\ord{0}_\mkq} \nnnl
    =& S_{nn\mb{k};\nu\mb{q}}
    - \sum_{\substack{m=1 \\ \mkq \in \mcW_\mathrm{Lower}}}^{\NWan}
    \frac{\left( g^\ord{H}_{mn\mb{k};\nu\mb{q}} \right)^* g^\ord{H}_{mn\mb{k};\nu\mb{q}}}{\veps^\ord{H0}_\nk - \veps^\ord{H0}_\mkq} \nnnl
    =& s^\ord{H}_{nn\mb{k};\nu\mb{q}}
    + \sum_{\substack{m=1 \\ \mkq \notin \mcW_\mathrm{Lower}}}^{\NWan}
    \frac{\left( \widetilde{g}^\ord{H}_{mn\mb{k};\nu\mb{q}} \right)^* g^\ord{H}_{mn\mb{k};\nu\mb{q}}}{\veps^\ord{H0}_\nk - \veps^\ord{H0}_\mkq}.
\end{align}
Equation \eqref{eq:eph_interpolation} is the WFPT interpolation formula for the upper Fan self-energy.
In the second equality, we use the fact that in the limit of exact interpolation $\delta g^\ord{H}_{mn\mb{k}}=0$ and $g^\ord{H}_{mn\mb{k}} = \widetilde{g}^\ord{H}_{mn\mb{k}}$ if $\mk, \nk \in \WP$, as explained in Sec.~\ref{sec:theory_example}.

We calculate the upper Fan and DW terms on a coarse $6\times6\times6$ $q$-point grid, as it converges much faster than the lower Fan term.
We develop and use an improved double-grid method to speed up the convergence with respect to the $q$ grid used to calculate the upper Fan and the DW self-energies.
The improved double-grid method utilizes the fact that the low-energy acoustic phonons should give a small contribution to the total self-energy due to the translational invariance.
With this improved double-grid method, one can achieve convergence at a considerably coarser $q$ grid than the conventional double-grid method~\cite{2020BrownAltvaterAHC,2020LihmAHC} (see Appendix~\ref{sec:app_doublegrid} for details).

Figures \ref{fig:si_indabs}(a) and \ref{fig:si_indabs}(b) show the renormalized band structure of silicon at $T$=1000~K computed using WFPT.
For comparison, we also show the renormalized band energies calculated directly in the plane-wave basis.
WFPT gives an accurate result at a much smaller computational cost.

Since WFPT enables a cheap calculation of the eigenvalue renormalization, we can now calculate quantities that require renormalized eigenvalues at a large number of $k$ points.
As a prototypical example, we study the temperature dependence of the indirect absorption spectrum.
Without the phonon-induced band-structure renormalization, the perturbative expression for the phonon-assisted absorption coefficient is~\cite{2010Kioupakis,2012NoffsingerIndabs}
\begin{align} \label{eq:s_indabs_alpha}
    \alpha(\Omega) &= \frac{\pi q^2}{c \epsilon_0 V \Omega n(\Omega)}
    \sum_{\nu,m,n,\mb{k},\mb{q},\pm} \left( n_\qnu + \frac{1}{2} \pm \frac{1}{2} \right) \nnnl
    &\times (f_\nk - f_\mkq) \abs{\mb{\lambda} \cdot (\mb{S}_1 + \mb{S}_2)}^2 \nnnl
    &\times \delta(\veps_\mkq - \veps_\nk - \hbar \Omega \pm \hbar \omega_\qnu),
\end{align}
where
\begin{equation} \label{eq:s_indabs_S1}
    \mb{S}_1 = \sum_{p=1}^{\infty} \frac{\mb{v}_{np\mb{k}} g_{pm\mb{k};-\qnu}}{\veps_\pk - \veps_\mkq \mp \hbar\omega_\qnu + i\Gamma_\pk}
\end{equation}
and
\begin{equation} \label{eq:s_indabs_S2}
    \mb{S}_2 = \sum_{p=1}^{\infty} \frac{g_{np\mb{k};-\qnu} \mb{v}_{pm\mb{k+q}}}{\veps_{p\mb{k+q}} - \veps_\nk \pm \hbar\omega_\qnu + i\Gamma_{p\mb{k+q}}}.
\end{equation}
Here, $\epsilon_0$ is the vacuum permittivity, $\Omega$ the frequency of the incident photon, $c$ the speed of light, $\mb{\lambda}$ the light polarization, and $n(\Omega)$ the frequency-dependent refractive index of the material.
Note that we use the energy conservation imposed by the delta function to rewrite the denominator of \myeqref{eq:s_indabs_S1} in a form different from Eq.~(2) of Ref.~\cite{2010Kioupakis}.

When the phonon-induced band-structure renormalization is included, the absorption coefficient becomes~\cite{2014PatrickIndabs}
\begin{align} \label{eq:s_indabs_alpha_Sigma}
    &\alpha(\Omega) = 2 \frac{4\pi^2q^2}{\Omega c\, n(\Omega)} \frac{1}{V}
    \sum_{\nu,m,n,\mb{k},\mb{q},\pm}
    \left( n_\qnu + \frac{1}{2} \pm \frac{1}{2} \right) \\
    &\times (f_\mk - f_\nkq) \abs{\mb{\lambda} \cdot (\mb{S}_1 + \mb{S}_2)}^2 \nnnl
    &\times \delta(\veps_\mkq + \Re\Sigma_\mkq - \veps_\nk - \Re\Sigma_\nk - \hbar \Omega \pm \hbar \omega_\qnu). \nonumber
\end{align}
The phonon-induced real part of the self-energy appears only in the delta function while $\mb{S}_1$ and $\mb{S}_2$ are unaffected~\cite{2014PatrickIndabs,2017GiustinoRMP}.

Often, the absorption spectra are insensitive to the imaginary part of the self-energy of the intermediate state (see Fig.~\ref{fig:s_conv_indabs}(d)).
Following Ref.~\cite{2012NoffsingerIndabs}, we approximate $\Gamma_\pk$ as a constant:
\begin{equation} \label{eq:s_indabs_Gamma}
    \Gamma_\pk = \Gamma_{p\mb{k+q}} = \Gamma.
\end{equation}
Also, we broaden the delta function by a Gaussian:
\begin{equation} \label{eq:s_delta_gaussian}
    \delta(\veps)
    \approx \frac{1}{\gamma\sqrt{\pi}} e^{-\veps^2/\gamma^2},
\end{equation}
where $\gamma$ is the broadening parameter.

We correct the band structure and the velocity matrix elements using the $G_0W_0$ method while calculating the EPC and the phonon-induced renormalization at the DFT level.
The $G_0W_0$-corrected band gap is 1.28~eV, which is consistent with a previous calculation~\cite{1986HybertsenGW}.

Figures \ref{fig:si_indabs}(c) and \ref{fig:si_indabs}(d) show the indirect absorption spectra near the onset calculated without [Fig.~\ref{fig:si_indabs}(c)] and with [Fig.~\ref{fig:si_indabs}(d)] the phonon-induced band renormalization.
The spectra calculated using the WFPT-interpolated band renormalization successfully reproduce the experimental spectra for the entire range of temperatures, from 20~K to 415~K.
Previous perturbative calculation of the indirect absorption spectra used an arbitrary temperature-dependent shift of the calculated spectra~\cite{2012NoffsingerIndabs} because the phonon contribution to the electron self-energy was not considered.
Our calculation is free from such temperature-dependent parameters and thus is an accurate and predictive calculation of the phonon-assisted indirect absorption.

An alternative method to study EPC is the adiabatic supercell method, where supercells with distorted lattice structures are used~\cite{2015ZachariasZG,2016ZachariasZG,2016MonserratPRL,2020GorelovOptical,2020Zacharias}.
The adiabatic supercell methods are limited in that the phonon emission and absorption processes are not separated.
Consequently, such methods cannot reproduce the characteristic two-slope behavior near the onset of indirect absorption due to the different onsets of phonon emission and absorption~\cite{2012NoffsingerIndabs}.
Also, the absorption onset is hard to resolve due to the large computational cost of supercell calculations and the need for a fine $q$-point sampling.
For infrared-active materials, adiabatic supercell methods also miss the nonadiabatic effect of the Fr\"ohlich interaction, which is the predominant mechanism for band renormalization in many materials~\cite{2020MiglioAHC}.

We note that we used two empirical parameters regarding quasiparticle energies to compute the spectra in Figs.~\ref{fig:si_indabs}(c) and \ref{fig:si_indabs}(d).
First, we shifted all theoretical spectra by $-54$~meV to match the experimental ones, which reflects the typical accuracy of $G_0W_0$ calculations~\cite{2020RangelGW}.
The spectra computed without EPC were additionally shifted by $-93$~meV to match the experimental ones at low temperatures.
Second, we scaled the EPC-induced self-energy by a factor of 1.36, correcting the underestimation due to the neglect of electron-electron interactions in calculating EPC~\cite{2014AntoniusZPR}.
The scaling factor was obtained by dividing the experimental slope of the temperature-dependent indirect band-gap renormalization at the high-temperature region~\cite{1958MacfarlaneExpZPR,1974BludauExpZPR} by the calculated one.
We emphasize that these parameters are temperature independent.
The need for these parameters is the limitation of semilocal DFT and the $G_0W_0$ method.
More sophisticated methods to treat electron-electron interactions may allow a completely parameter-free calculation but this is beyond the scope of this study.

The real-space matrix elements calculated using WFs and WFPs are well localized, as shown in Fig.~\ref{fig:si_decay}.
This fact explains the success of the WF- and WFP-based interpolation.

We summarized the computational cost for calculating the phonon-induced electron self-energy in Table~\ref{table:si_time}.
We find a speedup of several orders of magnitude.
The major speedup of the interpolation scheme comes from the calculation of the lower Fan term, where both the $k$ and $q$ points are finely sampled using electron and phonon WFs.
Although this speedup comes from the conventional WF interpolation, WFPT interpolation enables one to consistently calculate the self-energy in the WF basis.
If WFPT is not used, the upper and lower Fan parts of the self-energy should have been calculated in the plane-wave and the WF bases, respectively.
Assigning band indices to those two calculations when the bands are {\it nearly} degenerate is a nontrivial issue.
The WFPT interpolation of the upper Fan self-energy for the fine $k$ points (23 cpu$\cdot$h) also gives a considerable speedup of above 50 times compared to solving the Sternheimer equation in the plane-wave basis (1,241 cpu$\cdot$h).
In our calculation, the number of fine $k$ points is quite small (379), as we sampled only the states inside a small energy window and the crystal is highly symmetric.
The speedup of the WFPT interpolation will be much greater if a wider energy window is considered, or if the valence and conduction bands have a larger effective mass.
Even for silicon, the speedup will reach the above-mentioned factor of over 50 as the number of points in the fine $k$ grid is increased.

We note that the WFP interpolation of EPC-induced electron self-energy is conceptually completely different from the Wannier interpolation of the $GW$ quasiparticle band structure~\cite{2009HamannGWWannier}.
In the latter, the $GW$ quasiparticle energy is first calculated for all bands to be Wannierized and then Wannier interpolated.
However, in the WFP interpolation, the matrix elements are Wannier interpolated, and then the self-energy is calculated.
We briefly discuss why one should not directly interpolate the phonon-induced electron self-energy.
First, one may try to calculate the self-energy on the coarse $k$ grid before Wannierization.
However, doing so is very inefficient because one needs to sample $q$ points on a very fine grid to converge the self-energy.
Second, one may try to interpolate only the sum of the upper Fan and DW terms, as it converges at a relatively coarse $q$ grid.
However, to do so, one must set $\mcW_\mathrm{Lower}$ to be outside the {\it outer} window $\WD$ because the upper Fan term should be calculated for all states inside $\WD$ for Wannierization.
Then, the lower Fan terms cannot be calculated using Wannier interpolation because the contribution from the states that are excluded from Wannierization, which are outside $\WF$ but inside $\mcW_\mathrm{Lower}$, is not captured.
In our method, we get around all these problems by using WFPT to interpolate the upper Fan self-energy and by using electron and phonon WFs to efficiently calculate the EPC on a fine $k$ and $q$ grid.

\subsection{Application 2: Shift spin current}

\begin{figure}[htbp]
\includegraphics[width=0.8\columnwidth]{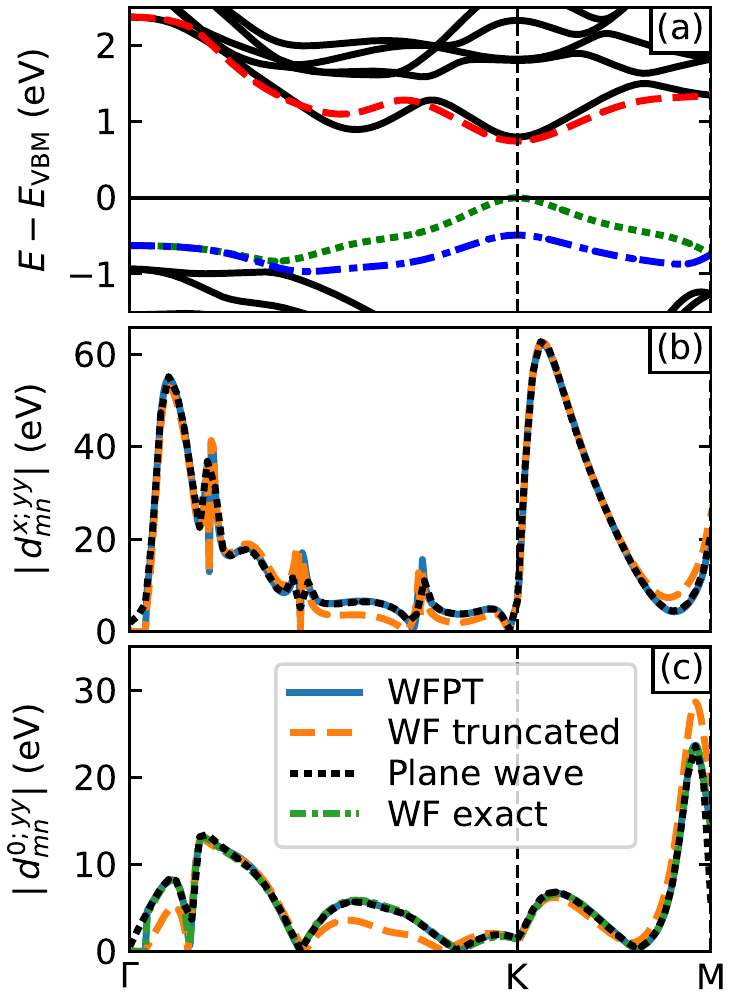}
\caption{
(a) Band structure of monolayer WTe$_2$.
(b, c) Matrix element of the (b) generalized spin-velocity derivative and (c) generalized derivative of the velocity operator.
The conduction band $m$ is indicated by the red dashed curve in panel (a).
The valence bands $n$ for panels (b) and (c) are indicated by the green dotted and blue dash-dotted curves in panel (a), respectively.
}
\label{fig:wte2_mel}
\end{figure}

\begin{figure*}[htbp]
\includegraphics[width=1.0\textwidth]{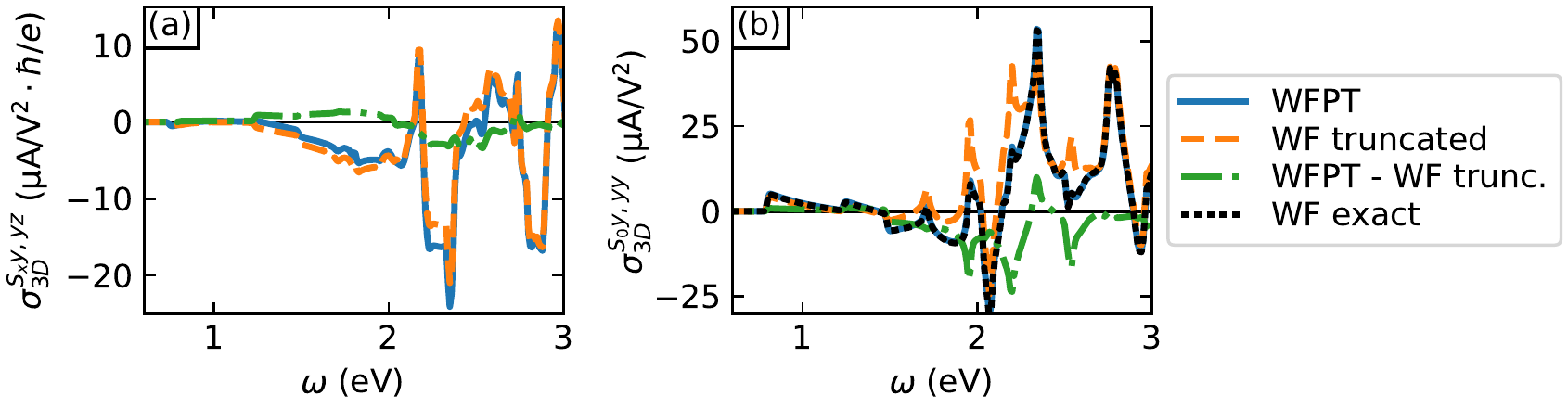}
\caption{
(a) Shift spin and (b) shift charge current of monolayer WTe$_2$ calculated using WFPT, WF with band truncation, and the difference between the two.
In panel (b), the shift current spectrum calculated using an exact WF-based interpolation~\cite{2018IbanezAzpirozShift} is also shown.
}
\label{fig:wte2_shift}
\end{figure*}

\begin{figure}[htbp]
\includegraphics[width=0.8\columnwidth]{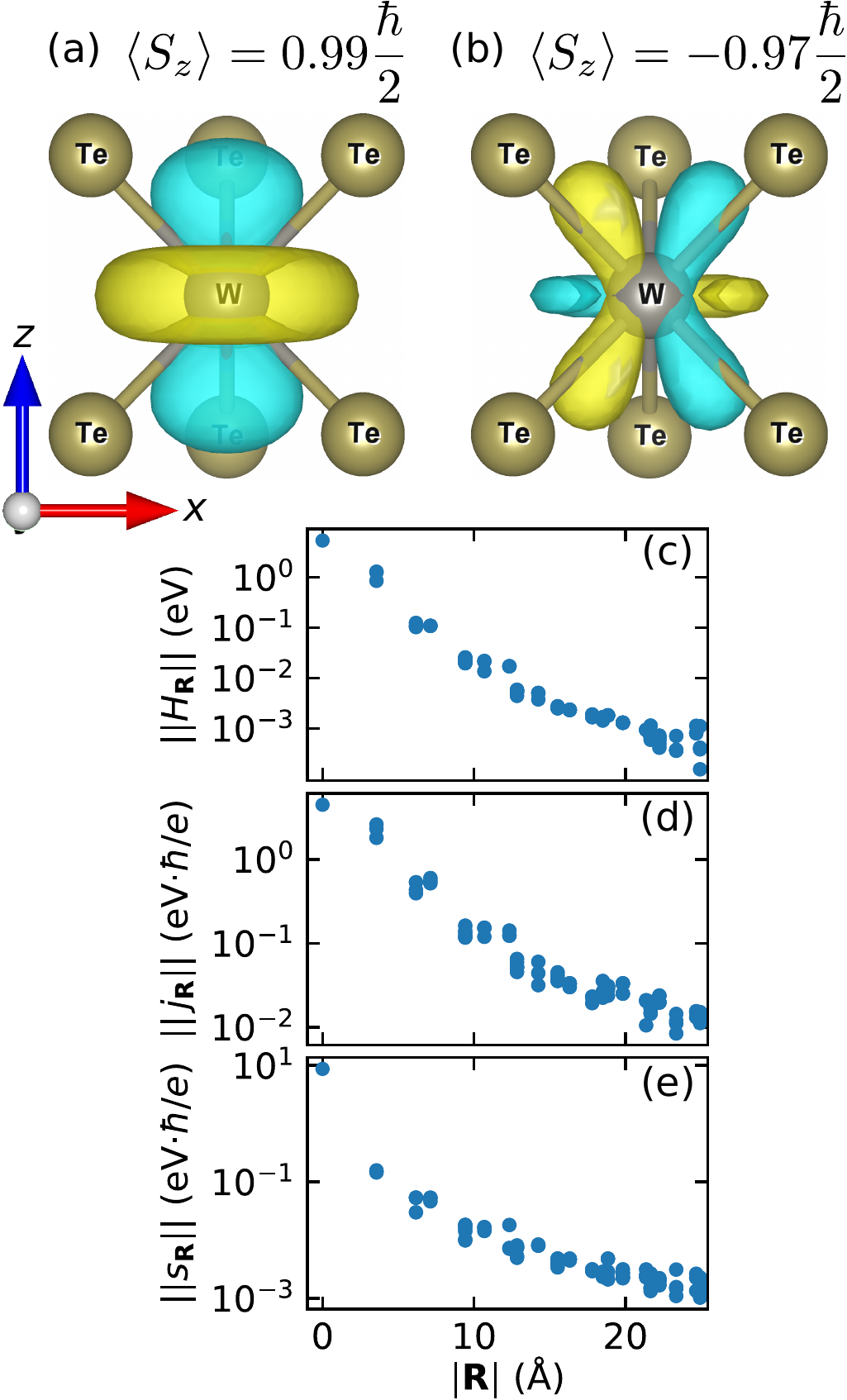}
\caption{
(a) Isosurface of the spin-up part of a WF.
Isovalues are $+0.087$ and $-0.087$ times the maximum absolute value for the yellow and cyan isosurfaces, respectively.
(b) Isosurface of the spin-down part of a WFP for the spin-velocity perturbation $j^{x;x}$.
Isovalues are $+0.25$ and $-0.25$ times the maximum absolute value for the yellow and cyan isosurfaces, respectively.
(c)-(e) Spatial decay of the (c) Hamiltonian, (d) spin-velocity, and (e) WFP matrix elements.
Each data point corresponds to the maximum absolute value over all WF pairs and the directions $a$, $b$, and $s$ for each $\mb{R}$.
The $y$ axes are on a logarithmic scale.
}
\label{fig:wte2_decay}
\end{figure}

\begin{table}[htbp]
\centering
\caption{Time required to calculate the shift spin current of the WTe$_2$ monolayer using direct plane-wave calculations and WFPT interpolation.}
\begin{tabular}{c|c|c} 
    \hline
    & \multicolumn{2}{c}{Time (cpu$\cdot$h)} \\ \cline{2-3}
    & Plane wave & WFPT \\ \hline
    SCF & 1.9 & 1.9 \\ 
    NSCF + Sternheimer (coarse $k$) & - & 47.2  \\
    Wannierization & - & 0.2 \\
    WFPT interpolation (fine $k$) & - & 3.0 \\
    NSCF + Sternheimer (fine $k$) & 52454.9$^*$ & - \\
    \hline
    Total & 52456.8$^*$ & 52.4 \\
    \hline
\end{tabular} \\
\footnotesize{$^*$ Estimated from a calculation with a smaller number of $k$ points.}
\label{table:wte2_time}
\end{table}

The second application of WFPT is the shift spin current, which is a second-order dc spin current response to an ac electric field.
The shift spin and shift charge conductivity can be written as~\cite{2021LihmSpin}
\begin{align} \label{eq:s_shift_formula}
    \sigma^{S_s,a;bc}_\mathrm{shift}(\Omega)
    &= \frac{i \pi q^3}{2\hbar^2 V\Omega^2} \sum_{\mb{k},m,n} (d^{s,b;a}_{mn\mb{k}} v^c_{nm\mb{k}} - d^{s,c;a}_{nm\mb{k}} v^b_{mn\mb{k}}) \nnnl
    &\times (f_\mk - f_\nk) \delta(\Omega + (\veps_\mk - \veps_\nk) / \hbar)
\end{align}
where $\Omega$ is the frequency of the incident photon, $s$ and $a$ the spin polarization axis and flow direction of the current, and $b$ and $c$ the directions of the external electric field.
In an actual calculation, we broaden the delta function by a Gaussian as in \myeqref{eq:s_delta_gaussian}.
The matrix element $d^{s,b;a}_{mn\mb{k}}$ is called the generalized spin-velocity derivative~\cite{2021LihmSpin} and is defined as
\begin{equation} \label{eq:shift_d_def}
    d^{s;b,a}_{mn\mb{k}}
    = j^{s;ab}_{mn\mb{k}}
    + \primesum{p} \frac{j^{s;a}_{mp\mb{k}} v^{b}_{pn\mb{k}}}{\veps^\ord{0}_\mk - \veps^\ord{0}_{p\mb{k}}}
    + \primesum{p} \frac{v^{b}_{mp\mb{k}} j^{s;a}_{pn}}{\veps^\ord{0}_\nk - \veps^\ord{0}_{p\mb{k}}},
\end{equation}
where $\hat{j}^{s;ab} = - \acomm{\hat{S}_s}{[\hat{r}^a,[\hat{r}^b, \opH\oord{0}]]} / 2\hbar^2$.
We define $\hat{S}_0 = \hat{\one}$ so that $s=0$ represents the charge current.
In contrast to its charge counterpart, the generalized spin-velocity derivative is not geometric: It cannot be written as a derivative with respect to $\mb{k}$.
Thus, the shift spin current cannot be calculated using the conventional WF interpolation method.

By choosing $\opV_{\mb{q}} = \hat{j}^{s;a} \delta_{\mb{q},\mb{0}}$ and $\opV'_{\mb{q}} = \hat{v}^{b} \delta_{\mb{q},\mb{0}}$ in the definition of $S_{mn\mb{k}}$ [\myeqref{eq:int_s_def}], we find
\begin{equation} \label{eq:shift_d_interpol}
    d^{s;b, a}_{mn\mb{k}}
    = j^{s;ab}_{mn\mb{k}}
    + S_{mn\mb{k}} + S_{nm\mb{k}}^*.
\end{equation}
The matrix elements of $\hat{j}^{s;ab}$ can easily be Wannier interpolated via a direct calculation of the matrix elements in the plane-wave basis for coarse $k$ points~\cite{\citeSupp}.
WFPT interpolation of $S_{mn\mb{k}}$ enables the interpolation of $d^{s;b,a}_{mn\mb{k}}$.
With the matrix elements interpolated to a dense grid of $k$ points, one can calculate the shift spin and shift charge current spectra using \myeqref{eq:s_shift_formula}.

We calculate the shift spin and shift charge current of monolayer WTe$_2$ in the 2H phase~\cite{2016HuangWTe2}.
In Fig.~\ref{fig:wte2_mel}, we show the interpolated matrix elements for the shift spin and shift charge currents.
For comparison, we also evaluate \myeqref{eq:shift_d_def} exactly by solving the Sternheimer equation in the plane-wave basis and approximately by truncating \myeqref{eq:shift_d_def} to include only the Wannierized bands.
We find that this band truncation gives incorrect matrix elements.
In contrast, the WFPT interpolation accurately reproduces the exact plane-wave basis result.
We also find that the matrix element calculated with an exact WF-based interpolation of the generalized derivative of the interband dipole matrix element~\cite{2018IbanezAzpirozShift} agrees with the result of the WFPT interpolation for the charge case.
Such an exact WF-based interpolation is not possible for the spin case.

By solving the Sternheimer equation, one can compute the shift spin current without any band-truncation error.
WFPT enables such a band-truncation-error-free calculation at a much lower computational cost.
To our knowledge, the shift spin current has not been studied using the Sternheimer equation-based method, which is devoid of the band-truncation error.
Here, we use WFPT to calculate the shift spin conductivity without the truncation error and study the effect of band truncation.

Figure \ref{fig:wte2_shift} shows the calculated shift spin and shift charge conductivity.
We again find that the truncation of \myeqref{eq:shift_d_def} gives considerable error in the shift current spectra.
Also, for the shift charge current, the result of the exact WF-based interpolation~\cite{2018IbanezAzpirozShift} agrees with the result of our WFPT interpolation.

The WF and WFP for WTe$_2$ are shown in Figs.~\ref{fig:wte2_decay}(a) and \ref{fig:wte2_decay}(b).
The WF shown is a $d_{z^2}$-orbital-like state of the W atom, with its spin almost fully polarized (99\%) along the $+z$ direction.
The WFP shown is for the spin-velocity perturbation $\hat{j}^{x;x}$.
The result can be qualitatively understood by considering the spin and orbital parts separately.
The spin operator $\hat{S}_x$ rotates the spin polarization by 180$^\circ$ around the $x$ axis so that the spin of the WFP is almost fully polarized (97\%) along the $-z$ direction.
The velocity operator acts on the orbital part.
Wavefunction perturbation due to the velocity operator is equivalent to the application of the position operator to the wavefunction~\cite{1986Baroni}.
Hence, the orbital part of the WFP will approximately be the orbital part of WF times $x$, which is indeed the case in Fig.~\ref{fig:wte2_decay}(b).
Figures \ref{fig:wte2_decay}(c)-\ref{fig:wte2_decay}(e) shows that the real-space matrix elements are spatially localized.

We summarize the computational cost for calculating the shift current in Table~\ref{table:wte2_time}.
The WFPT interpolation method speeds up the shift current calculation by 3 orders of magnitude.
We note that the calculation can be further optimized by using crystal symmetry to compute the Sternheimer matrix elements only in the irreducible wedge of the Brillouin zone, for both the coarse and the fine $k$ grids.
We leave such an optimization for future work.

\subsection{Application 3: Spin Hall conductivity}

Finally, we apply WFPT to the calculation of spin Hall conductivity.
The key quantity for calculating the spin Hall conductivity is the spin Berry curvature~\cite{2008GuoSHE}
\begin{equation} \label{eq:shc_berry}
    \Omega^{s,ab}_{\mk} = 2\sum_{n \neq m} \frac{\Im \left[ j^{s,a}_{mn\mb{k}} v^b_{nm\mb{k}} \right]}{(\veps_\mk - \veps_\nk)^2 + \eta^2}.
\end{equation}
Here, $\eta$ is a positive infinitesimal, and $a$, $b$, and $s$ are the directions of the external field, current flow, and the spin polarization, respectively.
The intrinsic dc spin Hall conductivity is given by the Kubo formula as~\cite{2005GuoSHE,2008GuoSHE}
\begin{equation} \label{eq:shc_formula}
    \sigma^{s}_{ab} = \frac{q}{\hbar V} \sum_{\mb{k}, m} f_\mk \Omega^{s,ab}_{\mk}.
\end{equation}
Here, $V$ is the volume of the unit cell, $q$ the electron charge, and $f_\mk$ the Fermi-Dirac occupation factor.
Contrary to the Berry curvature, the spin Berry curvature is not a geometric quantity.
Thus, in principle, an infinite number of bands should be included to evaluate \myeqref{eq:shc_berry}.
For example, Ref.~\cite{2005GuoSHE} included states up to 50~eV above the valence band maximum to converge the spin Hall conductivity of Si, Ge, GaAs, and AlAs.
Moreover, a very fine $k$-point grid is required to achieve convergence, which calls for efficient interpolation methods such as Wannier interpolation.
However, conventional Wannier interpolation methods for computing spin Hall conductivity~\cite{2018QiaoSHC,2019RyooSHC} truncate the sum over band $n$ in \myeqref{eq:shc_berry} to include only the Wannier-interpolated bands.

By solving the Sternheimer equation, one can accurately calculate the spin Berry curvature and the spin Hall conductivity without explicitly including the high-energy bands.
To our knowledge, the spin Hall conductivity has not been studied using the Sternheimer-equation-based method.
However, the computational cost of the Sternheimer approach is much higher than Wannier interpolation for the same $k$-point grid.
WFPT enables an efficient Wannier-interpolation calculation of the spin Hall conductivity without any band-truncation error, taking the advantages of both the Sternheimer method and the WF method.

The spin Berry curvature can be calculated by WFPT interpolation similarly to the case of the spin-velocity derivative.
We set $\opV_\mb{q} = \hat{j}^{s,a} \delta_{\mb{q},\mb{0}}$ and $\opV'_\mb{q} = \hat{v}^{b} \delta_{\mb{q},\mb{0}}$ and use \myeqref{eq:kubo_k_wfp} to find
\begin{align} \label{eq:s_shc_berry_interp}
    \frac{1}{2} \Omega^{s,ab}_{\mk}
    \approx& K_{mm\mb{k}} - \beta_{\mk}(0) + \beta_{\mk}(\eta)
    = k^\ord{H}_{mm\mb{k}} + \beta_{\mk}(\eta)
\end{align}
where
\begin{equation} \label{eq:shc_wf_term}
    \beta_{\mk}(\eta) = \primesum{n=1}^{\NWan}\frac{\Im \left[ \widetilde{j}^{s,a\rm (H)}_{mn\mb{k}} \widetilde{v}^{b\rm (H)}_{nm\mb{k}} \right]}{\left( \veps^\ord{H0}_\mk - \veps^\ord{H0}_\nk \right)^2 + \eta^2}.
\end{equation}
Here, we approximate that for $\mk \in \WP$ and $\nk \notin \WF$, $\abs{\veps_\mk - \veps_\nk} \gg \eta$, so the $\eta$ in the denominator can be ignored.
In practice, we avoid the sums over pairs of fully occupied states $\mk$ and $\nk$ in \myeqref{eq:shc_berry} and \myeqref{eq:shc_wf_term} since such contributions get canceled in \myeqref{eq:shc_formula}~\cite{2006WangAHC}.

\begin{figure}[htbp]
\centering
\includegraphics[width=1.0\columnwidth]{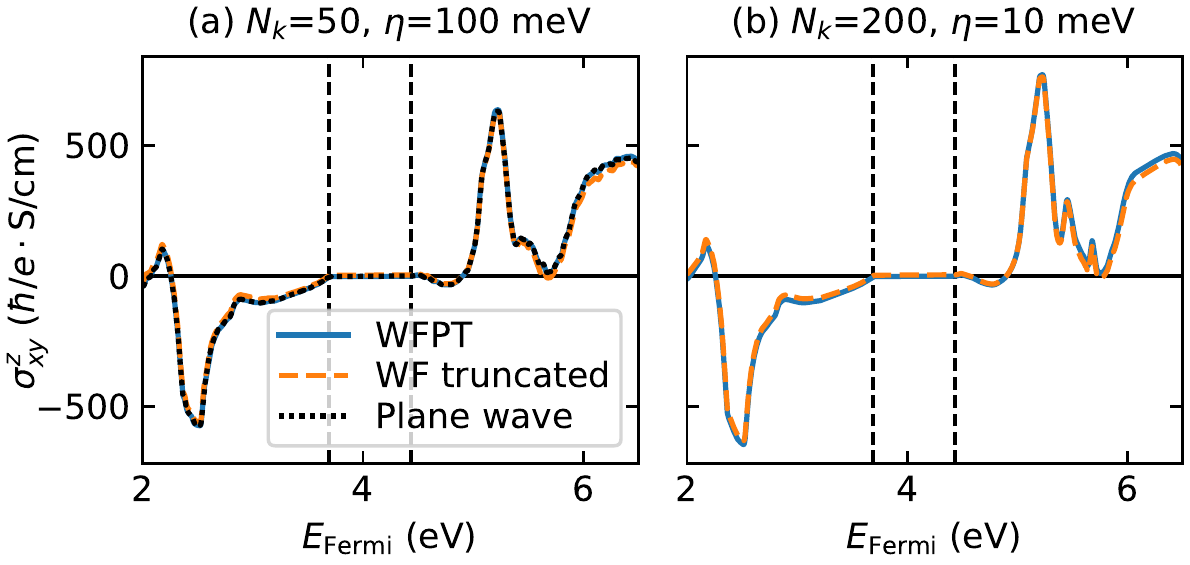}
\caption{
Spin Hall conductivity of monolayer WTe$_2$ computed using WFPT, using WF with band truncation, and by solving the Sternheimer equation in the plane-wave basis.
The dashed vertical lines indicate the valence band maximum and the conduction band minimum energies.
The temperature is set to 50~K.
The sum over $k$ points in Eq.~\eqref{eq:shc_formula} is performed over an $N_k\times N_k\times1$ grid, and $\eta$ is the broadening parameter in Eq.~\eqref{eq:shc_berry}.
The plane-wave calculation is not performed for $N_k=200$ due to the high computational cost.
}
\label{fig:wte2_shc}
\end{figure}

We calculate the spin Hall conductivity of monolayer WTe$_2$ in the 2H phase.
In Fig.~\ref{fig:wte2_shc}, we compare spin Hall conductivity at $T=50$~K computed using the WFPT interpolation method, conventional Wannier interpolation with band truncation~\cite{2019RyooSHC}, and the slow but exact Sternheimer method in the plane-wave basis.
The WFPT result agrees perfectly with the exact results, while the WF interpolation with band truncation shows some discrepancy, albeit small.
Interestingly, the band-truncation error is present even for the hole-doped case, where all the relevant bands are separated from the boundary of the inner window by more than 3~eV.
Table~\ref{table:s_wte2_shc_time} shows the speedup of WFPT interpolation compared to the plane-wave calculation for a $200\times200\times1$ $k$-point grid.
WFPT gives a speedup of over 2 orders of magnitude, without lowering the accuracy by band truncation.

\begin{table}[htbp]
\centering
\caption{Time required to calculate the spin Hall conductivity of the WTe$_2$ monolayer using a $200\times200\times1$ $k$-point grid by direct plane-wave calculations and WFPT interpolation.}
\begin{tabular}{c|c|c} 
    \hline
    & \multicolumn{2}{c}{Time (cpu$\cdot$h)} \\ \cline{2-3}
    & Plane wave & WFPT \\ \hline
    SCF & 1.9 & 1.9 \\ 
    NSCF + Sternheimer (coarse $k$) & - & 47.2  \\
    Wannierization & - & 0.2 \\
    WFPT interpolation (fine $k$) & - & 0.1 \\
    NSCF + Sternheimer (fine $k$) & 13113.7$^*$ & - \\
    \hline
    Total & 13115.6$^*$ & 49.4 \\
    \hline
\end{tabular} \\
\footnotesize{$^*$ Estimated from a calculation with a smaller number of $k$ points.}
\label{table:s_wte2_shc_time}
\end{table}

\section{Discussion}
If one is only interested in the valence bands of an insulator, one may choose all the windows, $\WP$, $\WF$, and $\WD$, to include only the occupied bands.
If WFPT is applied to the electronic dielectric response of such insulators, the WFP derived in this work simplifies to the result of Ref.~\cite{2015GeLocalized}, i.e., \myeqref{eq:wfpt_WFP} without the second term on the right-hand side.
We note, however, that even the Wannierization of only the valence bands may not be possible in the case of narrow-gap semiconductors or topological insulators where the orbital characters of the valence bands change rapidly with {\bf k}.
In the case of such semiconductors, where the low-energy conduction bands need to be interpolated and thus disentanglement is required, or that of metals, the simplified results do not apply.
Our formulation of WFPT is applicable to both isolated and entangled bands and provides a clear physical understanding of why the WFPs are spatially localized.
The generalization of WFPT from isolated bands to entangled bands requires the deliberate use of energy windows and subspaces to separate the energy ranges.
In addition, the changes in the Wannier basis functions and the resulting changes in the Hamiltonian matrix [\myeqref{eq:int_hop_1}] must be addressed to correctly interpolate matrix elements involving the sum over an infinite number of bands as done in Eqs.~(\ref{eq:int_s_wfp_derivation},\ref{eq:kubo_k_wfp}).
For WFs constructed without disentanglement, the Hamiltonian matrix elements do not change because the WFPs are spanned by eigenstates outside the outer window so that the Hamiltonian matrix elements between the WFPs and the unperturbed WFs [\myeqref{eq:int_deltag_def}] are zero.
Only as these highly nontrivial problems are solved, the methods for accurately calculating the proper sum over an infinite number of bands can be developed (as summarized in Table~\ref{table:key_quantities}), and WFPT enables several new kinds of calculations that were beyond the computational reach.

WFPT could find wide applications.
For EPC, WFPT opens up a way to systematically study the effect of phonon-induced band renormalization to quantities that require integration over fine $k$ points.
Phonon-assisted optical properties are one example.
We emphasize that the predictive calculation of the indirect absorption onset (Fig.~\ref{fig:si_indabs}(d)) is currently possible only with the WFPT interpolation.
Another example is electronic transport.
In transport calculations, the EPC-induced band renormalization is ignored or included only partially as a correction to the effective mass~\cite{2018SchlipfMobility,2020DSouzaMobility}.
WFPT interpolation will allow the complete consideration of the renormalization by calculating the band renormalization at every $k$ point in the fine grid for transport calculations.
Also, a self-consistent calculation of the phonon-induced electron self-energy~\cite{2020BrownAltvaterAHC} without neglecting the $k$ dependence of the self-energy may become feasible.

Zero-point renormalization and temperature dependence of band structures are also interesting by themselves.
Calculation of the zero-point renormalization for a large set of materials began only very recently~\cite{2020MiglioAHC}.
Also, most calculations are limited to the correction of the valence band maximum and conduction band minimum, partly due to the large computational cost of calculations for a large number of $k$ points.
WFPT will considerably reduce the barrier of the calculation of the zero-point and temperature-dependent renormalization for a large number of materials or for many $k$ points.
Combined with the automatic generation of Wannier functions~\cite{2020VitaleWannier}, an efficient, automatized high-throughput calculation will also be possible.

Similar developments can also be made for the shift spin current.
The interpolation scheme based on WFPs is necessary to calculate converged shift spin current spectra at a reasonable computational cost.
Considering the non-geometric property of the spin-transport-related matrix elements, WFPT will be useful for studying a wide range of spin transport phenomena.

Another promising application of WFPT is the calculation of the nonadiabatic Born effective charge, a concept recently introduced in Refs.~\cite{2019BistoniBorn,2021BinciBorn,2021DreyerBorn}.
Calculation of the nonadiabatic Born effective charge of metals and doped semiconductors requires a fine $k$-point sampling of wavefunction perturbations for phonon perturbations~\cite{2021DreyerBorn}.
WFPT interpolation allows such a calculation for all kinds of materials, not just for the special case where the low-energy conduction bands are isolated from the high-energy bands.

Finally, we discuss two ways the WFPT formalism itself can be expanded.
First, in this work, we did not consider the localized representation of the perturbations.
The concept of WFPs may be generalized to localized perturbations, for example by using phonon Wannier functions~\cite{2007GiustinoEPW,2008Eiguren}.
Such a generalization may enable the study of phonon-induced band renormalization in complex heterostructures by stitching WF- and WFP-based real-space matrix elements.
Second, considering time-periodic perturbations will allow the localized representation of time-dependent wavefunction perturbations.
This generalization will be useful for the study of dynamic properties such as the dielectric function and the higher-harmonic generation.
Also, these two generalizations may enable the application of WFPT to $GW$ calculations, especially in the calculation of frequency- and momentum-dependent screened Coulomb interaction using the Sternheimer equation~\cite{2010GiustinoGW} or the Lanczos method~\cite{2010UmariGW}.

\section{Conclusion}
In conclusion, we introduced the concept of WFPT, which is the extension of the nearsightedness principle to wavefunction perturbations.
We defined the WFP and showed that it gives an accurate localized representation of wavefunction perturbations.
Using WFPs, various matrix elements involving the wavefunction perturbation can be efficiently interpolated.
We applied WFPT to the calculation of the temperature-dependent electron band structures, the shift spin conductivity, and the spin Hall conductivity.
The significant speedup gained thanks to WFPT enabled the study of phenomena such as the temperature-dependent onset of the indirect optical absorption and the shift spin currents, which were beyond the reach of previous methods.
As density-functional perturbation theory has significantly expanded the applications of density functional theory, WFPT will also widen the scope of Wannier function methods for the theoretical and computational study of condensed matter.

\section{Methods}

We used the \texttt{Quantum ESPRESSO} package to perform plane-wave pseudopotential DFT calculations~\cite{2009GiannozziQE,2017GiannozziQE}.
We used optimized norm-conserving Vanderbilt (ONCV) pseudopotentials~\cite{2013HamannONCVPSP} taken from the PseudoDojo library (v0.4)~\cite{2018VanSettenPseudoDojo}.
For silicon, we used the Perdew-Wang functional~\cite{1992PerdewLDA} and a scalar-relativistic pseudopotential.
We used a kinetic energy cutoff of 70~Ry and a 14$\times$14$\times$14 $k$-point grid for DFT and DFPT calculations.
DFPT calculation was done on a $6\times 6 \times6$ $q$-point grid.
The experimental lattice parameter at 300~K ($a$=5.43~\AA) was used~\cite{1996ReeberSiliconExperiment}.
Thermal expansion was not taken into account as its effect on the indirect band gap is negligible~\cite{1985LautenschlagerSilicon,2012NoffsingerIndabs}.
For the WTe$_2$ monolayer, we used the Perdew-Burke-Ernzerhof functional~\cite{1996PerdewPBE} and included spin-orbit coupling using a fully relativistic pseudopotential.
We used a kinetic energy cutoff of 70~Ry and a $12 \times 12 \times 1$ $k$-point grid.
The lattice structure was optimized until the internal stresses and forces were below 0.01~kbar and $2\times 10^5$~Ry/Bohr, respectively.
The optimized lattice parameters of a rectangular in-plane supercell were $a$=3.56~\AA\ and $b$=6.17~\AA, consistent with a previous calculation~\cite{2016HuangWTe2}.
The supercell length along the nonperiodic direction was set to 15~\AA.

To construct the MLWFs, we used the Wannier90 package~\cite{2020PizziWannier90}.
For silicon, we constructed 10 MLWFs from 18 bands to accurately interpolate the bands near the conduction band minimum.
We used $sp^3$ orbitals centered at the Si atom and $s$ orbitals centered at the two interstitial sites as the initial guesses.
The upper bounds of the perturbation window ($\WP$), the inner window of the unperturbed system ($\WF$), and the outer window ($\WD$) were 5.6~eV, 6.1~eV, and 18.5~eV above the conduction band minimum, respectively.
A $12 \times 12 \times 12$ $k$-point grid was used.
For WTe$_2$, we constructed 28 MLWFs from 32 bands, using W $s$ and $d$, and Te $s$ and $p$ orbitals as initial guesses.
We excluded the semicore states in the Wannierization.
The upper bound of the perturbation window and the inner window of the unperturbed system were 2.8~eV and 3.3~eV above the valence band maximum, respectively.
The outer window was not set so that all 32 bands are used for Wannierization.
A $12 \times 12 \times 1$ $k$-point grid was used for the shift current and spin Hall applications.
In the Berry curvature calculation, we used a $24 \times 24 \times 1$ grid to reduce the finite-difference error in the position matrix elements~\cite{1997Marzari}.
For plotting the isosurfaces of the WFs and WFPs for WTe$_2$, we used a coarser $6 \times 6 \times 6$ $k$-point grid due to the high computational cost.
For all Wannier interpolation, we used the minimal-distance replica selection based on the Wigner-Seitz supercell that improves the quality of interpolation~\cite{2007YatesWannier}.

The lower Fan term in the phonon-induced electron self-energy of silicon was calculated using a $125\times125\times125$ $q$-point grid.
The denominator was smoothed with $\eta=$10~meV [\myeqref{eq:s_lofan_def}].
We used the on-the-mass-shell approximation (Rayleigh-Schr\"odinger perturbation) which is known to give a better estimate of the zero-point renormalization than the Dyson-Migdal approach~\cite{2018NeryPRB,2020BrownAltvaterAHC}.
The upper Fan and the DW self-energies were calculated on a coarse $6\times 6 \times6$ $q$-point grid without interpolation over $q$ points.
To speed up the convergence of the sum of the upper Fan and the DW self-energies, we used the improved double-grid method as detailed in Appendix~\ref{sec:app_doublegrid}.
See Sec.~\ref{sec:s_conv_selfen} of the Supplementary Material~\cite{\citeSupp} for the convergence study of the lower Fan term.

The $G_0W_0$ calculation of the quasiparticle band structure of silicon was done using BerkeleyGW~\cite{2012DeslippeBerkeleyGW}, within the plasmon-pole approximation~\cite{1986HybertsenGW}.
We converged the direct and indirect band gaps to within 5~meV.
We used a $12\times12\times12$ $q$-point grid, a cutoff of 30~Ry and 402 bands for the dielectric matrix, and 308 bands with the static remainder correction~\cite{2013DeslippeGwRemainder} for the self-energy.
The quasiparticle energies were calculated on a $12\times12\times12$ $k$-point grid and then Wannier interpolated~\cite{2009HamannGWWannier} using the EPW code~\cite{2016PonceEPW}.
The velocity matrix elements for the indirect optical absorption were renormalized following Ref.~\cite{2000Rohlfing}.

The indirect absorption spectra of silicon were calculated using the EPW code~\cite{2016PonceEPW}.
We used a 16-band WF model to include a sufficient number of intermediate bands.
We checked that the indirect absorption spectra change only slightly when more bands are included.
We used $60\times60\times60$ $k$- and $q$-point grids.
We sampled the initial and final states only in the $\pm$1~eV window around the center of the band gap.
Using the lattice symmetry, we only sampled the $k$ points in the irreducible wedge of the Brillouin zone.
In total, 379 $k$ points were sampled.
The imaginary part of the self-energy of intermediate electronic states was set to 50~meV.
Delta functions for energy conservation were broadened using a Gaussian function of width 15~meV.
We used the experimental frequency-dependent refractive index at 300~K~\cite{1995GreenSiExperiment}.
See Sec.~\ref{sec:s_conv_indabs} of the Supplementary Material~\cite{\citeSupp} for the convergence study.

The shift conductivity of monolayer WTe$_2$ was calculated using a $400\times400\times1$ $k$-point grid.
Delta functions for energy conservation were broadened using a Gaussian function of width 10~meV.
Following the scheme of Refs.~\cite{2006NastosShift,2018IbanezAzpirozShift}, we regularized the denominators including an intermediate-state energy with $\eta=$10~meV, which is in the range where the spectrum calculated using the given $k$-point grid remains stable~\cite{2006NastosShift,2018IbanezAzpirozShift}.
See Sec.~\ref{sec:s_conv_shift} of the Supplementary Material~\cite{\citeSupp} for the convergence study.

The spin Hall conductivity of monolayer WTe$_2$ was calculated using a $200\times200\times1$ $k$-point grid.
Note that $\eta=$10~meV was used in the calculation of the spin Berry curvature [\myeqref{eq:shc_berry}].
We used the Smoother module of the WannierBerri package~\cite{2021TsirkinWBerri} to compute the spin Hall conductivity at a finite temperature.
The response was scaled using an effective single-layer thickness of 7.1~$\mathrm{\AA}$, which is taken from the bulk lattice parameter~\cite{2015LeeWTe2}.
See Sec.~\ref{sec:s_conv_shc} of the Supplementary Material~\cite{\citeSupp} for the convergence study.

The WFPs and the corresponding matrix elements in the coarse $k$ grid were calculated using an in-house modified version of the ph.x program of \texttt{Quantum ESPRESSO}~\cite{2017GiannozziQE}.
We used in-house modified versions of the EPW~\cite{2016PonceEPW} and Wannier90~\cite{2020PizziWannier90} code for the EPC, and the shift spin and spin Hall current applications, respectively.
In the EPC calculation, we used the crystal symmetry to efficiently unfold the electron-phonon, DW, and Sternheimer matrix elements calculated for the $q$ points in the irreducible wedge to the full coarse $q$-point grid.
We also implemented the optimization~\cite{2021TsirkinWBerri} of the minimal-distance replica selection method that significantly reduces the computational cost in the in-house versions of EPW and Wannier90.
Isosurface plots were created using VESTA~\cite{2011MommaVESTA}.

\begin{acknowledgments}
This work was supported by the Creative-Pioneering Research Program through Seoul National University, the NRF of Korea No-2020R1A2C1014760, and the Institute for Basic Science (No. IBSR009-D1).
Computational resources have been provided by KISTI (KSC-2020-INO-0078).
\end{acknowledgments}

\appendix
\section{Derivation of the Wannier function perturbation} \label{sec:app_wfp_derivation}

In this appendix, we derive \myeqref{eq:wfpt_WFP}, the formula for the WFP.
First, we find the disentanglement subspace following the initialization scheme of Ref.~\cite{2001Souza}.
Concretely, we project $\opPW$, the projection operator for the unperturbed Wannier subspace, onto the subspace $\WD - \WP$ and diagonalize it.
We select $N_\mb{k}\NWan - N_{\rm P}$ states with the largest eigenvalues~\cite{2001Souza} and use them as the basis states for the perturbed WFs, together with $N_\mathrm{P}$ states inside $\WP$.
\begin{widetext}
From Eqs.~(\ref{eq:wfpt_WF0_def}, \ref{eq:wfpt_perturb}, \ref{eq:wfpt_wf_series}), we find that for states $\nk, \npkp \in \WD$, the matrix element of $\opPW$ is
\begin{align} \label{eq:wfpt_proj_guess}
    P_{\npkp,\nk}
    =& \mel{\psi_\npkp}{\opPW}{\psi_\nk}
    = \sum_{\mb{R}}\sum_{i=1}^{\NWan} \braket{\psi_\npkp}{w_\iR\oord{0}} \braket{w_\iR\oord{0}}{\psi_\nk} \nnnl
    =& \frac{1}{N_k} \sum_{\mb{R}}\sum_{i=1}^{\NWan} U_{n'i;\mb{k'}}\oord{0} U_{in;\mb{k}}^{(0) \dagger} e^{i(\mb{k-k'})\cdot\mb{R}} \nnnl
    +& \lambda \frac{1}{N_k} \sum_\mb{R} \primesum{m,\mb{k''}} \sum_{i=1}^{\NWan}
    \left[ U_{n'i;\mb{k'}}\oord{0} U_{im;\mb{k''}}^{(0) \dagger} e^{i(\mb{k''-k'})\cdot\mb{R}} \frac{\mel{\psi_{m\mb{k''}}\oord{0}}{\opV}{\psi\oord{0}_\nk}}{\veps^\ord{0}_\nk - \veps^\ord{0}_{m\mb{k''}}} 
    + (\nk \leftrightarrow \npkp)^* \right]
    + \mcO(\lambda^2) \nnnl
    =& P_{{\rm W}, n'n;\mb{k}}\oord{0} \delta_{\mb{k},\mb{k'}}
    + \lambda \primesum{m} \left( P_{{\rm W}, n'm;\mb{k'}}\oord{0} \frac{\mel{\psi_\mkp^\ord{0}}{\opV}{\psi_\nk^\ord{0}}}{\veps^\ord{0}_\nk - \veps^\ord{0}_\mkp}
    - \frac{\mel{\psi_\npkp^\ord{0}}{\opV}{\psi_\mk^\ord{0}}}{\veps^\ord{0}_\mk - \veps^\ord{0}_\npkp}
    P_{{\rm W}, mn;\mb{k}}^\ord{0} \right)
    + \mcO(\lambda^2).
\end{align}
\end{widetext}
Here, we defined
\begin{equation} \label{eq:wfpt_PW0_def}
    P_{{\rm W}, n'n;\mb{k}}\oord{0}
    = \sum_{i=1}^{\NWan} U_{n'i;\mb{k}}\oord{0} U_{in;\mb{k}}^{(0) \dagger}.
\end{equation}

We can obtain the eigenstates of $P_{\npkp,\nk}$ using first-order perturbation theory.
The zeroth-order matrix is $P_{{\rm W},n'n;\mb{k}}^\ord{0} \delta_{\mb{k},\mb{k'}}$.
We write its eigenvectors as $\widetilde{U}\oord{0}_{np;\mb{k}}$, where the eigenvalues for $p=1,\cdots,\NWan$ are 1  (in subspace \circled{1} and \circled{2}) and those for $p=\NWan+1,\cdots,N_{\mathrm{D},\mb{k}}$ are 0 (in subspace \circled{3}).
Since $\WF$ is the inner window for unperturbed WFs, one can choose the eigenvectors in $\WF$ to satisfy
\begin{equation} \label{eq:wfpt_utilde0_conv}
\widetilde{U}\oord{0}_{np;\mb{k}} = \delta_{n,p} \quad{\rm for}\quad p=1,\cdots,\NFk\,.
\end{equation}
Since the basis states are the perturbed wavefunctions, the zeroth-order eigenstates of $P_{\npkp,\nk}$ are
\begin{equation} \label{eq:wfpt_psitilde_def}
    \ket{\phi\oord{0}_{p\mb{k}}}
    = \sum_{n=1}^{N_{\mathrm{D},\mb{k}}} \ket{\psi_{\nk}} \widetilde{U}\oord{0}_{np;\mb{k}}.
\end{equation}

The $N_{\rm P}$ energy eigenstates inside $\WP$, the inner window for the perturbed system, are chosen without further disentanglement:
\begin{equation} \label{eq:wfpt_dis_basis_noortho_inside_WP}
    \left\{ \ket{\phi\oord{0}_{p\mb{k}}} : p = 1, \cdots, \NPk \right\}\,.
\end{equation}

Next, we choose the remaining $N_\mb{k} \NWan- N_{\rm P}$ states for the disentangled subspace as follows.
We apply first-order perturbation theory to $\ket{\phi^\ord{0}_{p\mb{k}}}$ for $p=\NPk+1,\cdots,\NWan$,
which are the zeroth-order eigenstates of $P_{\npkp,\nk}$ inside $\WD-\WP$ with eigenvalue 1.
This choice maximizes the perturbed eigenvalues because the unperturbed eigenvalues of the states that are not selected are 0.
The mixing between the selected states is not important because our goal here is to determine the subspace spanned by them.
The WFs that are obtained by projection depend only on the selected subspace, not on its basis~\cite{2001Souza}.
Hence, we use nondegenerate perturbation theory and only consider the mixing of the eigenstates of $P_{{\rm W};\mb{k}}\oord{0}$ with eigenvalue 1 with those with eigenvalue 0.
\begin{widetext}
Writing the first-order matrix elements in \myeqref{eq:wfpt_proj_guess} as $P\oord{1}_{\npkp,\nk}$, we find that $\ket{\phi_{p\mb{k}}\oord{0}}$ is perturbed as
\begin{align} \label{eq:wfpt_psitilde_pert}
    &\ket{\phi_{p\mb{k}}} - \ket{\phi\oord{0}_{p\mb{k}}} \nnnl
    &= \lambda \sum_{\mb{k'}} \sum_{q=\NWan+1}^{N_{\mathrm{D},\mb{k'}}} \sum_{n,n'} \ket{\phi\oord{0}_{q\mb{k'}}} \braket{\phi\oord{0}_{q\mb{k'}}}{\psi_\npkp} P\oord{1}_{\npkp,\nk} \braket{\psi_\nk}{\phi\oord{0}_{p\mb{k}}} + \mcO(\lambda^2) \nnnl
    &= \lambda \primesum{n,m,n',\mb{k'}} \sum_{q=\NWan+1}^{N_{\mathrm{D},\mb{k'}}} \ket{\phi\oord{0}_{q\mb{k'}}} \widetilde{U}^{(0)\dagger}_{qn';\mb{k'}}
    \left( P_{{\rm W}, n'm;\mb{k'}}^\ord{0} \frac{\mel{\psizero_\mkp}{\opV}{\psizero_\nk}}{\veps^\ord{0}_\nk - \veps^\ord{0}_\mkp}
    - \frac{\mel{\psizero_\npkp}{\opV}{\psizero_\mk}}{\veps^\ord{0}_\mk - \veps^\ord{0}_\npkp}
    P_{{\rm W}, mn;\mb{k}}^\ord{0} \right)
    \widetilde{U}^\ord{0}_{np;\mb{k}}
    + \mcO(\lambda^2) \nnnl
    &= -\lambda \sum_{m} (\opPD - \opPW) \ket{\psi_\mk^\ord{1}} \widetilde{U}^\ord{0}_{mp;\mb{k}}
    + \mcO(\lambda^2).
\end{align}
In the third equality, we used the fact that for $p=\NPk+1, ..., \NWan$ and $q=\NWan+1, ..., \NDk$, $\widetilde{U}^\ord{0}_{np;\mb{k}}$ and $\widetilde{U}^{(0)\dagger}_{qn';\mb{k'}}$ are the right and left eigenvectors of $P_{\mathrm{W};\mb{k}}^\ord{0}$ with eigenvalues 1 and 0, respectively.

The unperturbed state $\ket{\phi_{p\mb{k}}\oord{0}}$ defined in \myeqref{eq:wfpt_psitilde_def} also contains an $\mcO(\lambda^1)$ contribution because the basis states are $\ket{\psi_\nk}$'s, the perturbed eigenstates.
Using \myeqref{eq:wfpt_wf_series}, the perturbative expansion of the disentangled state $\ket{\phi_{p\mb{k}}}$ becomes
\begin{equation} \label{eq:wfpt_psitilde_series}
    \ket{\phi_{p\mb{k}}}
    = \sum_{n} \ket{\psizero_\nk} \widetilde{U}\oord{0}_{np;\mb{k}}
    + \lambda \sum_{m}
    \left( \hat{\one} - (\opPD - \opPW) \right)
    \ket{\psi_\mk\oord{1}} \widetilde{U}\oord{0}_{mp;\mb{k}}
    + \mcO(\lambda^2)
\end{equation}
for $p=\NPk+1, ..., \NWan$.

Combining results in \myeqref{eq:wfpt_utilde0_conv} and \myeqref{eq:wfpt_psitilde_series}, the final disentangled subspace for the perturbed system is spanned by
\begin{equation} \label{eq:wfpt_dis_basis_noortho}
    \bigcup_{\mb{k}} \left( \left\{ \ket{\phi\oord{0}_{p\mb{k}}} : p = 1, \cdots, \NPk \right\} 
    \cup \left\{ \ket{\phi_{p\mb{k}}} : p = N_{{\rm P},\mb{k}} + 1, \cdots, \NWan \right\} \right).
\end{equation}
The dimension of the disentangled subspace is $N_\mb{k} \NWan$, as desired.

Next, to make computation simpler, we linearly combine the basis states and write them in a simpler form.
For $\lambda=0$, the disentangled subspace [\myeqref{eq:wfpt_dis_basis_noortho}] is identical to the unperturbed Wannier subspace.
Hence, we can linearly combine the basis states to make their first-order terms orthogonal to the unperturbed Wannier subspace.
Mathematically, this orthogonalization is equivalent to applying $\left(\hat{\one}-\opPW\right)$ to the first-order terms:
\begin{align} \label{eq:wfpt_wf_dis}
    \ket{\widetilde{\phi}_\pk}
    = \begin{dcases}
    \sum_m \left[ \ket{\psi_\mk\oord{0}} 
    + \lambda  \left( \hat{\one} - \opPW \right) \ket{\psi_\mk^\ord{1}}
    \right] \widetilde{U}\oord{0}_{mp;\mb{k}} + \mcO(\lambda^2)
    & \text{for } p=1,\cdots,\NPk, \\
    \sum_m \left[ \ket{\psi_\mk\oord{0}} 
    + \lambda  \left( \hat{\one} - \opPD \right) \ket{\psi_\mk^\ord{1}}
    \right] \widetilde{U}\oord{0}_{mp;\mb{k}} + \mcO(\lambda^2)
    & \text{for } p=\NPk+1,\cdots,\NWan.
    \end{dcases}
\end{align}
This application of $\left(\hat{\one}-\opPW\right)$ guarantees that the basis states for the disentangled subspace, $\ket{\widetilde{\phi}_\pk}$'s, are orthonormal to each other with errors of order $\mcO(\lambda^2)$. Although the final form of the WFs of the perturbed system is not affected by this orthonormalization, it makes the following analysis simpler.

Using \myeqref{eq:wfpt_utilde0_conv} and the fact that $1 \leq m \leq \NPk$ is equivalent to $\mk \in \WP$, we find
\begin{align} \label{eq:wfpt_wf_dis_rewrite}
    \ket{\widetilde{\phi}_\pk}
    =& \sum_m \Big[ \ket{\psi_\mk\oord{0}} + \lambda \opQD \ket{\psi^\ord{1}_\mk} \Big] \widetilde{U}\oord{0}_{mp;\mb{k}}
    + \lambda \sum_{\substack{m\\ \mk \in \WP}} \Big[ \left( \opPD - \opPW \right) \ket{\psi^\ord{1}_\mk}
    \Big] \widetilde{U}\oord{0}_{mp;\mb{k}} + \mcO(\lambda^2)\,.
\end{align}
\end{widetext}

Finally, we compute the perturbed WFs.
We follow the scheme of Ref.~\cite{1997Marzari}, where the initial guess functions are projected onto the disentangled subspace and then orthonormalized by L\"owdin orthogonalization.
This procedure does not change the subspace but fixes the unitary transformation within the subspace to give localized WFs.
The overlap matrix $A$ between the initial guesses and the disentangled basis states is
\begin{align} \label{eq:wfpt_dis_A}
    A_{\nk,\iR}
    \equiv \braket{\widetilde{\phi}_{n\mb{k}}}{w_\iR\oord{0}}
    = \frac{1}{\sqrt{N_k}} (\widetilde{U}^{(0)\dagger}_{\mb{k}} U\oord{0}_{\mb{k}})_{ni} \expmixy{\mb{k}}{\mb{R}} + \mcO(\lambda^2).
\end{align}
The first-order term of $A_{\nk,\iR}$ vanishes because we linearly combined the disentangled basis functions so that their first-order term is orthogonal to the unperturbed Wannier subspace [\myeqref{eq:wfpt_wf_dis}].
We perform L\"owdin orthogonalization to find the gauge matrix and multiply it by $\ket{\widetilde{\phi}_\nk}$ to obtain the WFs for the perturbed system:
\begin{align} \label{eq:wfpt_WF_using_A}
    \ket{w_\iR (\lambda)} = \sum_{\nk} \ket{\widetilde{\phi}_\nk} \left[ A (A^\dagger A)^{-1/2} \right]_{\nk,\iR}.
\end{align}
From \myeqref{eq:wfpt_dis_A}, one can show that the gauge matrix satisfies
\begin{align} \label{eq:wfpt_A_Lowdin}
    \left[ A (A^\dagger A)^{-1/2} \right]_{\nk,\iR}
    = A_{\nk,\iR} + \mcO(\lambda^2).
\end{align}
Using Eqs.~(\ref{eq:wfpt_wf_dis_rewrite}-\ref{eq:wfpt_A_Lowdin}), we obtain \myeqref{eq:wfpt_WF_result} and \myeqref{eq:wfpt_WFP}.

\section{Necessity of all terms in the WFP} \label{sec:app_cases}
In this section, we discuss four possible modifications to the WFP formula [\myeqref{eq:wfpt_WFP}] that might seem reasonable but turn out to be undesirable.
The two most important properties of the WFPs are that they accurately represent the wavefunction perturbations and that they are spatially localized.
We demonstrate that the modifications make the WFPs lose either one of these two properties.

\begin{widetext}
Let us consider the following cases. \\
\emph{Case 0}: The true WFPs as defined in \myeqref{eq:wfpt_WFP}. \\
\emph{Case 1}: Keep only the first term in \myeqref{eq:wfpt_WFP}. This case is equivalent to using no inner window for the Wannierization of the perturbed system, or, in other words, $\WP = \emptyset$.
\begin{equation} \label{eq:s_wfpt_WFP_case1}
    \ket{w^{(1),\ \mathrm{Case 1}}_\iR}
    = \frac{1}{\sqrt{N_k}} \sum_{m,\mb{k}} \opQD \ket{\psi^\ord{1}_\mk} U\oord{0}_{mi,\mb{k}} e^{-i\mb{k}\cdot\mb{R}}
\end{equation}
\emph{Case 2}: Keep only the second term in \myeqref{eq:wfpt_WFP}. In other words, ignore the contribution from states in subspace \circled{4}, which are outside the outer window $\WD$.
\begin{equation} \label{eq:s_wfpt_WFP_case2}
    \ket{w^{(1),\ \mathrm{Case 2}}_\iR}
    = \frac{1}{\sqrt{N_k}} \sum_{\substack{m,\mb{k}\\ \mk \in \WP}} \left( \opPD - \opPW \right) \ket{\psi^\ord{1}_\mk} U\oord{0}_{mi;\mb{k}} e^{-i\mb{k}\cdot\mb{R}}
\end{equation}
\emph{Case 3}: Do not impose the condition $\mk \in \WP$ in the second term of \myeqref{eq:wfpt_WFP}.
\begin{align} \label{eq:s_wfpt_WFP_case3}
    \ket{w^{(1),\ \mathrm{Case 3}}_\iR}
    =& \frac{1}{\sqrt{N_k}} \sum_{m,\mb{k}} \opQD \ket{\psi^\ord{1}_\mk} U\oord{0}_{mi,\mb{k}} e^{-i\mb{k}\cdot\mb{R}}
    + \frac{1}{\sqrt{N_k}} \sum_{m,\mb{k}} \left( \opPD - \opPW \right) \ket{\psi^\ord{1}_\mk} U\oord{0}_{mi;\mb{k}} e^{-i\mb{k}\cdot\mb{R}} \nnnl
    =& \frac{1}{\sqrt{N_k}} \sum_{m,\mb{k}} \left( \hat{\one} - \opPW \right) \ket{\psi^\ord{1}_\mk} U\oord{0}_{mi;\mb{k}} e^{-i\mb{k}\cdot\mb{R}}
\end{align}
\emph{Case 4}: Impose the condition $\mk \in \WP$ not only in the second term, but also in the first term of \myeqref{eq:wfpt_WFP}. This case is equivalent to not using an outer window for the Wannierization of the perturbed system to include all bands, or in other words, $\WD = (-\infty, \infty)$.
\begin{align} \label{eq:s_wfpt_WFP_case4}
    \ket{w^{(1),\ \mathrm{Case 4}}_\iR}
    =& \frac{1}{\sqrt{N_k}} \sum_{\substack{m,\mb{k}\\ \mk \in \WP}} \opQD \ket{\psi^\ord{1}_\mk} U\oord{0}_{mi,\mb{k}} e^{-i\mb{k}\cdot\mb{R}}
    + \frac{1}{\sqrt{N_k}} \sum_{\substack{m,\mb{k}\\ \mk \in \WP}} \left( \opPD - \opPW \right) \ket{\psi^\ord{1}_\mk} U\oord{0}_{mi;\mb{k}} e^{-i\mb{k}\cdot\mb{R}} \nnnl
    =& \frac{1}{\sqrt{N_k}} \sum_{\substack{m,\mb{k}\\ \mk \in \WP}} \left( \hat{\one} - \opPW \right) \ket{\psi^\ord{1}_\mk} U\oord{0}_{mi;\mb{k}} e^{-i\mb{k}\cdot\mb{R}}
\end{align}
\end{widetext}

To study whether the modified WFPs can accurately and efficiently interpolate the wavefunction perturbation, we calculate the following matrix element which we call the ``smoothed Sternheimer matrix element'':
\begin{align} \label{eq:s_Xmn_def}
    X_{mn\mb{k};\mb{q}}
    =& \sum_{\substack{p=1 \\ p\mb{k+q} \in \WP}}^{\infty} \frac{(h_{pm\mb{k};\mb{q}\nu})^* h_{pn\mb{k};\mb{q}\nu}}{\vepszero_\mk - \vepszero_{p\mb{k+q}}} \times f(\vepszero_\mk - \vepszero_{p\mb{k+q}}) \nnnl
    +& \sum_{\substack{p=1 \\ p\mb{k+q} \notin \WP}}^{\infty} \frac{(h_{pm\mb{k};\mb{q}\nu})^* h_{pn\mb{k};\mb{q}\nu}}{\vepszero_\mk - \vepszero_{p\mb{k+q}}},
\end{align}
where
\begin{equation}
    f(x) = 1 - e^{-(x / \sigma)^2}.
\end{equation}
Here, we defined
\begin{equation}
    h_{mn\mb{k},\mb{q}\nu} = \sum_{\kappa,a} \mel{u_{m\mb{k+q}}}{\partial_{\mb{q}\kappa a} \opH_\mathrm{KS}}{u_\nk} U_{\kappa a,\nu}(\mb{q}),
\end{equation}
where $\partial_{\mb{q}\kappa a} \opH_\mathrm{KS}$ is the derivative of the Kohn-Sham Hamiltonian with respect to a monochromatic displacement of atom $\kappa$ along Cartesian direction $a$ with wavevector $\mb{q}$ and $U_{\kappa a,\nu}(\mb{q})$ the displacement pattern for phonon eigenmode $\nu$ in units of the inverse square root of mass.
The $h$ matrix element is related to the electron-phonon matrix element $g$, which is in units of energy, as
\begin{equation} \label{eq:s_g_and_h}
    g_{mn\mb{k},\mb{q}\nu} = \frac{1}{\sqrt{2\omega_{\qnu}}} h_{mn\mb{k},\mb{q}\nu}.
\end{equation}

The form of the matrix element $X_{mn\mb{k};\mb{q}}$ in \myeqref{eq:s_Xmn_def} is similar to the static Fan self-energy except that the smoothing factor $f(\vepszero_\mk - \vepszero_{p\mb{k+q}})$ suppresses the contribution from states $p\mb{k+q}$ that are energetically close to $\mk$.
Since our goal is to study whether the contribution of the high-energy bands is accurately captured, we apply this factor to suppress the contribution from the low-energy bands.
In the following, we set $\sigma$=0.5~eV.

\begin{figure}[htbp]
\includegraphics[width=1.0\columnwidth]{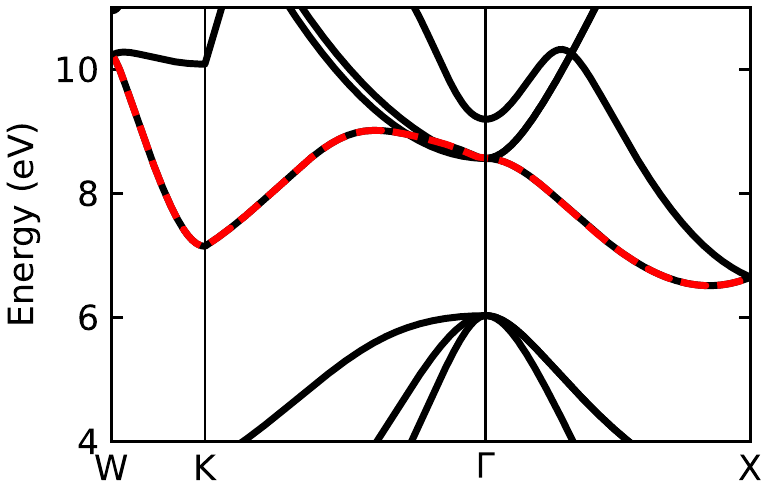}
\caption{
Band structure of silicon. Matrix elements shown in Fig.~\ref{fig:s_cases_mel} are for the states on the band denoted with the red dashed curve.
}
\label{fig:s_cases_band}
\end{figure}

\begin{figure*}[htbp]
\includegraphics[width=1.0\textwidth]{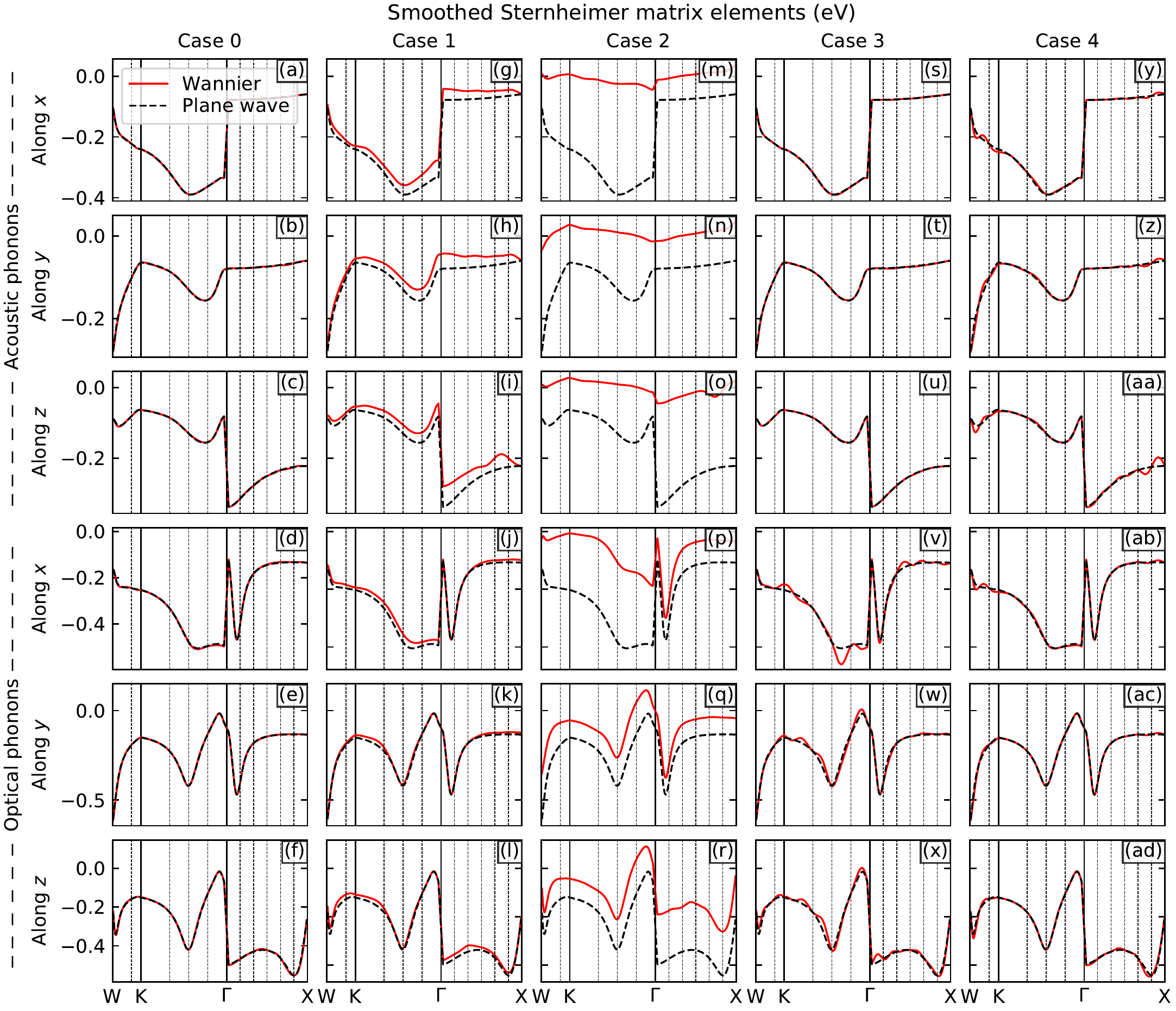}
\caption{
Smoothed Sternheimer matrix elements for silicon obtained from an exact calculation using the plane-wave basis and from interpolations using different modifications of the WFP formula.
The vertical dashed lines indicate $k$ points on the coarse $k$ grid where the interpolation should be exact.
}
\label{fig:s_cases_mel}
\end{figure*}

\begin{figure}[htbp]
\includegraphics[width=1.0\columnwidth]{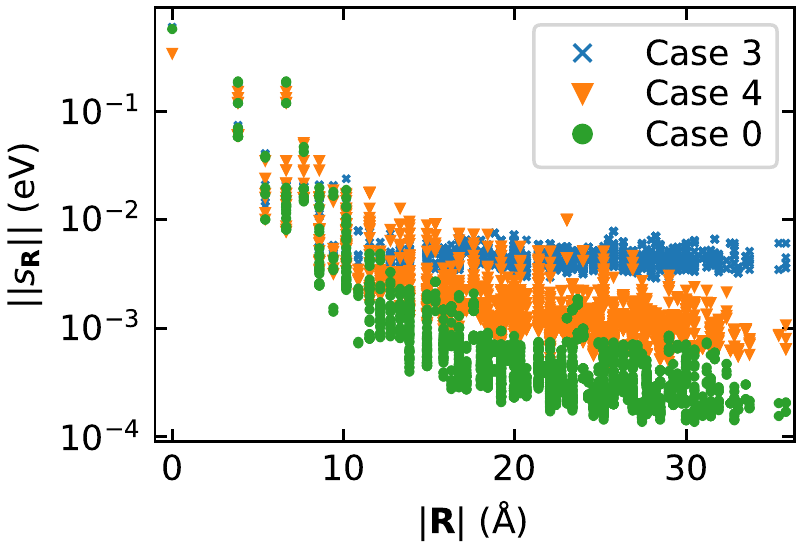}
\caption{
Spatial decay of the WFP matrix elements for silicon.
Each data point corresponds to the maximum absolute value over all WF pairs and atomic displacements for each lattice vector $\mb{R}$. The $y$ axes are on a logarithmic scale.
Only the $\mb{q}=0$ case is considered.
}
\label{fig:s_cases_decay}
\end{figure}

We calculate the smoothed Sternheimer matrix elements using the WFPT interpolation and the exact plane-wave evaluation of \myeqref{eq:s_Xmn_def}.
For the WFPT calculation, we use \myeqref{eq:int_s_wfp} to find
\begin{align} \label{eq:s_sth_wfpt}
    X_{mn\mb{k};\mb{q}}
    =& \sum_{\substack{p=1 \\ p\mb{k+q} \in \WP}}^{\NWan} \frac{(\widetilde{h}^\ord{H}_{pm\mb{k};\mb{q}\nu})^* h^\ord{H}_{pn\mb{k};\mb{q}\nu}}{\veps^\ord{H0}_\mk - \veps^\ord{H0}_{p\mb{k+q}}} \times f(\veps^\ord{H0}_\mk - \veps^\ord{H0}_{p\mb{k+q}}) \nnnl
    +& \sum_{\substack{p=1 \\ p\mb{k+q} \notin \WP}}^{\NWan} \frac{(\widetilde{h}^\ord{H}_{pm\mb{k};\mb{q}\nu})^* h^\ord{H}_{pn\mb{k};\mb{q}\nu}}{\veps^\ord{H0}_\mk - \veps^\ord{H0}_{p\mb{k+q}}}
    + s^\ord{H}_{mn\mb{k};\qnu}.
\end{align}
Note that $\widetilde{h}$ is equal to $h$ for states inside $\WP$, just as we discussed for $\widetilde{g}$ and $g$ in Sec.~\ref{sec:theory_example}.
For the plane-wave calculation, we solve the following Sternheimer equation to evaluate the infinite sum in the second term of \myeqref{eq:s_Xmn_def}, in which we use the fact that $p\mb{k+q} \notin \WP$ if $p \geq 19$.
\begin{align}
    &\left( \opH_\mathrm{KS} - \veps_\mk \right) \ket{\delta_{\mb{q}\kappa a} \psi_\mk} \nnnl
    =& -\left( \sum_{p=19}^{\infty} \ket{\psi_{p\mb{k+q}}}\bra{\psi_{p\mb{k+q}}} \right) \left( \partial_{\mb{q}\kappa a} \opH_\mathrm{KS} \right) \ket{\psi_\mk}
\end{align}
The contribution of the bands with $p \leq 18$ to \myeqref{eq:s_Xmn_def} is obtained by directly calculating the eigenvalues and the matrix elements.

We calculate the smoothed Sternheimer matrix element of silicon for $\mb{q}=0$ and $m=n$ along the band path shown in Fig.~\ref{fig:s_cases_band}.
Figure~\ref{fig:s_cases_mel} shows the comparison for the five cases and for all six phonon modes.
For case 0, the Wannier interpolation is accurate for all $k$ points and phonon modes.
For cases 1 and 2, the Wannier interpolation does not reproduce the exact result, even at the coarse $k$ points denoted by the vertical dashed lines.
This result indicates that the two terms in \myeqref{eq:wfpt_WFP} are both required to make the WFPs accurately represent the perturbation of the wavefunctions.
For cases 3 and 4, the Wannier interpolation is accurate at the coarse $k$ points, indicating that the WFPs at least represent the wavefunction perturbations.
This property can be easily shown from \myeqref{eq:s_wfpt_WFP_case3} and \myeqref{eq:s_wfpt_WFP_case4} as the summands are exact wavefunction perturbations for $\mk \in \WP$.
However, the interpolation displays wiggles [Figs.~\ref{fig:s_cases_mel}(v-ab)], which imply that the WFPs and the matrix elements are not localized enough.

The spatial decay of the WFP matrix elements for cases 0, 3, and 4 are shown in Fig.~\ref{fig:s_cases_decay}. (We did not plot the decay for cases 1 and 2 because the WFPs for cases 1 and 2 fail to represent the wavefunction perturbation so they are of no use even if they are spatially localized.)
We find that the matrix elements in cases 3 and 4 are much more delocalized than those in case 0, consistent with the interpolation quality in Fig.~\ref{fig:s_cases_mel}.

This result can be understood as follows.
For case 3, because the condition $\mk \in \WP$ is removed, the denominator in \myeqref{eq:wfpt_wfp_second} can be close to zero making the summand a rapidly varying and diverging function of $\mb{k}$.
Thus, the corresponding WFPs are highly delocalized.
For case 4, the outer window for Wannierization is set for the unperturbed system but not for the perturbed system.
Hence, the difference between the unperturbed and perturbed WFs will be greater than in case 0 where the outer window is consistently used.
Therefore, the WFPs constructed with case 4 are more delocalized than case 0.

In passing, we note that the good quality of the interpolation for the acoustic modes for case 3 [Figs.~\ref{fig:s_cases_mel}(s-u)] is an exceptional case resulting from the acoustic sum rule.
Because of the translational invariance, the matrix elements of the acoustic phonon modes at $\mb{q}=0$ satisfy
\begin{equation}
    h_{mn\mb{k};\mb{0}\nu}
    \sim \mel{\psi_\mk}{i\comm{\opH}{\hat{p}}}{\psi_\nk}
    = i(\vepszero_\mk - \vepszero_\nk) p_{mn\mb{k}}
\end{equation}
where $p_{mn\mb{k}}$ is the matrix element of the momentum operator~\cite{2020LihmAHC}.
Thus, when the denominator in \myeqref{eq:wfpt_wfp_second} becomes close to zero, the numerator also becomes small.

To summarize, all four modifications to the definition of the WFPs considered in this section either fail to produce WFPs that accurately represent the wavefunction perturbations or generate delocalized WFPs.

\section{Smoothed inner window for WFPT} \label{sec:app_smooth_window}
In this section, we discuss how to remove the discontinuity due to the $\mk \in \WP$ condition in \myeqref{eq:wfpt_WFP} by smoothing the inner window in the disentanglement step for constructing the projection-only WFs.
The issue we aim to solve is that the inner window is applied in an ``all-or-nothing'' manner.
So, the solution is to assign weights to the eigenstates that are 1 for states inside the inner window and continuously decay to 0 outside the inner window.

Concretely, we define a coefficient $l_\nk$ for each state $\nk$ inside $\WF - \WP$.
The coefficient should decay from 1 to 0 as the unperturbed eigenvalue moves away from the boundary energy for $\WP$.
In Appendix~\ref{sec:app_wfp_derivation}, in the disentanglement step, we projected $\opPW$ to the perturbed eigenstates inside $\WD - \WP$ and diagonalized the resulting matrix $P_{\npkp,\nk}$ [\myeqref{eq:wfpt_proj_guess}].
Here, instead of projecting to $\WD - \WP$, we use the following operator:
\begin{align} \label{eq:s_smooth_qdef}
    \opQ_\mathrm{smooth}
    =& \sum_{\nk \in \WD - \WF} \ket{\psi_\nk}\bra{\psi_\nk} \nnnl
    +& \sum_{\nk \in \WF - \WP} \frac{1}{1 - l_\nk} \ket{\psi_\nk}\bra{\psi_\nk}.
\end{align}
For each state $\nk$ inside $\WF - \WP$, the weight of the projector $\ket{\psi_\nk}\bra{\psi_\nk}$ in $\opQ_\mathrm{smooth}$ increases as $\vepszero_\nk$ approaches the boundary energy of $\WP$ and $l_\nk$ approaches 1.
Then, we diagonalize
\begin{equation}
    \opP_\mathrm{smooth} = \opQ_\mathrm{smooth} \opPW \opQ_\mathrm{smooth}
\end{equation}
and take the $N_\mb{k}\NWan - N_{\rm P}$ states with the largest eigenvalues.
If all $l_\nk=0$,
\begin{equation} \label{eq:s_smooth_q_f0}
    \opQ_\mathrm{smooth} = \sum_{\nk \in \WD - \WP} \ket{\psi_\nk}\bra{\psi_\nk}
\end{equation}
holds and the original scheme of Ref.~\cite{2001Souza} is obtained.

Now, we calculate the eigenstates of $\opP_\mathrm{smooth}$ using perturbation theory.
Writing the zeroth-order eigenstates as $\ket{\phi\oord{0}_\pk}$ for $p=\NPk+1,\cdots,\NDk$, and using \myeqref{eq:wfpt_utilde0_conv}, we have
\begin{equation}
    \opP_\mathrm{smooth}(\lambda=0) \ket{\phi\oord{0}_\pk} =
    \begin{dcases}
        \frac{\ket{\phi\oord{0}_\pk}}{(1-l_\pk)^2} & :\ \NPk < p \leq \NFk \\
        \ket{\phi\oord{0}_\pk} & :\ \NFk < p \leq \NWan \\
        0 & :\ \NWan < p \leq \NDk.
    \end{dcases}
\end{equation}
The total number of states with nonzero unperturbed eigenvalues of $\opP_\mathrm{smooth}$ is $N_\mb{k}\NWan - N_{\rm P}$, and the perturbed version of these states will be selected for Wannierization.
Since the mixing between the selected states is not important because our goal here is to determine the subspace spanned by them, we only consider the mixing between the eigenstates of $\opP_\mathrm{smooth}(\lambda=0)$ with nonzero eigenvalues and those with zero eigenvalues.

\begin{widetext}
Using first-order perturbation theory in a similar manner to \myeqref{eq:wfpt_psitilde_pert}, the perturbed eigenstates of $\opP_\mathrm{smooth}$ are
\begin{align} \label{eq:s_smooth_phi_perturb}
    \ket{\phi_\pk} - \ket{\phi\oord{0}_\pk} =
    \begin{dcases}
        -\lambda (1-l_\pk) \sum_{m} (\opPD - \opPW) \ket{\psi_\mk^\ord{1}} \widetilde{U}^\ord{0}_{mp;\mb{k}}
    + \mcO(\lambda^2) & \text{for } p = \NPk+1, \cdots, \NFk \\
        -\lambda \sum_{m} (\opPD - \opPW) \ket{\psi_\mk^\ord{1}} \widetilde{U}^\ord{0}_{mp;\mb{k}}
    + \mcO(\lambda^2) & \text{for } p = \NFk+1, \cdots, \NWan
    \end{dcases}
\end{align}
The only difference with \myeqref{eq:wfpt_psitilde_pert} is the $\left(1-l_\pk\right)$ factor for $p = \NPk+1, \cdots, \NFk$.

Finally, we include the states inside $\WP$ ($\ket{\phi\oord{0}_\pk}$ for $p=1,\cdots,\NPk$) and orthogonalize them to $\mathcal{O}(\lambda)$ as done in \myeqref{eq:wfpt_wf_dis}.
With \myeqref{eq:wfpt_utilde0_conv}, the resulting basis states for the disentangled subspace are
\begin{align} \label{eq:s_smooth_wf_dis}
    \ket{\widetilde{\phi}_\pk} = \begin{dcases}
        \ket{\psi_\pk\oord{0}} + \lambda  \left( \hat{\one} - \opPW \right) \ket{\psi_\pk^\ord{1}} + \mcO(\lambda^2)
    & \text{for } p=1,\cdots,\NPk, \\
        \ket{\psi_\pk\oord{0}} + \lambda  \left( \hat{\one} - \opPD + l_\pk (\opPD - \opPW) \right) \ket{\psi_\pk^\ord{1}} + \mcO(\lambda^2)
    & \text{for } p=\NPk+1,\cdots,\NFk \\
        \sum_m \left[ \ket{\psi_\mk\oord{0}}
    + \lambda  \left( \hat{\one} - \opPD \right) \ket{\psi_\mk^\ord{1}}
    \right] \widetilde{U}\oord{0}_{mp;\mb{k}} + \mcO(\lambda^2)
    & \text{for } p=\NFk+1,\cdots,\NWan.
    \end{dcases}
\end{align}
Note the additional $l_\pk (\opPD - \opPW)$ factor for the $p = \NPk+1, \cdots, \NFk$ case, which is absent in \myeqref{eq:wfpt_wf_dis}.

By projecting and orthogonalizing the unperturbed WFs to the disentangled subspace, the WFPs become
\begin{align} \label{eq:s_smooth_WFP}
    \ket{w^{(1)}_\iR}
    =& \frac{1}{\sqrt{N_k}}
    \sum_{m,\mb{k}}
    \opQD \ket{\psi^\ord{1}_\mk}
    U\oord{0}_{mi,\mb{k}} e^{-i\mb{k}\cdot\mb{R}}
    + \frac{1}{\sqrt{N_k}} \sum_{\substack{m,\mb{k}\\ \mk \in \WP}}
    \left( \opPD - \opPW \right) \ket{\psi^\ord{1}_\mk}
    U\oord{0}_{mi;\mb{k}} e^{-i\mb{k}\cdot\mb{R}} \nnnl
    &+ \frac{1}{\sqrt{N_k}} \sum_{\substack{m,\mb{k}\\ \mk \in \WF-\WP}}
    l_\mk \left( \opPD - \opPW \right) \ket{\psi^\ord{1}_\mk}
    U\oord{0}_{mi;\mb{k}} e^{-i\mb{k}\cdot\mb{R}}.
\end{align}
The only difference between \myeqref{eq:s_smooth_WFP} and the original formula for WFPs [\myeqref{eq:wfpt_WFP}] is that \myeqref{eq:s_smooth_WFP} has the third term.
The factor $l_\mk$ in this term smoothly changes from 1 to 0 as $\vepszero_\mk$ moves from the boundary energy of $\WP$ to that of $\WF$.
Hence, this third term eliminates the discontinuity generated by the second term of \myeqref{eq:s_smooth_WFP} while maintaining the desired property that the WFPs should accurately represent the wavefunction perturbation of the states inside $\WP$.
Therefore, \myeqref{eq:s_smooth_WFP} is an alternative way to create WFPs and is expected to give more localized WFPs than \myeqref{eq:wfpt_WFP}.
We detail how the real-space matrix elements can be calculated with this modified definition of WFPs in Appendix~\ref{sec:app_real_space_smooth}.

For the systems studied in this work, we find that the WFPs constructed without smoothing the inner window are already sufficiently localized.
Smoothing the inner window will be useful in systems where the WFP delocalization due to the discontinuity at the boundary of the inner window is a problem.

\section{WFP correction to the Hamiltonian matrix element} \label{sec:app_wf_interpol}
In this appendix, we prove \myeqref{eq:int_HW1}.
\begin{align} \label{eq:int_HW1_proof}
    H^\ord{\rm W1}_{i'\mb{k'},i\mb{k}}
    =& \frac{1}{N_k} \sum_{\mb{R},\mb{R'},\mb{q}} 
    \Bigg[ \mel{w^\ord{0}_{i'\mb{R'}}}{\opV_\mb{q}}{w^\ord{0}_{i\mb{R}}}
    + \mel{w^\ord{0}_{i'\mb{R'}}}{\opH^\ord{0}}{w^\ord{1}_{\iR;\mb{q}}}
    + \mel{w^\ord{1}_{i'\mb{R'};-\mb{q}}}{\opH^\ord{0}}{w^\ord{0}_\iR} \Bigg]  \expmixy{\mb{k'}}{\mb{R'}} e^{i\mb{k}\cdot\mb{R}} \nnnl
    =& \frac{1}{N_k} \sum_{\mb{R},\mb{R'},\mb{q}} 
    \Bigg[ \mel{w^\ord{0}_{i'\mb{R'}}}{\opV_\mb{q}}{w^\ord{0}_{i\mb{R+R'}}}
    + \mel{w^\ord{0}_{i'\mb{R'}}}{\opH^\ord{0}}{w^\ord{1}_{i\mb{R+R'};\mb{q}}}
    + \mel{w^\ord{1}_{i'\mb{R'};-\mb{q}}}{\opH^\ord{0}}{w^\ord{0}_{i\mb{R+R'}}} \Bigg]  \expmixy{\mb{k'}}{\mb{R'}} e^{i\mb{k}\cdot(\mb{R+R'})} \nnnl
    =& \frac{1}{N_k} \sum_{\mb{R},\mb{R'},\mb{q}} 
    \left( g_{i'i\mb{R};\mb{q}} + \delta g_{i'i\mb{R};\mb{q}} \right)
    e^{i\mb{q}\cdot\mb{R'}} \expmixy{\mb{k'}}{\mb{R'}} e^{i\mb{k}\cdot(\mb{R+R'})} \nnnl
    =& \sum_{\mb{R},\mb{q}}
    \left( g_{i'i\mb{R};\mb{q}} + \delta g_{i'i\mb{R};\mb{q}} \right)
    e^{i\mb{k}\cdot\mb{R}} \delta_{\mb{k}+\mb{q},\mb{k'}} \nnnl
    =& \widetilde{g}^\mathrm{(W)}_{i'i\mb{k};\mb{k'-k}}
\end{align}

\section{Interpolation of matrix elements for Kubo formula} \label{sec:app_interp_kubo_formula}
In this section, we discuss the interpolation of the $K_{mn\mb{k}}$ matrix element defined in \myeqref{eq:kubo_k_def}.
Wannier interpolation of $K_{mn\mb{k}}$ can be done analogously to that of $S_{mn\mb{k}}$ in Eqs.~(\ref{eq:int_s_R_def}-\ref{eq:int_s_wfp}).
For $\nk,\mk \in \WP$, one can use the Wannier interpolation of the wavefunction perturbation [\myeqref{eq:int_wf_H1}] to find
\begin{align} \label{eq:s_kubo_k_derivation}
    K_{mn\mb{k}}
    =& \braket{\psi^\ord{\rm H1}_{\mk;\mb{q}}}{\psi'^{\rm (H1)}_{\nk;\mb{q}}} \nnnl
    =& \frac{1}{N_k} \sum_{i,i',\mb{R},\mb{R'}}
    e^{i\mb{k}\cdot(\mb{R}-\mb{R'})}
    V^{\dagger\ord{0}}_{mi';\mb{k}} \braket{w^{(1)}_{i'\mb{R'};\mb{q}}} {w'^{(1)}_{\iR;\mb{q}}} V^\ord{0}_{in;\mb{k}}
    + \primesum{p=1}^{\NWan}\frac{\left( \widetilde{g}^{\rm (H)}_{pm\mb{k};\mb{q}} \right)^* \widetilde{g}'^{\rm (H)}_{pn\mb{k};\mb{q}}}{(\veps^\ord{H0}_\mk - \veps^\ord{H0}_{p\mb{k+q}})(\veps^\ord{H0}_\nk - \veps^\ord{H0}_{p\mb{k+q}})} \nnnl
    =& \sum_{i,i',\mb{R}} V^{\dagger\ord{0}}_{mi';\mb{k}} k_{i'i\mb{R};\mb{q}} V^\ord{0}_{in;\mb{k}} e^{i\mb{k}\cdot\mb{R}}
    + \primesum{p=1}^{\NWan}\frac{\left( \widetilde{g}^{\rm (H)}_{pm\mb{k};\mb{q}} \right)^* \widetilde{g}'^{\rm (H)}_{pn\mb{k};\mb{q}}}{(\veps^\ord{H0}_\mk - \veps^\ord{H0}_{p\mb{k+q}})(\veps^\ord{H0}_\nk - \veps^\ord{H0}_{p\mb{k+q}})}.
\end{align}
In the third equality, we used the discrete translation properties of the WFP [\myeqref{eq:rs_WFP_trans}] and renamed $\mb{R}-\mb{R'}$ to $\mb{R}$.
We also used the real-space matrix element defined in \myeqref{eq:kubo_k_R_def}.

Using \myeqref{eq:kubo_k_RtoW} and \myeqref{eq:kubo_k_WtoH}, one can rewrite \myeqref{eq:s_kubo_k_derivation} as \myeqref{eq:kubo_k_wfp}.
Since all quantities appearing in \myeqref{eq:kubo_k_wfp} can be calculated using a WF- or WFP-based interpolation, $K_{mn\mb{k}}$ can also be interpolated.

\section{Calculation of real-space matrix elements} \label{sec:app_real_space}
\subsection{Without the smoothed inner window}
In this appendix, we describe how to calculate the real-space matrix elements using the Bloch eigenstates computed at a coarse $k$ grid.
For the potential matrix element $g_{\mb{R};\mb{q}}$, Ref.~\cite{2007GiustinoEPW} gives
\begin{equation} \label{eq:k2r_g}
    g_{ij\mb{R};\mb{q}}
    = \frac{1}{N_k} \sum_{m,n,\mb{k}} U^{(0)\dagger}_{im;\mb{k+q}} \mel{\psizero_\mkq}{\opV_\mb{q}}{\psizero_\nk} U^\ord{0}_{nj;\mb{k}} e^{-i\mb{k}\cdot\mb{R}},
\end{equation}
which easily follows from the definition of $g_{\mb{R};\mb{q}}$ [\myeqref{eq:int_g_def}].
Here, we derive analogous expressions for the newly introduced matrices $\delta g_{\mb{R};\mb{q}}$ and $s_{\mb{R};\mb{q}}$.

First, for $\delta g_{\mb{R};\mb{q}}$, we start from its definition [Eq.~\eqref{eq:int_deltag_def}] and use Eqs.~(\ref{eq:wfpt_WF0_def}, \ref{eq:wfpt_WFP}) to find
\begin{align} \label{eq:k2r_deltag}
    \delta g_{ij\mb{R};\mb{q}}
    =& \frac{1}{N_k} \sum_{\substack{m,n,p,\mb{k} \\ \nk \in \WP}} U^{(0)\dagger}_{im;\mb{k+q}}
    \frac{\veps^\ord{0}_\mkq (P\oord{0}_\mathrm{D} - P\oord{0}_{\mathrm{W}})_{mp;\mb{k+q}} \mel{\psizero_{p\mb{k+q}}}{\opV_\mb{q}}{\psizero_\nk}}{\veps^\ord{0}_\nk - \veps^\ord{0}_{p\mb{k+q}}} U^\ord{0}_{nj;\mb{k}} e^{-i\mb{k}\cdot\mb{R}} \nnnl
    +& \frac{1}{N_k} \sum_{\substack{m,n,p,\mb{k} \\ \mkq \in \WP}}
    U^{(0)\dagger}_{im;\mb{k+q}}
    \frac{\mel{\psizero_\mkq}{\opV_\mb{q}}{\psizero_{p\mb{k}}} (P\oord{0}_\mathrm{D} - P^\ord{0}_{\mathrm{W}})_{pn;\mb{k}} \veps^\ord{0}_\nk}{\veps^\ord{0}_\mkq - \veps^\ord{0}_{p\mb{k}}}
    U^\ord{0}_{nj;\mb{k}} e^{-i\mb{k}\cdot\mb{R}}.
\end{align}
In \myeqref{eq:k2r_deltag}, all band indices are limited to the bands inside $\WD$.
The $\opQD \ket{\psi_\mk^\ord{1}}$ term in \myeqref{eq:wfpt_WFP} does not appear because the WFs are orthogonal to the $\opQD$ subspace.
Thus, $\delta g_{ij\mb{R};\mb{q}}$ can be evaluated using only the energies and matrix elements for the states inside $\WD$.

For $s_{\mb{R};\mb{q}}$, using the definitions [Eqs.~(\ref{eq:wfpt_WF0_def}, \ref{eq:wfpt_WFP}, \ref{eq:int_s_R_def})], we find
\begin{align} \label{eq:k2r_s_R}
    s_{ij\mb{R};\mb{q}}
    =& \frac{1}{N_k} \sum_{m,n,\mb{k}}
    U^{(0)\dagger}_{im,\mb{k}}
    \mel{\psi^\ord{1}_{\mk;\mb{q}}}{\opQD\opV'_\mb{q}}{\psizero_\nk}
    U^\ord{0}_{nj,\mb{k}} e^{-i\mb{k}\cdot\mb{R}} \nnnl
    +& \frac{1}{N_k} \primesum{\substack{m,n,p,q,\mb{k}\\ \mk \in \WP}}
    U^{(0)\dagger}_{im,\mb{k}}
    \frac{\mel{\psizero_\mk}{\opV_\mb{q}^\dagger}{\psizero_\pkq}}{\veps^\ord{0}_\mk - \veps^\ord{0}_\pkq}
    (P\oord{0}_{\rm D} - P\oord{0}_{{\rm W}})_{pq;\mb{k+q}}
    \mel{\psizero_{q\mb{k+q}}}{\opV'_\mb{q}}{\psizero_\nk}
    U^\ord{0}_{nj;\mb{k}} e^{-i\mb{k}\cdot\mb{R}}.
\end{align}
The first term involves the wavefunction perturbation, which can be calculated by solving the Sternheimer equation.
The second term refers only to the bands inside the outer window.

For $k_{ij\mb{R};\mb{q}}$, we similarly find
\begin{align} \label{eq:s_kubo_k_k2r}
    k_{ij\mb{R};\mb{q}}
    =& \sum_{m,n,\mb{k}}
    U^{(0)\dagger}_{im;\mb{k}}
    \mel{\psi\oord{1}_{\mk;\mb{q}}}{\opQD}{\psi'^{(1)}_{\nk;\mb{q}}}
    U\oord{0}_{nj;\mb{k}} \expmixy{\mb{k}}{\mb{R}} \nnnl
    +& \frac{1}{N_k} \sum_{\substack{m,n,p,q,\mb{k}\\ \nk, \mk \in \WP}}
    U^{(0)\dagger}_{im,\mb{k}}
    \frac{\mel{\psizero_\mk}{\opV_\mb{q}^\dagger}{\psizero_\pkq}}{\veps^\ord{0}_\mk - \veps^\ord{0}_\pkq}
    (P\oord{0}_{{\rm D}} - P\oord{0}_{{\rm W}})_{pq;\mb{k+q}}
    \frac{\mel{\psizero_{q\mb{k+q}}}{\opV'_\mb{q}}{\psizero_\nk}}{\veps^\ord{0}_\nk - \veps^\ord{0}_{q\mb{k+q}}}
    U^\ord{0}_{nj;\mb{k}} e^{-i\mb{k}\cdot\mb{R}}.
\end{align}

\subsection{With the smoothed inner window} \label{sec:app_real_space_smooth}
In this subsection, we detail how the formulas for the real-space matrix elements change if the WFPs are constructed with the smoothed inner window as described in Appendix~\ref{sec:app_smooth_window}.
For brevity, we define
\begin{equation}
    \widetilde{l}_\mk =
    \begin{cases}
    1 & \text{: $\mk \in \WP$} \\
    l_\mk & \text{: $\mk \in \WF - \WP$}
    \end{cases}
\end{equation}
and rewrite \myeqref{eq:s_smooth_WFP} as
\begin{align} \label{eq:s_smooth_WFP_2}
    \ket{w^{(1)}_\iR}
    = \frac{1}{\sqrt{N_k}}
    \sum_{m,\mb{k}}
    \opQD \ket{\psi^\ord{1}_\mk}
    U\oord{0}_{mi,\mb{k}} e^{-i\mb{k}\cdot\mb{R}}
    + \frac{1}{\sqrt{N_k}} \sum_{\substack{m,\mb{k}\\ \mk \in \WF}}
    \widetilde{l}_\mk \left( \opPD - \opPW \right) \ket{\psi^\ord{1}_\mk}
    U\oord{0}_{mi;\mb{k}} e^{-i\mb{k}\cdot\mb{R}}.
\end{align}
Then, one finds that the real-space matrix elements are
\begin{align} \label{eq:s_smooth_k2r_deltag}
    \delta g_{ij\mb{R};\mb{q}}
    =& \frac{1}{N_k} \sum_{\substack{m,n,p,\mb{k} \\ \nk \in \WF}} U^{(0)\dagger}_{im;\mb{k+q}}
    \frac{\veps^\ord{0}_\mkq (P\oord{0}_\mathrm{D} - P\oord{0}_{\mathrm{W}})_{mp;\mb{k+q}} \mel{\psizero_{p\mb{k+q}}}{\opV_\mb{q}}{\psizero_\nk}}{\veps^\ord{0}_\nk - \veps^\ord{0}_{p\mb{k+q}}} \widetilde{l}_\nk U^\ord{0}_{nj;\mb{k}} e^{-i\mb{k}\cdot\mb{R}} \nnnl
    +& \frac{1}{N_k} \sum_{\substack{m,n,p,\mb{k} \\ \mkq \in \WF}}
    U^{(0)\dagger}_{im;\mb{k+q}} \widetilde{l}_\mkq
    \frac{\mel{\psizero_\mkq}{\opV_\mb{q}}{\psizero_{p\mb{k}}} (P\oord{0}_\mathrm{D} - P^\ord{0}_{\mathrm{W}})_{pm;\mb{k}} \veps^\ord{0}_\nk}{\veps^\ord{0}_\mkq - \veps^\ord{0}_{p\mb{k}}}
    U^\ord{0}_{nj;\mb{k}} e^{-i\mb{k}\cdot\mb{R}},
\end{align}
\begin{align} \label{eq:s_smooth_k2r_s}
    s_{ij\mb{R};\mb{q}}
    =& \frac{1}{N_k} \sum_{m,n,\mb{k}}
    U^{(0)\dagger}_{im,\mb{k}}
    \mel{\psi^\ord{1}_{\mk;\mb{q}}}{\opQD\opV'_\mb{q}}{\psizero_\nk}
    U^\ord{0}_{nj,\mb{k}} e^{-i\mb{k}\cdot\mb{R}} \nnnl
    +& \frac{1}{N_k} \primesum{\substack{m,n,p,q,\mb{k}\\ \mk \in \WF}}
    U^{(0)\dagger}_{im,\mb{k}} \widetilde{l}_\mk
    \frac{\mel{\psizero_\mk}{\opV_\mb{q}^\dagger}{\psizero_\pkq}}{\veps^\ord{0}_\mk - \veps^\ord{0}_\pkq}
    (P\oord{0}_{\rm D} - P\oord{0}_{{\rm W}})_{pq;\mb{k+q}}
    \mel{\psizero_{q\mb{k+q}}}{\opV'_\mb{q}}{\psizero_\nk}
    U^\ord{0}_{nj;\mb{k}} e^{-i\mb{k}\cdot\mb{R}},
\end{align}
and
\begin{align} \label{eq:s_smooth_k2r_k}
    k_{ij\mb{R};\mb{q}}
    =& \sum_{m,n,\mb{k}}
    U^{(0)\dagger}_{im;\mb{k}}
    \mel{\psi\oord{1}_{\mk;\mb{q}}}{\opQD}{\psi'^{(1)}_{\nk;\mb{q}}}
    U\oord{0}_{nj;\mb{k}} \expmixy{\mb{k}}{\mb{R}} \nnnl
    +& \frac{1}{N_k} \sum_{\substack{m,n,p,q,\mb{k}\\ \nk, \mk \in \WF}}
    U^{(0)\dagger}_{im,\mb{k}} \widetilde{l}_\mk
    \frac{\mel{\psizero_\mk}{\opV_\mb{q}^\dagger}{\psizero_\pkq}}{\veps^\ord{0}_\mk - \veps^\ord{0}_\pkq}
    (P\oord{0}_{{\rm D}} - P\oord{0}_{{\rm W}})_{pq;\mb{k+q}}
    \frac{\mel{\psizero_{q\mb{k+q}}}{\opV'_\mb{q}}{\psizero_\nk}}{\veps^\ord{0}_\nk - \veps^\ord{0}_{q\mb{k+q}}}
    \widetilde{l}_\nk U^\ord{0}_{nj;\mb{k}} e^{-i\mb{k}\cdot\mb{R}}.
\end{align}
When $l_\pk=0$, which means that the inner window is not smoothed at all, \myeqref{eq:s_smooth_k2r_deltag}, \myeqref{eq:s_smooth_k2r_s}, and \myeqref{eq:s_smooth_k2r_k} reduce to \myeqref{eq:k2r_deltag}, \myeqref{eq:k2r_s_R}, and \myeqref{eq:s_kubo_k_k2r}, respectively.
\end{widetext}

\section{Improved double-grid method for phonon-induced electron self-energy} \label{sec:app_doublegrid}
In this section, we discuss the improved double-grid method for accelerating the convergence of the phonon-induced electron self-energy with respect to the coarse $k$ grid size.
Here, we focus on the diagonal part of the self-energy, but the results can also be applied to the off-diagonal one~\cite{2020LihmAHC}.
Our aim is to calculate $\Sigma_\nk$ for all $\nk \in \WP$.

The convergence of the self-energy with respect to the $q$-point sampling is slow.
This convergence rate is dominated by the convergence of the lower Fan self-energy~\cite{2015PonceJCP}.
The convergence of the upper Fan term with respect to the $q$-point sampling is much faster because the denominator $\veps_\nk - \veps_\mkq$ in \myeqref{eq:s_upfan_def} is never small.
Therefore, one can save considerable computational cost by calculating the rapidly convergent upper Fan self-energy at a coarse $q$-point grid and the slowly convergent lower Fan self-energy at a finer $q$-point grid.
\begin{equation} \label{eq:s_dg_simple}
    \Sigma_\nk(\eta)
    = \sum_{\mb{q},\nu}^\mathrm{coarse} \left[ \Sigma^{\rm upper\ Fan}_\nk + \Sigma^{\rm DW}_\nk \right]
    + \sum_{\mb{q},\nu}^\mathrm{fine} \left[ \Sigma^{\rm lower\ Fan}_\nk(\eta) \right]
\end{equation}
This method is the double-grid method~\cite{2020BrownAltvaterAHC}.
To distinguish this method from the improved version we propose later, we call it the ``conventional double-grid method.''
Note that one should also calculate the DW self-energy at the coarse grid when using the double-grid method.
The reason for this is that the convergence of the sum of the DW and the upper Fan self-energies is much faster than the convergence of each term because of the approximate cancellation between the two terms~\cite{1976Allen,2019QueralesFlores,2020LihmAHC}.

\begin{figure*}[htbp]
\includegraphics[width=1.0\textwidth]{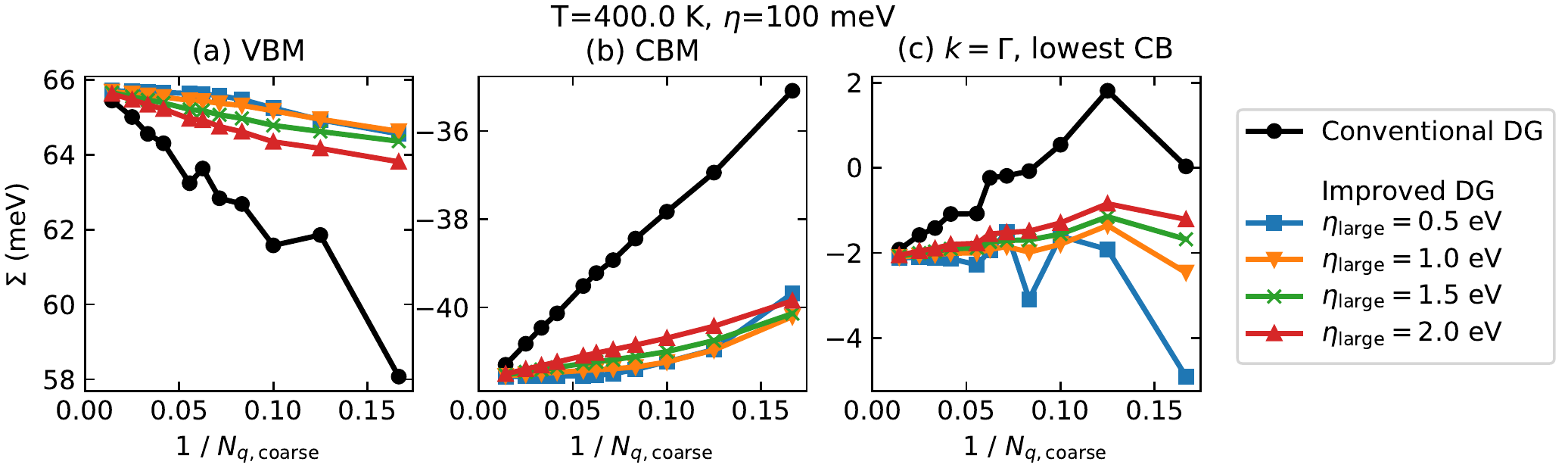}
\caption{
Phonon-induced electron self-energy of silicon at $T$=400~K calculated using the conventional and improved double-grid (DG) methods for the (a) VBM, (b) CBM, and the lowest conduction band (CB) at $\Gamma$.
The fine $q$ grid size and $\eta$ for the self-energy are fixed to $125\times125\times125$ and 100~meV, respectively.
}
\label{fig:s_dg_doublegrid}
\end{figure*}

However, we found that the conventional double-grid method is not very successful in the case of the finite-temperature self-energy of silicon.
Figure~\ref{fig:s_dg_doublegrid} shows that the sum of the upper Fan and DW self-energy (black dots) converges as $1/N_q$ if an $N_q \times N_q \times N_q$ $q$-point grid is used.
When a $6\times6\times6$ coarse grid is used, the error in the band-gap renormalization at 400~K is 14~meV, around 13\% of the converged value.
Therefore, we need to speed up the convergence.

\begin{figure*}[htbp]
\includegraphics[width=1.0\textwidth]{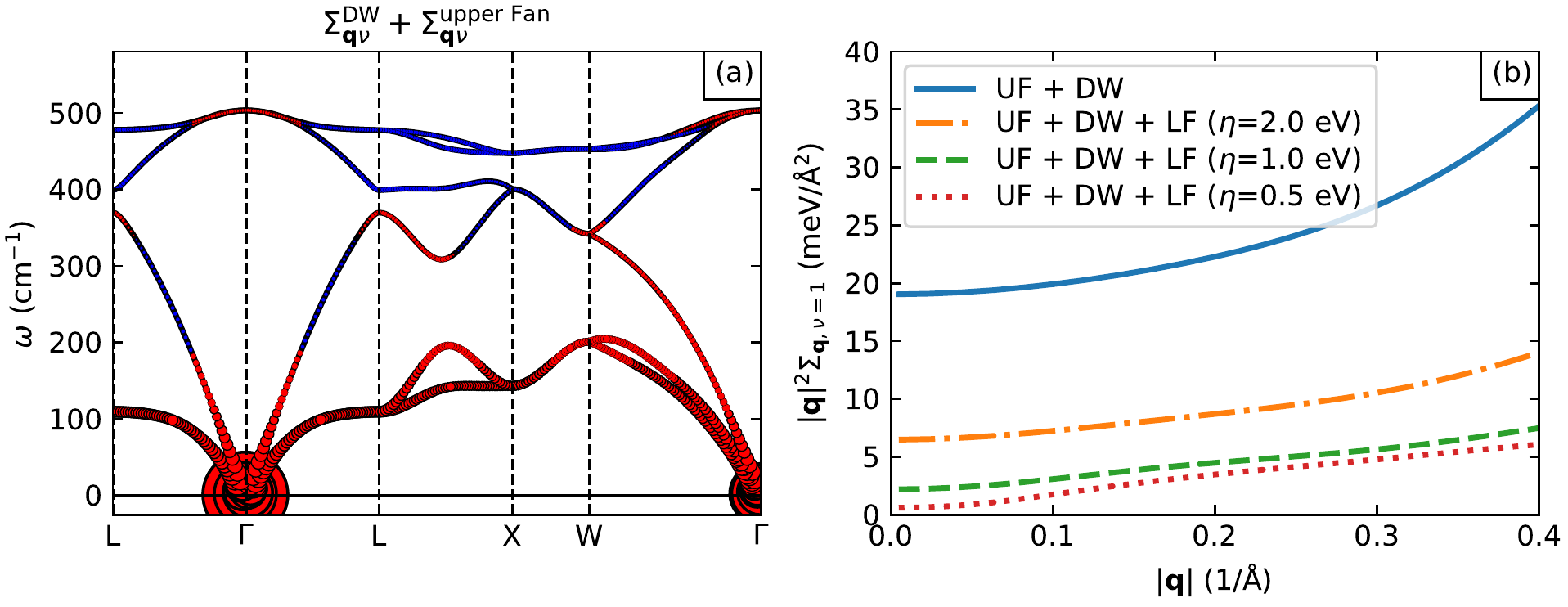}
\caption{
(a) Contribution of each phonon mode to the sum of the DW and upper Fan self-energies of silicon for the electron state at the valence band maximum. The size and color of the circles indicate the magnitude and sign of the phonon-mode-resolved electron self-energy, where red (blue) means a positive (negative) contribution.
(b) Sum of the upper Fan (UF), DW, and optionally lower Fan (LF) self-energies for the lowest-frequency ($\nu$=1) acoustic mode multiplied by $\abs{\mb{q}}^2$ for $q$ points along the $\mathrm{\Gamma}$-L line.
}
\label{fig:s_dg_mode}
\end{figure*}

By investigating the contribution of each phonon mode to the self-energy, we find that the contribution of the low-energy acoustic phonons dominates the slow convergence rate as shown in Fig.~\ref{fig:s_dg_mode}(a).
The slow $1/N_q$ convergence can be understood as follows.
Let us focus on the acoustic modes at small $|{\bf q}|$ satisfying $\omega_\qnu \approx v \abs{\mathbf{q}}$.
We set the momentum cutoff $q_{\rm cut} \ll 2\pi/a$ for the low-energy modes where $a$ is the lattice parameter.
We assume that the temperature is high enough to satisfy $v q_{\rm cut} \ll k_\mathrm{B} T$ so that $2n_\qnu + 1 \approx 2k_\mathrm{B}T / \hbar \omega_\qnu$.
To make the $q$ dependence explicit, we write $g_{mn\mb{k},\mb{q}\nu}$ in terms of $h_{mn\mb{k},\mb{q}\nu}$ using \myeqref{eq:s_g_and_h} and approximate $h_{mn\mb{k},\mb{q}\nu}$ by $h_{mn\mb{k},\mb{0}\nu}$.
Note that $h$, not $g$, should be approximated because $g$ has an additional $|{\bf q}|$ dependence in the $1/\sqrt{\omega_\qnu}$ factor in \myeqref{eq:s_g_and_h}.
Then, the sum of \myeqref{eq:s_upfan_def} for $\abs{\mathbf{q}} < q_{\rm cut}$ becomes
\begin{align} \label{eq:s_upfan_acoustic}
    \Sigma^{\rm upper\ Fan,\ acoustic}_{n\mb{k}}
    \approx& \sum_{m \notin \WP} \frac{\abs{h_{mn\mb{k};\mb{0}\nu}}^2}{\veps_\nk - \veps_\mk} \int_{\abs{\mathbf{q}}<q_{\rm cut}} d\mb{q} \frac{k_\mathrm{B}T}{\hbar v^2 \abs{\mathbf{q}}^2} \nnnl
    =& q_{\rm cut} \frac{4\pi k_\mathrm{B}T}{\hbar v^2} \sum_{m \notin \WP} \frac{\abs{h_{mn\mb{k};\mb{0}\nu}}^2}{\veps_\nk - \veps_\mk}.
\end{align}
The acoustic phonon contribution to the upper Fan self-energy is proportional to $q_{\rm cut}$.
Using a coarse $q$ grid gives an effective cutoff $q_{\rm cut} \sim 1/N_q$ because the acoustic phonon modes with $q < 1/N_q \times 2\pi/a$ are not sampled.
Therefore, the conventional double-grid method converges as $1/N_q$.
Note that the convergence is much faster at $T=0$ because then $2n_\qnu + 1 = 1$ does not give a $1/q$ factor, so $\Sigma^{\rm upper\ Fan,\ acoustic}_{n\mb{k}}$ becomes proportional to $q_{\rm cut}^2 \sim 1/N_q^2$.

To speed up the convergence, we utilize the fact that the acoustic modes around $\mathbf{q}=0$ actually have little effect on the total self-energy because of the translational symmetry.
Since the acoustic modes at $\mathbf{q}=0$ describe a uniform translation of the whole lattice, the band eigenvalues should not change.
However, the sum of the phonon-mode-resolved upper Fan term and the DW term is not small and even diverges at $\mathbf{q}=0$.
This divergence is canceled by the divergence of the lower Fan term in the opposite sign, so their sum, the total self-energy, is zero at $\mathbf{q}=0$.

To make use of this cancellation, we rewrite the total self-energy as follows.
\begin{align} \label{eq:s_dg_improved}
    \Sigma_\nk(\eta)
    =& \sum_{\mb{q},\nu}^\mathrm{coarse} \left[ \Sigma^{\rm upper\ Fan}_\nk + \Sigma^{\rm DW}_\nk + \Sigma^{\rm lower\ Fan}_\nk(\eta_\mathrm{large}) \right] \nnnl
    +& \sum_{\mb{q},\nu}^\mathrm{fine} \left[ \Sigma^{\rm lower\ Fan}_\nk(\eta) - \Sigma^{\rm lower\ Fan}_\nk(\eta_\mathrm{large}) \right]
\end{align}
Here, $\eta$ is the small parameter that should be converged to 0, while $\eta_\mathrm{large}$ is a large value around 1~eV and does not need to be converged.
Equation \myeqref{eq:s_dg_improved} is the ``improved double-grid method'' we propose in this work.
The key property of \myeqref{eq:s_dg_improved} is that the first term converges much faster with respect to the $q$ grid than the first term of \myeqref{eq:s_dg_simple}, which is just the sum of the upper Fan and DW self-energies.
The reason for this is that the diverging contribution of the low-energy acoustic modes in the upper Fan and DW terms is suppressed by the additional lower Fan term.
Figure~\ref{fig:s_dg_mode}(b) shows the self-energy contributions multiplied by the phase-space volume factor $\abs{\mathbf{q}}^2$.
We find that the contribution of modes with small $\abs{\mathbf{q}}$ is indeed suppressed by adding the lower Fan term.

The cancellation becomes exact as $\eta_\mathrm{large}$ is decreased.
However, we need to choose $\eta_\mathrm{large}$ to be large enough so that the denominator in the lower Fan term~[\myeqref{eq:s_lofan_def}] is always far from zero and does not create additional convergence issues.
For example, Fig.~\ref{fig:s_dg_doublegrid}(c) shows a slow convergence of the self-energy for the lowest conduction band at $\mathrm{\Gamma}$ for $\eta_\mathrm{large}=0.5$~eV.
The slow convergence is more pronounced for states far from band extrema because there are more resonant couplings mediated by phonons to other electron states, leading to a rapid variation in the phonon-mode-resolved lower Fan self-energy.
Practically, we find that $\eta_\mathrm{large}$ in [1~eV, 2~eV] is a good choice for silicon.

Figure~\ref{fig:s_dg_doublegrid} shows that the total self-energy converges much faster with respect to the coarse $q$ grid when using the improved double-grid method.
By using the improved double-grid method with $\eta_\mathrm{\large} = 1~\mathrm{eV}$, the absolute (relative) error in the band-gap renormalization at 400~K is only 2.4~meV (2\%) with a $6\times6\times6$ coarse $q$ grid, in contrast to 14~meV (13\%) for the conventional double-grid method.
Also, the choice of $\eta_\mathrm{\large}$ gives only a minor effect on the convergence rate, and more importantly, the convergence is consistently faster than the conventional double-grid method for a wide range of $\eta_\mathrm{\large}$ in [1~eV, 2~eV].

\FloatBarrier 
\makeatletter\@input{xy.tex}\makeatother
\bibliography{main}

\begin{thebibliography}{89}%
\makeatletter
\providecommand \@ifxundefined [1]{%
 \@ifx{#1\undefined}
}%
\providecommand \@ifnum [1]{%
 \ifnum #1\expandafter \@firstoftwo
 \else \expandafter \@secondoftwo
 \fi
}%
\providecommand \@ifx [1]{%
 \ifx #1\expandafter \@firstoftwo
 \else \expandafter \@secondoftwo
 \fi
}%
\providecommand \natexlab [1]{#1}%
\providecommand \enquote  [1]{``#1''}%
\providecommand \bibnamefont  [1]{#1}%
\providecommand \bibfnamefont [1]{#1}%
\providecommand \citenamefont [1]{#1}%
\providecommand \href@noop [0]{\@secondoftwo}%
\providecommand \href [0]{\begingroup \@sanitize@url \@href}%
\providecommand \@href[1]{\@@startlink{#1}\@@href}%
\providecommand \@@href[1]{\endgroup#1\@@endlink}%
\providecommand \@sanitize@url [0]{\catcode `\\12\catcode `\$12\catcode
  `\&12\catcode `\#12\catcode `\^12\catcode `\_12\catcode `\%12\relax}%
\providecommand \@@startlink[1]{}%
\providecommand \@@endlink[0]{}%
\providecommand \url  [0]{\begingroup\@sanitize@url \@url }%
\providecommand \@url [1]{\endgroup\@href {#1}{\urlprefix }}%
\providecommand \urlprefix  [0]{URL }%
\providecommand \Eprint [0]{\href }%
\providecommand \doibase [0]{http://dx.doi.org/}%
\providecommand \selectlanguage [0]{\@gobble}%
\providecommand \bibinfo  [0]{\@secondoftwo}%
\providecommand \bibfield  [0]{\@secondoftwo}%
\providecommand \translation [1]{[#1]}%
\providecommand \BibitemOpen [0]{}%
\providecommand \bibitemStop [0]{}%
\providecommand \bibitemNoStop [0]{.\EOS\space}%
\providecommand \EOS [0]{\spacefactor3000\relax}%
\providecommand \BibitemShut  [1]{\csname bibitem#1\endcsname}%
\let\auto@bib@innerbib\@empty
\bibitem [{\citenamefont {Kohn}(1996)}]{1996KohnNearsighted}%
  \BibitemOpen
  \bibfield  {author} {\bibinfo {author} {\bibfnamefont {W.}~\bibnamefont
  {Kohn}},\ }\bibfield  {title} {\enquote {\bibinfo {title} {Density functional
  and density matrix method scaling linearly with the number of atoms},}\
  }\href {\doibase 10.1103/PhysRevLett.76.3168} {\bibfield  {journal} {\bibinfo
   {journal} {Phys. Rev. Lett.}\ }\textbf {\bibinfo {volume} {76}},\ \bibinfo
  {pages} {3168--3171} (\bibinfo {year} {1996})}\BibitemShut {NoStop}%
\bibitem [{\citenamefont {Prodan}\ and\ \citenamefont
  {Kohn}(2005)}]{2005ProdanNearsighted}%
  \BibitemOpen
  \bibfield  {author} {\bibinfo {author} {\bibfnamefont {E.}~\bibnamefont
  {Prodan}}\ and\ \bibinfo {author} {\bibfnamefont {W.}~\bibnamefont {Kohn}},\
  }\bibfield  {title} {\enquote {\bibinfo {title} {Nearsightedness of
  electronic matter},}\ }\href {\doibase 10.1073/pnas.0505436102} {\bibfield
  {journal} {\bibinfo  {journal} {Proceedings of the National Academy of
  Sciences}\ }\textbf {\bibinfo {volume} {102}},\ \bibinfo {pages}
  {11635--11638} (\bibinfo {year} {2005})}\BibitemShut {NoStop}%
\bibitem [{\citenamefont {Wannier}(1937)}]{1937Wannier}%
  \BibitemOpen
  \bibfield  {author} {\bibinfo {author} {\bibfnamefont {Gregory~H.}\
  \bibnamefont {Wannier}},\ }\bibfield  {title} {\enquote {\bibinfo {title}
  {The structure of electronic excitation levels in insulating crystals},}\
  }\href {\doibase 10.1103/PhysRev.52.191} {\bibfield  {journal} {\bibinfo
  {journal} {Phys. Rev.}\ }\textbf {\bibinfo {volume} {52}},\ \bibinfo {pages}
  {191--197} (\bibinfo {year} {1937})}\BibitemShut {NoStop}%
\bibitem [{\citenamefont {Kohn}(1959)}]{1959Kohn}%
  \BibitemOpen
  \bibfield  {author} {\bibinfo {author} {\bibfnamefont {W.}~\bibnamefont
  {Kohn}},\ }\bibfield  {title} {\enquote {\bibinfo {title} {Analytic
  properties of {Bloch} waves and {Wannier} functions},}\ }\href {\doibase
  10.1103/PhysRev.115.809} {\bibfield  {journal} {\bibinfo  {journal} {Phys.
  Rev.}\ }\textbf {\bibinfo {volume} {115}},\ \bibinfo {pages} {809--821}
  (\bibinfo {year} {1959})}\BibitemShut {NoStop}%
\bibitem [{\citenamefont {Marzari}\ and\ \citenamefont
  {Vanderbilt}(1997)}]{1997Marzari}%
  \BibitemOpen
  \bibfield  {author} {\bibinfo {author} {\bibfnamefont {Nicola}\ \bibnamefont
  {Marzari}}\ and\ \bibinfo {author} {\bibfnamefont {David}\ \bibnamefont
  {Vanderbilt}},\ }\bibfield  {title} {\enquote {\bibinfo {title}
  {Maximally-localized generalized {{Wannier}} functions for composite energy
  bands},}\ }\href {\doibase 10.1103/PhysRevB.56.12847} {\bibfield  {journal}
  {\bibinfo  {journal} {Physical Review B}\ }\textbf {\bibinfo {volume} {56}},\
  \bibinfo {pages} {12847--12865} (\bibinfo {year} {1997})}\BibitemShut
  {NoStop}%
\bibitem [{\citenamefont {Souza}\ \emph {et~al.}(2001)\citenamefont {Souza},
  \citenamefont {Marzari},\ and\ \citenamefont {Vanderbilt}}]{2001Souza}%
  \BibitemOpen
  \bibfield  {author} {\bibinfo {author} {\bibfnamefont {Ivo}\ \bibnamefont
  {Souza}}, \bibinfo {author} {\bibfnamefont {Nicola}\ \bibnamefont {Marzari}},
  \ and\ \bibinfo {author} {\bibfnamefont {David}\ \bibnamefont {Vanderbilt}},\
  }\bibfield  {title} {\enquote {\bibinfo {title} {Maximally localized
  {{Wannier}} functions for entangled energy bands},}\ }\href {\doibase
  10.1103/PhysRevB.65.035109} {\bibfield  {journal} {\bibinfo  {journal}
  {Physical Review B}\ }\textbf {\bibinfo {volume} {65}},\ \bibinfo {pages}
  {035109} (\bibinfo {year} {2001})}\BibitemShut {NoStop}%
\bibitem [{\citenamefont {Marzari}\ \emph {et~al.}(2012)\citenamefont
  {Marzari}, \citenamefont {Mostofi}, \citenamefont {Yates}, \citenamefont
  {Souza},\ and\ \citenamefont {Vanderbilt}}]{2012MarzariRMP}%
  \BibitemOpen
  \bibfield  {author} {\bibinfo {author} {\bibfnamefont {Nicola}\ \bibnamefont
  {Marzari}}, \bibinfo {author} {\bibfnamefont {Arash~A.}\ \bibnamefont
  {Mostofi}}, \bibinfo {author} {\bibfnamefont {Jonathan~R.}\ \bibnamefont
  {Yates}}, \bibinfo {author} {\bibfnamefont {Ivo}\ \bibnamefont {Souza}}, \
  and\ \bibinfo {author} {\bibfnamefont {David}\ \bibnamefont {Vanderbilt}},\
  }\bibfield  {title} {\enquote {\bibinfo {title} {Maximally localized
  {Wannier} functions: Theory and applications},}\ }\href {\doibase
  10.1103/RevModPhys.84.1419} {\bibfield  {journal} {\bibinfo  {journal} {Rev.
  Mod. Phys.}\ }\textbf {\bibinfo {volume} {84}},\ \bibinfo {pages}
  {1419--1475} (\bibinfo {year} {2012})}\BibitemShut {NoStop}%
\bibitem [{\citenamefont {Giustino}\ \emph {et~al.}(2007)\citenamefont
  {Giustino}, \citenamefont {Cohen},\ and\ \citenamefont
  {Louie}}]{2007GiustinoEPW}%
  \BibitemOpen
  \bibfield  {author} {\bibinfo {author} {\bibfnamefont {Feliciano}\
  \bibnamefont {Giustino}}, \bibinfo {author} {\bibfnamefont {Marvin~L.}\
  \bibnamefont {Cohen}}, \ and\ \bibinfo {author} {\bibfnamefont {Steven~G.}\
  \bibnamefont {Louie}},\ }\bibfield  {title} {\enquote {\bibinfo {title}
  {Electron-phonon interaction using {{Wannier}} functions},}\ }\href {\doibase
  10.1103/PhysRevB.76.165108} {\bibfield  {journal} {\bibinfo  {journal}
  {Physical Review B}\ }\textbf {\bibinfo {volume} {76}},\ \bibinfo {pages}
  {165108} (\bibinfo {year} {2007})}\BibitemShut {NoStop}%
\bibitem [{\citenamefont {Zhou}\ and\ \citenamefont
  {Bernardi}(2016)}]{2016ZhouTransport}%
  \BibitemOpen
  \bibfield  {author} {\bibinfo {author} {\bibfnamefont {Jin-Jian}\
  \bibnamefont {Zhou}}\ and\ \bibinfo {author} {\bibfnamefont {Marco}\
  \bibnamefont {Bernardi}},\ }\bibfield  {title} {\enquote {\bibinfo {title}
  {{\emph{Ab Initio}} electron mobility and polar phonon scattering in
  {{GaAs}}},}\ }\href {\doibase 10.1103/PhysRevB.94.201201} {\bibfield
  {journal} {\bibinfo  {journal} {Physical Review B}\ }\textbf {\bibinfo
  {volume} {94}},\ \bibinfo {pages} {201201} (\bibinfo {year}
  {2016})}\BibitemShut {NoStop}%
\bibitem [{\citenamefont {Ponc\'e}\ \emph {et~al.}(2018)\citenamefont
  {Ponc\'e}, \citenamefont {Margine},\ and\ \citenamefont
  {Giustino}}]{2018PonceTransport}%
  \BibitemOpen
  \bibfield  {author} {\bibinfo {author} {\bibfnamefont {Samuel}\ \bibnamefont
  {Ponc\'e}}, \bibinfo {author} {\bibfnamefont {Elena~R.}\ \bibnamefont
  {Margine}}, \ and\ \bibinfo {author} {\bibfnamefont {Feliciano}\ \bibnamefont
  {Giustino}},\ }\bibfield  {title} {\enquote {\bibinfo {title} {Towards
  predictive many-body calculations of phonon-limited carrier mobilities in
  semiconductors},}\ }\href {\doibase 10.1103/PhysRevB.97.121201} {\bibfield
  {journal} {\bibinfo  {journal} {Phys. Rev. B}\ }\textbf {\bibinfo {volume}
  {97}},\ \bibinfo {pages} {121201} (\bibinfo {year} {2018})}\BibitemShut
  {NoStop}%
\bibitem [{\citenamefont {Noffsinger}\ \emph {et~al.}(2012)\citenamefont
  {Noffsinger}, \citenamefont {Kioupakis}, \citenamefont {{Van de Walle}},
  \citenamefont {Louie},\ and\ \citenamefont {Cohen}}]{2012NoffsingerIndabs}%
  \BibitemOpen
  \bibfield  {author} {\bibinfo {author} {\bibfnamefont {Jesse}\ \bibnamefont
  {Noffsinger}}, \bibinfo {author} {\bibfnamefont {Emmanouil}\ \bibnamefont
  {Kioupakis}}, \bibinfo {author} {\bibfnamefont {Chris~G.}\ \bibnamefont {{Van
  de Walle}}}, \bibinfo {author} {\bibfnamefont {Steven~G.}\ \bibnamefont
  {Louie}}, \ and\ \bibinfo {author} {\bibfnamefont {Marvin~L.}\ \bibnamefont
  {Cohen}},\ }\bibfield  {title} {\enquote {\bibinfo {title} {Phonon-{{Assisted
  Optical Absorption}} in {{Silicon}} from {{First Principles}}},}\ }\href
  {\doibase 10.1103/PhysRevLett.108.167402} {\bibfield  {journal} {\bibinfo
  {journal} {Physical Review Letters}\ }\textbf {\bibinfo {volume} {108}},\
  \bibinfo {pages} {167402} (\bibinfo {year} {2012})}\BibitemShut {NoStop}%
\bibitem [{\citenamefont {Margine}\ and\ \citenamefont
  {Giustino}(2013)}]{2013MargineSC}%
  \BibitemOpen
  \bibfield  {author} {\bibinfo {author} {\bibfnamefont {E.~R.}\ \bibnamefont
  {Margine}}\ and\ \bibinfo {author} {\bibfnamefont {F.}~\bibnamefont
  {Giustino}},\ }\bibfield  {title} {\enquote {\bibinfo {title} {Anisotropic
  migdal-eliashberg theory using {Wannier} functions},}\ }\href {\doibase
  10.1103/PhysRevB.87.024505} {\bibfield  {journal} {\bibinfo  {journal} {Phys.
  Rev. B}\ }\textbf {\bibinfo {volume} {87}},\ \bibinfo {pages} {024505}
  (\bibinfo {year} {2013})}\BibitemShut {NoStop}%
\bibitem [{\citenamefont {Jhalani}\ \emph {et~al.}(2017)\citenamefont
  {Jhalani}, \citenamefont {Zhou},\ and\ \citenamefont
  {Bernardi}}]{2017JhalaniUltrafast}%
  \BibitemOpen
  \bibfield  {author} {\bibinfo {author} {\bibfnamefont {Vatsal~A.}\
  \bibnamefont {Jhalani}}, \bibinfo {author} {\bibfnamefont {Jin-Jian}\
  \bibnamefont {Zhou}}, \ and\ \bibinfo {author} {\bibfnamefont {Marco}\
  \bibnamefont {Bernardi}},\ }\bibfield  {title} {\enquote {\bibinfo {title}
  {Ultrafast {{Hot Carrier Dynamics}} in {{GaN}} and {{Its Impact}} on the
  {{Efficiency Droop}}},}\ }\href {\doibase 10.1021/acs.nanolett.7b02212}
  {\bibfield  {journal} {\bibinfo  {journal} {Nano Letters}\ }\textbf {\bibinfo
  {volume} {17}},\ \bibinfo {pages} {5012--5019} (\bibinfo {year}
  {2017})}\BibitemShut {NoStop}%
\bibitem [{\citenamefont {Wang}\ \emph {et~al.}(2006)\citenamefont {Wang},
  \citenamefont {Yates}, \citenamefont {Souza},\ and\ \citenamefont
  {Vanderbilt}}]{2006WangAHC}%
  \BibitemOpen
  \bibfield  {author} {\bibinfo {author} {\bibfnamefont {Xinjie}\ \bibnamefont
  {Wang}}, \bibinfo {author} {\bibfnamefont {Jonathan~R.}\ \bibnamefont
  {Yates}}, \bibinfo {author} {\bibfnamefont {Ivo}\ \bibnamefont {Souza}}, \
  and\ \bibinfo {author} {\bibfnamefont {David}\ \bibnamefont {Vanderbilt}},\
  }\bibfield  {title} {\enquote {\bibinfo {title} {{\emph{Ab Initio}}
  calculation of the anomalous {{Hall}} conductivity by {{Wannier}}
  interpolation},}\ }\href {\doibase 10.1103/PhysRevB.74.195118} {\bibfield
  {journal} {\bibinfo  {journal} {Physical Review B}\ }\textbf {\bibinfo
  {volume} {74}},\ \bibinfo {pages} {195118} (\bibinfo {year}
  {2006})}\BibitemShut {NoStop}%
\bibitem [{\citenamefont {Yates}\ \emph {et~al.}(2007)\citenamefont {Yates},
  \citenamefont {Wang}, \citenamefont {Vanderbilt},\ and\ \citenamefont
  {Souza}}]{2007YatesWannier}%
  \BibitemOpen
  \bibfield  {author} {\bibinfo {author} {\bibfnamefont {Jonathan~R.}\
  \bibnamefont {Yates}}, \bibinfo {author} {\bibfnamefont {Xinjie}\
  \bibnamefont {Wang}}, \bibinfo {author} {\bibfnamefont {David}\ \bibnamefont
  {Vanderbilt}}, \ and\ \bibinfo {author} {\bibfnamefont {Ivo}\ \bibnamefont
  {Souza}},\ }\bibfield  {title} {\enquote {\bibinfo {title} {Spectral and
  {{Fermi}} surface properties from {{Wannier}} interpolation},}\ }\href
  {\doibase 10.1103/PhysRevB.75.195121} {\bibfield  {journal} {\bibinfo
  {journal} {Physical Review B}\ }\textbf {\bibinfo {volume} {75}},\ \bibinfo
  {pages} {195121} (\bibinfo {year} {2007})}\BibitemShut {NoStop}%
\bibitem [{\citenamefont {{Iba{\~n}ez-Azpiroz}}\ \emph
  {et~al.}(2018)\citenamefont {{Iba{\~n}ez-Azpiroz}}, \citenamefont {Tsirkin},\
  and\ \citenamefont {Souza}}]{2018IbanezAzpirozShift}%
  \BibitemOpen
  \bibfield  {author} {\bibinfo {author} {\bibfnamefont {Julen}\ \bibnamefont
  {{Iba{\~n}ez-Azpiroz}}}, \bibinfo {author} {\bibfnamefont {Stepan~S.}\
  \bibnamefont {Tsirkin}}, \ and\ \bibinfo {author} {\bibfnamefont {Ivo}\
  \bibnamefont {Souza}},\ }\bibfield  {title} {\enquote {\bibinfo {title}
  {{\emph{Ab Initio}} calculation of the shift photocurrent by {{Wannier}}
  interpolation},}\ }\href {\doibase 10.1103/PhysRevB.97.245143} {\bibfield
  {journal} {\bibinfo  {journal} {Physical Review B}\ }\textbf {\bibinfo
  {volume} {97}},\ \bibinfo {pages} {245143} (\bibinfo {year}
  {2018})}\BibitemShut {NoStop}%
\bibitem [{\citenamefont {Tsirkin}\ \emph {et~al.}(2018)\citenamefont
  {Tsirkin}, \citenamefont {Puente},\ and\ \citenamefont
  {Souza}}]{2018TsirkinGyro}%
  \BibitemOpen
  \bibfield  {author} {\bibinfo {author} {\bibfnamefont {Stepan~S.}\
  \bibnamefont {Tsirkin}}, \bibinfo {author} {\bibfnamefont {Pablo~Aguado}\
  \bibnamefont {Puente}}, \ and\ \bibinfo {author} {\bibfnamefont {Ivo}\
  \bibnamefont {Souza}},\ }\bibfield  {title} {\enquote {\bibinfo {title}
  {Gyrotropic effects in trigonal tellurium studied from first principles},}\
  }\href {\doibase 10.1103/PhysRevB.97.035158} {\bibfield  {journal} {\bibinfo
  {journal} {Physical Review B}\ }\textbf {\bibinfo {volume} {97}},\ \bibinfo
  {pages} {035158} (\bibinfo {year} {2018})}\BibitemShut {NoStop}%
\bibitem [{\citenamefont {Qiao}\ \emph {et~al.}(2018)\citenamefont {Qiao},
  \citenamefont {Zhou}, \citenamefont {Yuan},\ and\ \citenamefont
  {Zhao}}]{2018QiaoSHC}%
  \BibitemOpen
  \bibfield  {author} {\bibinfo {author} {\bibfnamefont {Junfeng}\ \bibnamefont
  {Qiao}}, \bibinfo {author} {\bibfnamefont {Jiaqi}\ \bibnamefont {Zhou}},
  \bibinfo {author} {\bibfnamefont {Zhe}\ \bibnamefont {Yuan}}, \ and\ \bibinfo
  {author} {\bibfnamefont {Weisheng}\ \bibnamefont {Zhao}},\ }\bibfield
  {title} {\enquote {\bibinfo {title} {Calculation of intrinsic spin {{Hall}}
  conductivity by {{Wannier}} interpolation},}\ }\href {\doibase
  10.1103/PhysRevB.98.214402} {\bibfield  {journal} {\bibinfo  {journal}
  {Physical Review B}\ }\textbf {\bibinfo {volume} {98}},\ \bibinfo {pages}
  {214402} (\bibinfo {year} {2018})}\BibitemShut {NoStop}%
\bibitem [{\citenamefont {Ryoo}\ \emph {et~al.}(2019)\citenamefont {Ryoo},
  \citenamefont {Park},\ and\ \citenamefont {Souza}}]{2019RyooSHC}%
  \BibitemOpen
  \bibfield  {author} {\bibinfo {author} {\bibfnamefont {Ji~Hoon}\ \bibnamefont
  {Ryoo}}, \bibinfo {author} {\bibfnamefont {Cheol-Hwan}\ \bibnamefont {Park}},
  \ and\ \bibinfo {author} {\bibfnamefont {Ivo}\ \bibnamefont {Souza}},\
  }\bibfield  {title} {\enquote {\bibinfo {title} {Computation of intrinsic
  spin {{Hall}} conductivities from first principles using maximally localized
  {{Wannier}} functions},}\ }\href {\doibase 10.1103/PhysRevB.99.235113}
  {\bibfield  {journal} {\bibinfo  {journal} {Physical Review B}\ }\textbf
  {\bibinfo {volume} {99}},\ \bibinfo {pages} {235113} (\bibinfo {year}
  {2019})}\BibitemShut {NoStop}%
\bibitem [{\citenamefont {Sternheimer}(1954)}]{1954Sternheimer}%
  \BibitemOpen
  \bibfield  {author} {\bibinfo {author} {\bibfnamefont {R.~M.}\ \bibnamefont
  {Sternheimer}},\ }\bibfield  {title} {\enquote {\bibinfo {title} {Electronic
  polarizabilities of ions from the {Hartree-Fock} wave functions},}\ }\href
  {\doibase 10.1103/PhysRev.96.951} {\bibfield  {journal} {\bibinfo  {journal}
  {Phys. Rev.}\ }\textbf {\bibinfo {volume} {96}},\ \bibinfo {pages} {951--968}
  (\bibinfo {year} {1954})}\BibitemShut {NoStop}%
\bibitem [{\citenamefont {Baroni}\ \emph {et~al.}(2001)\citenamefont {Baroni},
  \citenamefont {de~Gironcoli}, \citenamefont {Dal~Corso},\ and\ \citenamefont
  {Giannozzi}}]{2001BaroniRMP}%
  \BibitemOpen
  \bibfield  {author} {\bibinfo {author} {\bibfnamefont {Stefano}\ \bibnamefont
  {Baroni}}, \bibinfo {author} {\bibfnamefont {Stefano}\ \bibnamefont
  {de~Gironcoli}}, \bibinfo {author} {\bibfnamefont {Andrea}\ \bibnamefont
  {Dal~Corso}}, \ and\ \bibinfo {author} {\bibfnamefont {Paolo}\ \bibnamefont
  {Giannozzi}},\ }\bibfield  {title} {\enquote {\bibinfo {title} {Phonons and
  related crystal properties from density-functional perturbation theory},}\
  }\href {\doibase 10.1103/RevModPhys.73.515} {\bibfield  {journal} {\bibinfo
  {journal} {Rev. Mod. Phys.}\ }\textbf {\bibinfo {volume} {73}},\ \bibinfo
  {pages} {515--562} (\bibinfo {year} {2001})}\BibitemShut {NoStop}%
\bibitem [{\citenamefont {Ge}\ and\ \citenamefont
  {Lu}(2015)}]{2015GeLocalized}%
  \BibitemOpen
  \bibfield  {author} {\bibinfo {author} {\bibfnamefont {Xiaochuan}\
  \bibnamefont {Ge}}\ and\ \bibinfo {author} {\bibfnamefont {Deyu}\
  \bibnamefont {Lu}},\ }\bibfield  {title} {\enquote {\bibinfo {title} {Local
  representation of the electronic dielectric response function},}\ }\href
  {\doibase 10.1103/PhysRevB.92.241107} {\bibfield  {journal} {\bibinfo
  {journal} {Physical Review B}\ }\textbf {\bibinfo {volume} {92}},\ \bibinfo
  {pages} {241107} (\bibinfo {year} {2015})}\BibitemShut {NoStop}%
\bibitem [{\citenamefont {Giustino}\ \emph
  {et~al.}(2010{\natexlab{a}})\citenamefont {Giustino}, \citenamefont {Louie},\
  and\ \citenamefont {Cohen}}]{2010GiustinoPRL}%
  \BibitemOpen
  \bibfield  {author} {\bibinfo {author} {\bibfnamefont {Feliciano}\
  \bibnamefont {Giustino}}, \bibinfo {author} {\bibfnamefont {Steven~G.}\
  \bibnamefont {Louie}}, \ and\ \bibinfo {author} {\bibfnamefont {Marvin~L.}\
  \bibnamefont {Cohen}},\ }\bibfield  {title} {\enquote {\bibinfo {title}
  {Electron-phonon renormalization of the direct band gap of diamond},}\ }\href
  {\doibase 10.1103/PhysRevLett.105.265501} {\bibfield  {journal} {\bibinfo
  {journal} {Physical Review Letters}\ }\textbf {\bibinfo {volume} {105}},\
  \bibinfo {pages} {265501} (\bibinfo {year} {2010}{\natexlab{a}})}\BibitemShut
  {NoStop}%
\bibitem [{\citenamefont {Gonze}\ \emph {et~al.}(2011)\citenamefont {Gonze},
  \citenamefont {Boulanger},\ and\ \citenamefont {Côté}}]{2011GonzeAnnPhys}%
  \BibitemOpen
  \bibfield  {author} {\bibinfo {author} {\bibfnamefont {X.}~\bibnamefont
  {Gonze}}, \bibinfo {author} {\bibfnamefont {P.}~\bibnamefont {Boulanger}}, \
  and\ \bibinfo {author} {\bibfnamefont {M.}~\bibnamefont {Côté}},\
  }\bibfield  {title} {\enquote {\bibinfo {title} {Theoretical approaches to
  the temperature and zero-point motion effects on the electronic band
  structure},}\ }\href {\doibase 10.1002/andp.201000100} {\bibfield  {journal}
  {\bibinfo  {journal} {Ann. Phys.}\ }\textbf {\bibinfo {volume} {523}},\
  \bibinfo {pages} {168--178} (\bibinfo {year} {2011})}\BibitemShut {NoStop}%
\bibitem [{\citenamefont {Sipe}\ and\ \citenamefont
  {Shkrebtii}(2000)}]{2000SipeBPVE}%
  \BibitemOpen
  \bibfield  {author} {\bibinfo {author} {\bibfnamefont {J.~E.}\ \bibnamefont
  {Sipe}}\ and\ \bibinfo {author} {\bibfnamefont {A.~I.}\ \bibnamefont
  {Shkrebtii}},\ }\bibfield  {title} {\enquote {\bibinfo {title} {Second-order
  optical response in semiconductors},}\ }\href {\doibase
  10.1103/PhysRevB.61.5337} {\bibfield  {journal} {\bibinfo  {journal}
  {Physical Review B}\ }\textbf {\bibinfo {volume} {61}},\ \bibinfo {pages}
  {5337--5352} (\bibinfo {year} {2000})}\BibitemShut {NoStop}%
\bibitem [{\citenamefont {Lihm}\ and\ \citenamefont {Park}()}]{2021LihmSpin}%
  \BibitemOpen
  \bibfield  {author} {\bibinfo {author} {\bibfnamefont {Jae-Mo}\ \bibnamefont
  {Lihm}}\ and\ \bibinfo {author} {\bibfnamefont {Cheol-Hwan}\ \bibnamefont
  {Park}},\ }\href@noop {} {\bibinfo  {journal} {unpublished}\ }\BibitemShut
  {NoStop}%
\bibitem [{\citenamefont {Xu}\ \emph {et~al.}(2021)\citenamefont {Xu},
  \citenamefont {Wang}, \citenamefont {Zhou},\ and\ \citenamefont
  {Li}}]{2020XuSpinPhotocurrent}%
  \BibitemOpen
\bibfield  {journal} {  }\bibfield  {author} {\bibinfo {author} {\bibfnamefont
  {Haowei}\ \bibnamefont {Xu}}, \bibinfo {author} {\bibfnamefont {Hua}\
  \bibnamefont {Wang}}, \bibinfo {author} {\bibfnamefont {Jian}\ \bibnamefont
  {Zhou}}, \ and\ \bibinfo {author} {\bibfnamefont {Ju}~\bibnamefont {Li}},\
  }\bibfield  {title} {\enquote {\bibinfo {title} {Pure spin photocurrent in
  non-centrosymmetric crystals: Bulk spin photovoltaic effect},}\ }\href
  {\doibase 10.1038/s41467-021-24541-7} {\bibfield  {journal} {\bibinfo
  {journal} {Nature Communications}\ }\textbf {\bibinfo {volume} {12}},\
  \bibinfo {pages} {4330} (\bibinfo {year} {2021})}\BibitemShut {NoStop}%
\bibitem [{\citenamefont {Mu}\ \emph {et~al.}(2021)\citenamefont {Mu},
  \citenamefont {Pan},\ and\ \citenamefont {Zhou}}]{2020MuSpinPhotocurrent}%
  \BibitemOpen
  \bibfield  {author} {\bibinfo {author} {\bibfnamefont {Xingchi}\ \bibnamefont
  {Mu}}, \bibinfo {author} {\bibfnamefont {Yiming}\ \bibnamefont {Pan}}, \ and\
  \bibinfo {author} {\bibfnamefont {Jian}\ \bibnamefont {Zhou}},\ }\bibfield
  {title} {\enquote {\bibinfo {title} {Pure bulk orbital and spin photocurrent
  in two-dimensional ferroelectric materials},}\ }\href {\doibase
  10.1038/s41524-021-00531-7} {\bibfield  {journal} {\bibinfo  {journal} {npj
  Computational Materials}\ }\textbf {\bibinfo {volume} {7}},\ \bibinfo {pages}
  {61} (\bibinfo {year} {2021})}\BibitemShut {NoStop}%
\bibitem [{\citenamefont {He}\ and\ \citenamefont
  {Vanderbilt}(2001)}]{2001HeWannier}%
  \BibitemOpen
  \bibfield  {author} {\bibinfo {author} {\bibfnamefont {Lixin}\ \bibnamefont
  {He}}\ and\ \bibinfo {author} {\bibfnamefont {David}\ \bibnamefont
  {Vanderbilt}},\ }\bibfield  {title} {\enquote {\bibinfo {title} {Exponential
  decay properties of wannier functions and related quantities},}\ }\href
  {\doibase 10.1103/PhysRevLett.86.5341} {\bibfield  {journal} {\bibinfo
  {journal} {Phys. Rev. Lett.}\ }\textbf {\bibinfo {volume} {86}},\ \bibinfo
  {pages} {5341--5344} (\bibinfo {year} {2001})}\BibitemShut {NoStop}%
\bibitem [{\citenamefont {Cornean}\ \emph {et~al.}(2019)\citenamefont
  {Cornean}, \citenamefont {Gontier}, \citenamefont {Levitt},\ and\
  \citenamefont {Monaco}}]{2019CorneanWannier}%
  \BibitemOpen
  \bibfield  {author} {\bibinfo {author} {\bibfnamefont {Horia~D}\ \bibnamefont
  {Cornean}}, \bibinfo {author} {\bibfnamefont {David}\ \bibnamefont
  {Gontier}}, \bibinfo {author} {\bibfnamefont {Antoine}\ \bibnamefont
  {Levitt}}, \ and\ \bibinfo {author} {\bibfnamefont {Domenico}\ \bibnamefont
  {Monaco}},\ }\bibfield  {title} {\enquote {\bibinfo {title} {Localised
  {Wannier} functions in metallic systems},}\ }in\ \href {\doibase
  10.1007/s00023-019-00767-6} {\emph {\bibinfo {booktitle} {Annales Henri
  Poincar{\'e}}}},\ Vol.~\bibinfo {volume} {20}\ (\bibinfo {organization}
  {Springer},\ \bibinfo {year} {2019})\ pp.\ \bibinfo {pages}
  {1367--1391}\BibitemShut {NoStop}%
\bibitem [{\citenamefont {Macfarlane}\ \emph {et~al.}(1958)\citenamefont
  {Macfarlane}, \citenamefont {McLean}, \citenamefont {Quarrington},\ and\
  \citenamefont {Roberts}}]{1958MacfarlaneExpZPR}%
  \BibitemOpen
  \bibfield  {author} {\bibinfo {author} {\bibfnamefont {G.~G.}\ \bibnamefont
  {Macfarlane}}, \bibinfo {author} {\bibfnamefont {T.~P.}\ \bibnamefont
  {McLean}}, \bibinfo {author} {\bibfnamefont {J.~E.}\ \bibnamefont
  {Quarrington}}, \ and\ \bibinfo {author} {\bibfnamefont {V.}~\bibnamefont
  {Roberts}},\ }\bibfield  {title} {\enquote {\bibinfo {title} {Fine structure
  in the absorption-edge spectrum of si},}\ }\href {\doibase
  10.1103/PhysRev.111.1245} {\bibfield  {journal} {\bibinfo  {journal} {Phys.
  Rev.}\ }\textbf {\bibinfo {volume} {111}},\ \bibinfo {pages} {1245--1254}
  (\bibinfo {year} {1958})}\BibitemShut {NoStop}%
\bibitem [{\citenamefont {Allen}\ and\ \citenamefont
  {Heine}(1976)}]{1976Allen}%
  \BibitemOpen
  \bibfield  {author} {\bibinfo {author} {\bibfnamefont {P~B}\ \bibnamefont
  {Allen}}\ and\ \bibinfo {author} {\bibfnamefont {V}~\bibnamefont {Heine}},\
  }\bibfield  {title} {\enquote {\bibinfo {title} {Theory of the temperature
  dependence of electronic band structures},}\ }\href {\doibase
  10.1088/0022-3719/9/12/013} {\bibfield  {journal} {\bibinfo  {journal}
  {Journal of Physics C: Solid State Physics}\ }\textbf {\bibinfo {volume}
  {9}},\ \bibinfo {pages} {2305--2312} (\bibinfo {year} {1976})}\BibitemShut
  {NoStop}%
\bibitem [{\citenamefont {Allen}\ and\ \citenamefont
  {Cardona}(1981)}]{1981Allen}%
  \BibitemOpen
  \bibfield  {author} {\bibinfo {author} {\bibfnamefont {P.~B.}\ \bibnamefont
  {Allen}}\ and\ \bibinfo {author} {\bibfnamefont {M.}~\bibnamefont
  {Cardona}},\ }\bibfield  {title} {\enquote {\bibinfo {title} {Theory of the
  temperature dependence of the direct gap of germanium},}\ }\href {\doibase
  10.1103/PhysRevB.23.1495} {\bibfield  {journal} {\bibinfo  {journal}
  {Physical Review B}\ }\textbf {\bibinfo {volume} {23}},\ \bibinfo {pages}
  {1495--1505} (\bibinfo {year} {1981})}\BibitemShut {NoStop}%
\bibitem [{\citenamefont {Allen}\ and\ \citenamefont
  {Cardona}(1983)}]{1983Allen}%
  \BibitemOpen
  \bibfield  {author} {\bibinfo {author} {\bibfnamefont {P.~B.}\ \bibnamefont
  {Allen}}\ and\ \bibinfo {author} {\bibfnamefont {M.}~\bibnamefont
  {Cardona}},\ }\bibfield  {title} {\enquote {\bibinfo {title} {Temperature
  dependence of the direct gap of {Si} and {Ge}},}\ }\href {\doibase
  10.1103/PhysRevB.27.4760} {\bibfield  {journal} {\bibinfo  {journal} {Phys.
  Rev. B}\ }\textbf {\bibinfo {volume} {27}},\ \bibinfo {pages} {4760--4769}
  (\bibinfo {year} {1983})}\BibitemShut {NoStop}%
\bibitem [{\citenamefont {Giustino}(2017)}]{2017GiustinoRMP}%
  \BibitemOpen
  \bibfield  {author} {\bibinfo {author} {\bibfnamefont {Feliciano}\
  \bibnamefont {Giustino}},\ }\bibfield  {title} {\enquote {\bibinfo {title}
  {Electron-phonon interactions from first principles},}\ }\href {\doibase
  10.1103/RevModPhys.89.015003} {\bibfield  {journal} {\bibinfo  {journal}
  {Rev. Mod. Phys.}\ }\textbf {\bibinfo {volume} {89}},\ \bibinfo {pages}
  {015003} (\bibinfo {year} {2017})}\BibitemShut {NoStop}%
\bibitem [{\citenamefont {Lihm}\ and\ \citenamefont
  {Park}(2020)}]{2020LihmAHC}%
  \BibitemOpen
  \bibfield  {author} {\bibinfo {author} {\bibfnamefont {Jae-Mo}\ \bibnamefont
  {Lihm}}\ and\ \bibinfo {author} {\bibfnamefont {Cheol-Hwan}\ \bibnamefont
  {Park}},\ }\bibfield  {title} {\enquote {\bibinfo {title} {Phonon-induced
  renormalization of electron wave functions},}\ }\href {\doibase
  10.1103/PhysRevB.101.121102} {\bibfield  {journal} {\bibinfo  {journal}
  {Physical Review B}\ }\textbf {\bibinfo {volume} {101}},\ \bibinfo {pages}
  {121102} (\bibinfo {year} {2020})}\BibitemShut {NoStop}%
\bibitem [{\citenamefont {Poncé}\ \emph {et~al.}(2014)\citenamefont {Poncé},
  \citenamefont {Antonius}, \citenamefont {Gillet}, \citenamefont {Boulanger},
  \citenamefont {Laflamme~Janssen}, \citenamefont {Marini}, \citenamefont
  {Côté},\ and\ \citenamefont {Gonze}}]{2014PoncePRB}%
  \BibitemOpen
  \bibfield  {author} {\bibinfo {author} {\bibfnamefont {S.}~\bibnamefont
  {Poncé}}, \bibinfo {author} {\bibfnamefont {G.}~\bibnamefont {Antonius}},
  \bibinfo {author} {\bibfnamefont {Y.}~\bibnamefont {Gillet}}, \bibinfo
  {author} {\bibfnamefont {P.}~\bibnamefont {Boulanger}}, \bibinfo {author}
  {\bibfnamefont {J.}~\bibnamefont {Laflamme~Janssen}}, \bibinfo {author}
  {\bibfnamefont {A.}~\bibnamefont {Marini}}, \bibinfo {author} {\bibfnamefont
  {M.}~\bibnamefont {Côté}}, \ and\ \bibinfo {author} {\bibfnamefont
  {X.}~\bibnamefont {Gonze}},\ }\bibfield  {title} {\enquote {\bibinfo {title}
  {Temperature dependence of electronic eigenenergies in the adiabatic harmonic
  approximation},}\ }\href {\doibase 10.1103/PhysRevB.90.214304} {\bibfield
  {journal} {\bibinfo  {journal} {Phys. Rev. B}\ }\textbf {\bibinfo {volume}
  {90}},\ \bibinfo {pages} {214304} (\bibinfo {year} {2014})}\BibitemShut
  {NoStop}%
\bibitem [{\citenamefont {Nery}\ \emph {et~al.}(2018)\citenamefont {Nery},
  \citenamefont {Allen}, \citenamefont {Antonius}, \citenamefont {Reining},
  \citenamefont {Miglio},\ and\ \citenamefont {Gonze}}]{2018NeryPRB}%
  \BibitemOpen
  \bibfield  {author} {\bibinfo {author} {\bibfnamefont {Jean~Paul}\
  \bibnamefont {Nery}}, \bibinfo {author} {\bibfnamefont {Philip~B.}\
  \bibnamefont {Allen}}, \bibinfo {author} {\bibfnamefont {Gabriel}\
  \bibnamefont {Antonius}}, \bibinfo {author} {\bibfnamefont {Lucia}\
  \bibnamefont {Reining}}, \bibinfo {author} {\bibfnamefont {Anna}\
  \bibnamefont {Miglio}}, \ and\ \bibinfo {author} {\bibfnamefont {Xavier}\
  \bibnamefont {Gonze}},\ }\bibfield  {title} {\enquote {\bibinfo {title}
  {Quasiparticles and phonon satellites in spectral functions of semiconductors
  and insulators: {Cumulants} applied to the full first-principles theory and
  the {Fr\"ohlich} polaron},}\ }\href {\doibase 10.1103/PhysRevB.97.115145}
  {\bibfield  {journal} {\bibinfo  {journal} {Phys. Rev. B}\ }\textbf {\bibinfo
  {volume} {97}},\ \bibinfo {pages} {115145} (\bibinfo {year}
  {2018})}\BibitemShut {NoStop}%
\bibitem [{\citenamefont {Miglio}\ \emph {et~al.}(2020)\citenamefont {Miglio},
  \citenamefont {{Brousseau-Couture}}, \citenamefont {Godbout}, \citenamefont
  {Antonius}, \citenamefont {Chan}, \citenamefont {Louie}, \citenamefont
  {C{\^o}t{\'e}}, \citenamefont {Giantomassi},\ and\ \citenamefont
  {Gonze}}]{2020MiglioAHC}%
  \BibitemOpen
  \bibfield  {author} {\bibinfo {author} {\bibfnamefont {Anna}\ \bibnamefont
  {Miglio}}, \bibinfo {author} {\bibfnamefont {V{\'e}ronique}\ \bibnamefont
  {{Brousseau-Couture}}}, \bibinfo {author} {\bibfnamefont {Emile}\
  \bibnamefont {Godbout}}, \bibinfo {author} {\bibfnamefont {Gabriel}\
  \bibnamefont {Antonius}}, \bibinfo {author} {\bibfnamefont {Yang-Hao}\
  \bibnamefont {Chan}}, \bibinfo {author} {\bibfnamefont {Steven~G.}\
  \bibnamefont {Louie}}, \bibinfo {author} {\bibfnamefont {Michel}\
  \bibnamefont {C{\^o}t{\'e}}}, \bibinfo {author} {\bibfnamefont {Matteo}\
  \bibnamefont {Giantomassi}}, \ and\ \bibinfo {author} {\bibfnamefont
  {Xavier}\ \bibnamefont {Gonze}},\ }\bibfield  {title} {\enquote {\bibinfo
  {title} {Predominance of non-adiabatic effects in zero-point renormalization
  of the electronic band gap},}\ }\href {\doibase 10.1038/s41524-020-00434-z}
  {\bibfield  {journal} {\bibinfo  {journal} {npj Computational Materials}\
  }\textbf {\bibinfo {volume} {6}},\ \bibinfo {pages} {167} (\bibinfo {year}
  {2020})}\BibitemShut {NoStop}%
\bibitem [{\citenamefont {{Brown-Altvater}}\ \emph {et~al.}(2020)\citenamefont
  {{Brown-Altvater}}, \citenamefont {Antonius}, \citenamefont {Rangel},
  \citenamefont {Giantomassi}, \citenamefont {Draxl}, \citenamefont {Gonze},
  \citenamefont {Louie},\ and\ \citenamefont {Neaton}}]{2020BrownAltvaterAHC}%
  \BibitemOpen
  \bibfield  {author} {\bibinfo {author} {\bibfnamefont {Florian}\ \bibnamefont
  {{Brown-Altvater}}}, \bibinfo {author} {\bibfnamefont {Gabriel}\ \bibnamefont
  {Antonius}}, \bibinfo {author} {\bibfnamefont {Tonatiuh}\ \bibnamefont
  {Rangel}}, \bibinfo {author} {\bibfnamefont {Matteo}\ \bibnamefont
  {Giantomassi}}, \bibinfo {author} {\bibfnamefont {Claudia}\ \bibnamefont
  {Draxl}}, \bibinfo {author} {\bibfnamefont {Xavier}\ \bibnamefont {Gonze}},
  \bibinfo {author} {\bibfnamefont {Steven~G.}\ \bibnamefont {Louie}}, \ and\
  \bibinfo {author} {\bibfnamefont {Jeffrey~B.}\ \bibnamefont {Neaton}},\
  }\bibfield  {title} {\enquote {\bibinfo {title} {Band gap renormalization,
  carrier mobilities, and the electron-phonon self-energy in crystalline
  naphthalene},}\ }\href {\doibase 10.1103/PhysRevB.101.165102} {\bibfield
  {journal} {\bibinfo  {journal} {Physical Review B}\ }\textbf {\bibinfo
  {volume} {101}},\ \bibinfo {pages} {165102} (\bibinfo {year}
  {2020})}\BibitemShut {NoStop}%
\bibitem [{\citenamefont {Kioupakis}\ \emph {et~al.}(2010)\citenamefont
  {Kioupakis}, \citenamefont {Rinke}, \citenamefont {Schleife}, \citenamefont
  {Bechstedt},\ and\ \citenamefont {Van~de Walle}}]{2010Kioupakis}%
  \BibitemOpen
  \bibfield  {author} {\bibinfo {author} {\bibfnamefont {Emmanouil}\
  \bibnamefont {Kioupakis}}, \bibinfo {author} {\bibfnamefont {Patrick}\
  \bibnamefont {Rinke}}, \bibinfo {author} {\bibfnamefont {André}\
  \bibnamefont {Schleife}}, \bibinfo {author} {\bibfnamefont {Friedhelm}\
  \bibnamefont {Bechstedt}}, \ and\ \bibinfo {author} {\bibfnamefont
  {Chris~G.}\ \bibnamefont {Van~de Walle}},\ }\bibfield  {title} {\enquote
  {\bibinfo {title} {Free-carrier absorption in nitrides from first
  principles},}\ }\href {\doibase 10.1103/PhysRevB.81.241201} {\bibfield
  {journal} {\bibinfo  {journal} {Physical Review B}\ }\textbf {\bibinfo
  {volume} {81}},\ \bibinfo {pages} {241201} (\bibinfo {year}
  {2010})}\BibitemShut {NoStop}%
\bibitem [{\citenamefont {Patrick}\ and\ \citenamefont
  {Giustino}(2014)}]{2014PatrickIndabs}%
  \BibitemOpen
  \bibfield  {author} {\bibinfo {author} {\bibfnamefont {Christopher~E.}\
  \bibnamefont {Patrick}}\ and\ \bibinfo {author} {\bibfnamefont {Feliciano}\
  \bibnamefont {Giustino}},\ }\bibfield  {title} {\enquote {\bibinfo {title}
  {Unified theory of electron-phonon renormalization and phonon-assisted
  optical absorption},}\ }\href {\doibase 10.1088/0953-8984/26/36/365503}
  {\bibfield  {journal} {\bibinfo  {journal} {Journal of Physics: Condensed
  Matter}\ }\textbf {\bibinfo {volume} {26}},\ \bibinfo {pages} {365503}
  (\bibinfo {year} {2014})}\BibitemShut {NoStop}%
\bibitem [{\citenamefont {Hybertsen}\ and\ \citenamefont
  {Louie}(1986)}]{1986HybertsenGW}%
  \BibitemOpen
  \bibfield  {author} {\bibinfo {author} {\bibfnamefont {Mark~S.}\ \bibnamefont
  {Hybertsen}}\ and\ \bibinfo {author} {\bibfnamefont {Steven~G.}\ \bibnamefont
  {Louie}},\ }\bibfield  {title} {\enquote {\bibinfo {title} {Electron
  correlation in semiconductors and insulators: {{Band}} gaps and quasiparticle
  energies},}\ }\href {\doibase 10.1103/PhysRevB.34.5390} {\bibfield  {journal}
  {\bibinfo  {journal} {Physical Review B}\ }\textbf {\bibinfo {volume} {34}},\
  \bibinfo {pages} {5390--5413} (\bibinfo {year} {1986})}\BibitemShut {NoStop}%
\bibitem [{\citenamefont {Zacharias}\ \emph {et~al.}(2015)\citenamefont
  {Zacharias}, \citenamefont {Patrick},\ and\ \citenamefont
  {Giustino}}]{2015ZachariasZG}%
  \BibitemOpen
  \bibfield  {author} {\bibinfo {author} {\bibfnamefont {Marios}\ \bibnamefont
  {Zacharias}}, \bibinfo {author} {\bibfnamefont {Christopher~E.}\ \bibnamefont
  {Patrick}}, \ and\ \bibinfo {author} {\bibfnamefont {Feliciano}\ \bibnamefont
  {Giustino}},\ }\bibfield  {title} {\enquote {\bibinfo {title} {Stochastic
  approach to phonon-assisted optical absorption},}\ }\href {\doibase
  10.1103/PhysRevLett.115.177401} {\bibfield  {journal} {\bibinfo  {journal}
  {Physical Review Letters}\ }\textbf {\bibinfo {volume} {115}},\ \bibinfo
  {pages} {177401} (\bibinfo {year} {2015})}\BibitemShut {NoStop}%
\bibitem [{\citenamefont {Zacharias}\ and\ \citenamefont
  {Giustino}(2016)}]{2016ZachariasZG}%
  \BibitemOpen
  \bibfield  {author} {\bibinfo {author} {\bibfnamefont {Marios}\ \bibnamefont
  {Zacharias}}\ and\ \bibinfo {author} {\bibfnamefont {Feliciano}\ \bibnamefont
  {Giustino}},\ }\bibfield  {title} {\enquote {\bibinfo {title} {One-shot
  calculation of temperature-dependent optical spectra and phonon-induced
  band-gap renormalization},}\ }\href {\doibase 10.1103/PhysRevB.94.075125}
  {\bibfield  {journal} {\bibinfo  {journal} {Physical Review B}\ }\textbf
  {\bibinfo {volume} {94}},\ \bibinfo {pages} {075125} (\bibinfo {year}
  {2016})}\BibitemShut {NoStop}%
\bibitem [{\citenamefont {Monserrat}\ and\ \citenamefont
  {Vanderbilt}(2016)}]{2016MonserratPRL}%
  \BibitemOpen
  \bibfield  {author} {\bibinfo {author} {\bibfnamefont {Bartomeu}\
  \bibnamefont {Monserrat}}\ and\ \bibinfo {author} {\bibfnamefont {David}\
  \bibnamefont {Vanderbilt}},\ }\bibfield  {title} {\enquote {\bibinfo {title}
  {Temperature effects in the band structure of topological insulators},}\
  }\href {\doibase 10.1103/PhysRevLett.117.226801} {\bibfield  {journal}
  {\bibinfo  {journal} {Physical Review Letters}\ }\textbf {\bibinfo {volume}
  {117}},\ \bibinfo {pages} {226801} (\bibinfo {year} {2016})}\BibitemShut
  {NoStop}%
\bibitem [{\citenamefont {Gorelov}\ \emph {et~al.}(2020)\citenamefont
  {Gorelov}, \citenamefont {Ceperley}, \citenamefont {Holzmann},\ and\
  \citenamefont {Pierleoni}}]{2020GorelovOptical}%
  \BibitemOpen
  \bibfield  {author} {\bibinfo {author} {\bibfnamefont {Vitaly}\ \bibnamefont
  {Gorelov}}, \bibinfo {author} {\bibfnamefont {David~M.}\ \bibnamefont
  {Ceperley}}, \bibinfo {author} {\bibfnamefont {Markus}\ \bibnamefont
  {Holzmann}}, \ and\ \bibinfo {author} {\bibfnamefont {Carlo}\ \bibnamefont
  {Pierleoni}},\ }\bibfield  {title} {\enquote {\bibinfo {title} {Electronic
  structure and optical properties of quantum crystals from first principles
  calculations in the {Born–Oppenheimer} approximation},}\ }\href {\doibase
  10.1063/5.0031843} {\bibfield  {journal} {\bibinfo  {journal} {The Journal of
  Chemical Physics}\ }\textbf {\bibinfo {volume} {153}},\ \bibinfo {pages}
  {234117} (\bibinfo {year} {2020})}\BibitemShut {NoStop}%
\bibitem [{\citenamefont {Zacharias}\ \emph {et~al.}(2020)\citenamefont
  {Zacharias}, \citenamefont {Scheffler},\ and\ \citenamefont
  {Carbogno}}]{2020Zacharias}%
  \BibitemOpen
  \bibfield  {author} {\bibinfo {author} {\bibfnamefont {Marios}\ \bibnamefont
  {Zacharias}}, \bibinfo {author} {\bibfnamefont {Matthias}\ \bibnamefont
  {Scheffler}}, \ and\ \bibinfo {author} {\bibfnamefont {Christian}\
  \bibnamefont {Carbogno}},\ }\bibfield  {title} {\enquote {\bibinfo {title}
  {Fully anharmonic nonperturbative theory of vibronically renormalized
  electronic band structures},}\ }\href {\doibase 10.1103/PhysRevB.102.045126}
  {\bibfield  {journal} {\bibinfo  {journal} {Phys. Rev. B}\ }\textbf {\bibinfo
  {volume} {102}},\ \bibinfo {pages} {045126} (\bibinfo {year}
  {2020})}\BibitemShut {NoStop}%
\bibitem [{\citenamefont {Rangel}\ \emph {et~al.}(2020)\citenamefont {Rangel},
  \citenamefont {Del~Ben}, \citenamefont {Varsano}, \citenamefont {Antonius},
  \citenamefont {Bruneval}, \citenamefont {{da Jornada}}, \citenamefont {{van
  Setten}}, \citenamefont {Orhan}, \citenamefont {O'Regan}, \citenamefont
  {Canning}, \citenamefont {Ferretti}, \citenamefont {Marini}, \citenamefont
  {Rignanese}, \citenamefont {Deslippe}, \citenamefont {Louie},\ and\
  \citenamefont {Neaton}}]{2020RangelGW}%
  \BibitemOpen
  \bibfield  {author} {\bibinfo {author} {\bibfnamefont {Tonatiuh}\
  \bibnamefont {Rangel}}, \bibinfo {author} {\bibfnamefont {Mauro}\
  \bibnamefont {Del~Ben}}, \bibinfo {author} {\bibfnamefont {Daniele}\
  \bibnamefont {Varsano}}, \bibinfo {author} {\bibfnamefont {Gabriel}\
  \bibnamefont {Antonius}}, \bibinfo {author} {\bibfnamefont {Fabien}\
  \bibnamefont {Bruneval}}, \bibinfo {author} {\bibfnamefont {Felipe~H.}\
  \bibnamefont {{da Jornada}}}, \bibinfo {author} {\bibfnamefont {Michiel~J.}\
  \bibnamefont {{van Setten}}}, \bibinfo {author} {\bibfnamefont {Okan~K.}\
  \bibnamefont {Orhan}}, \bibinfo {author} {\bibfnamefont {David~D.}\
  \bibnamefont {O'Regan}}, \bibinfo {author} {\bibfnamefont {Andrew}\
  \bibnamefont {Canning}}, \bibinfo {author} {\bibfnamefont {Andrea}\
  \bibnamefont {Ferretti}}, \bibinfo {author} {\bibfnamefont {Andrea}\
  \bibnamefont {Marini}}, \bibinfo {author} {\bibfnamefont {Gian-Marco}\
  \bibnamefont {Rignanese}}, \bibinfo {author} {\bibfnamefont {Jack}\
  \bibnamefont {Deslippe}}, \bibinfo {author} {\bibfnamefont {Steven~G.}\
  \bibnamefont {Louie}}, \ and\ \bibinfo {author} {\bibfnamefont {Jeffrey~B.}\
  \bibnamefont {Neaton}},\ }\bibfield  {title} {\enquote {\bibinfo {title}
  {Reproducibility in {$G_0 W_0$} calculations for solids},}\ }\href {\doibase
  10.1016/j.cpc.2020.107242} {\bibfield  {journal} {\bibinfo  {journal}
  {Computer Physics Communications}\ }\textbf {\bibinfo {volume} {255}},\
  \bibinfo {pages} {107242} (\bibinfo {year} {2020})}\BibitemShut {NoStop}%
\bibitem [{\citenamefont {Antonius}\ \emph {et~al.}(2014)\citenamefont
  {Antonius}, \citenamefont {Ponc{\'e}}, \citenamefont {Boulanger},
  \citenamefont {C{\^o}t{\'e}},\ and\ \citenamefont {Gonze}}]{2014AntoniusZPR}%
  \BibitemOpen
  \bibfield  {author} {\bibinfo {author} {\bibfnamefont {G.}~\bibnamefont
  {Antonius}}, \bibinfo {author} {\bibfnamefont {S.}~\bibnamefont {Ponc{\'e}}},
  \bibinfo {author} {\bibfnamefont {P.}~\bibnamefont {Boulanger}}, \bibinfo
  {author} {\bibfnamefont {M.}~\bibnamefont {C{\^o}t{\'e}}}, \ and\ \bibinfo
  {author} {\bibfnamefont {X.}~\bibnamefont {Gonze}},\ }\bibfield  {title}
  {\enquote {\bibinfo {title} {Many-{{Body Effects}} on the {{Zero}}-{{Point
  Renormalization}} of the {{Band Structure}}},}\ }\href {\doibase
  10.1103/PhysRevLett.112.215501} {\bibfield  {journal} {\bibinfo  {journal}
  {Physical Review Letters}\ }\textbf {\bibinfo {volume} {112}},\ \bibinfo
  {pages} {215501} (\bibinfo {year} {2014})}\BibitemShut {NoStop}%
\bibitem [{\citenamefont {Bludau}\ \emph {et~al.}(1974)\citenamefont {Bludau},
  \citenamefont {Onton},\ and\ \citenamefont {Heinke}}]{1974BludauExpZPR}%
  \BibitemOpen
  \bibfield  {author} {\bibinfo {author} {\bibfnamefont {W.}~\bibnamefont
  {Bludau}}, \bibinfo {author} {\bibfnamefont {A.}~\bibnamefont {Onton}}, \
  and\ \bibinfo {author} {\bibfnamefont {W.}~\bibnamefont {Heinke}},\
  }\bibfield  {title} {\enquote {\bibinfo {title} {Temperature dependence of
  the band gap of silicon},}\ }\href {\doibase 10.1063/1.1663501} {\bibfield
  {journal} {\bibinfo  {journal} {Journal of Applied Physics}\ }\textbf
  {\bibinfo {volume} {45}},\ \bibinfo {pages} {1846--1848} (\bibinfo {year}
  {1974})}\BibitemShut {NoStop}%
\bibitem [{\citenamefont {Hamann}\ and\ \citenamefont
  {Vanderbilt}(2009)}]{2009HamannGWWannier}%
  \BibitemOpen
  \bibfield  {author} {\bibinfo {author} {\bibfnamefont {D.~R.}\ \bibnamefont
  {Hamann}}\ and\ \bibinfo {author} {\bibfnamefont {David}\ \bibnamefont
  {Vanderbilt}},\ }\bibfield  {title} {\enquote {\bibinfo {title} {Maximally
  localized {Wannier} functions for {$GW$} quasiparticles},}\ }\href {\doibase
  10.1103/PhysRevB.79.045109} {\bibfield  {journal} {\bibinfo  {journal} {Phys.
  Rev. B}\ }\textbf {\bibinfo {volume} {79}},\ \bibinfo {pages} {045109}
  (\bibinfo {year} {2009})}\BibitemShut {NoStop}%
\bibitem [{Note1()}]{Note1}%
  \BibitemOpen
  \bibinfo {note} {See Supplemental Material, which includes Refs.~\cite
  {2021PonceMobility, 1996AdolphVelocity, 2012LopezOrbitalMag}, at [URL will be
  inserted by publisher] for the details of the calculation of the velocity and
  related matrix elements and the convergence study.}\BibitemShut {Stop}%
\bibitem [{\citenamefont {Huang}\ \emph {et~al.}(2016)\citenamefont {Huang},
  \citenamefont {Fan}, \citenamefont {Singh}, \citenamefont {Chen},
  \citenamefont {Jiang},\ and\ \citenamefont {Zheng}}]{2016HuangWTe2}%
  \BibitemOpen
  \bibfield  {author} {\bibinfo {author} {\bibfnamefont {H.~H.}\ \bibnamefont
  {Huang}}, \bibinfo {author} {\bibfnamefont {Xiaofeng}\ \bibnamefont {Fan}},
  \bibinfo {author} {\bibfnamefont {David~J.}\ \bibnamefont {Singh}}, \bibinfo
  {author} {\bibfnamefont {Hong}\ \bibnamefont {Chen}}, \bibinfo {author}
  {\bibfnamefont {Q.}~\bibnamefont {Jiang}}, \ and\ \bibinfo {author}
  {\bibfnamefont {W.~T.}\ \bibnamefont {Zheng}},\ }\bibfield  {title} {\enquote
  {\bibinfo {title} {Controlling phase transition for single-layer {MTe2} ({M}
  = {Mo} and {W}): modulation of the potential barrier under strain},}\ }\href
  {\doibase 10.1039/C5CP06706E} {\bibfield  {journal} {\bibinfo  {journal}
  {Phys. Chem. Chem. Phys.}\ }\textbf {\bibinfo {volume} {18}},\ \bibinfo
  {pages} {4086--4094} (\bibinfo {year} {2016})}\BibitemShut {NoStop}%
\bibitem [{\citenamefont {Baroni}\ and\ \citenamefont
  {Resta}(1986)}]{1986Baroni}%
  \BibitemOpen
  \bibfield  {author} {\bibinfo {author} {\bibfnamefont {Stefano}\ \bibnamefont
  {Baroni}}\ and\ \bibinfo {author} {\bibfnamefont {Raffaele}\ \bibnamefont
  {Resta}},\ }\bibfield  {title} {\enquote {\bibinfo {title} {Ab initio
  calculation of the macroscopic dielectric constant in silicon},}\ }\href
  {\doibase 10.1103/PhysRevB.33.7017} {\bibfield  {journal} {\bibinfo
  {journal} {Phys. Rev. B}\ }\textbf {\bibinfo {volume} {33}},\ \bibinfo
  {pages} {7017--7021} (\bibinfo {year} {1986})}\BibitemShut {NoStop}%
\bibitem [{\citenamefont {Guo}\ \emph {et~al.}(2008)\citenamefont {Guo},
  \citenamefont {Murakami}, \citenamefont {Chen},\ and\ \citenamefont
  {Nagaosa}}]{2008GuoSHE}%
  \BibitemOpen
  \bibfield  {author} {\bibinfo {author} {\bibfnamefont {G.~Y.}\ \bibnamefont
  {Guo}}, \bibinfo {author} {\bibfnamefont {S.}~\bibnamefont {Murakami}},
  \bibinfo {author} {\bibfnamefont {T.-W.}\ \bibnamefont {Chen}}, \ and\
  \bibinfo {author} {\bibfnamefont {N.}~\bibnamefont {Nagaosa}},\ }\bibfield
  {title} {\enquote {\bibinfo {title} {Intrinsic {{Spin Hall Effect}} in
  {{Platinum}}: {{First}}-{{Principles Calculations}}},}\ }\href {\doibase
  10.1103/PhysRevLett.100.096401} {\bibfield  {journal} {\bibinfo  {journal}
  {Physical Review Letters}\ }\textbf {\bibinfo {volume} {100}},\ \bibinfo
  {pages} {096401} (\bibinfo {year} {2008})}\BibitemShut {NoStop}%
\bibitem [{\citenamefont {Guo}\ \emph {et~al.}(2005)\citenamefont {Guo},
  \citenamefont {Yao},\ and\ \citenamefont {Niu}}]{2005GuoSHE}%
  \BibitemOpen
  \bibfield  {author} {\bibinfo {author} {\bibfnamefont {G.~Y.}\ \bibnamefont
  {Guo}}, \bibinfo {author} {\bibfnamefont {Yugui}\ \bibnamefont {Yao}}, \ and\
  \bibinfo {author} {\bibfnamefont {Qian}\ \bibnamefont {Niu}},\ }\bibfield
  {title} {\enquote {\bibinfo {title} {Ab initio calculation of the intrinsic
  spin hall effect in semiconductors},}\ }\href {\doibase
  10.1103/PhysRevLett.94.226601} {\bibfield  {journal} {\bibinfo  {journal}
  {Phys. Rev. Lett.}\ }\textbf {\bibinfo {volume} {94}},\ \bibinfo {pages}
  {226601} (\bibinfo {year} {2005})}\BibitemShut {NoStop}%
\bibitem [{\citenamefont {Schlipf}\ \emph {et~al.}(2018)\citenamefont
  {Schlipf}, \citenamefont {Ponc{\'e}},\ and\ \citenamefont
  {Giustino}}]{2018SchlipfMobility}%
  \BibitemOpen
  \bibfield  {author} {\bibinfo {author} {\bibfnamefont {Martin}\ \bibnamefont
  {Schlipf}}, \bibinfo {author} {\bibfnamefont {Samuel}\ \bibnamefont
  {Ponc{\'e}}}, \ and\ \bibinfo {author} {\bibfnamefont {Feliciano}\
  \bibnamefont {Giustino}},\ }\bibfield  {title} {\enquote {\bibinfo {title}
  {Carrier {{Lifetimes}} and {{Polaronic Mass Enhancement}} in the {{Hybrid
  Halide Perovskite CH$_3$NH$_3$PbI$_3$}} from {{Multiphonon Fr\"ohlich
  Coupling}}},}\ }\href {\doibase 10.1103/PhysRevLett.121.086402} {\bibfield
  {journal} {\bibinfo  {journal} {Physical Review Letters}\ }\textbf {\bibinfo
  {volume} {121}},\ \bibinfo {pages} {086402} (\bibinfo {year}
  {2018})}\BibitemShut {NoStop}%
\bibitem [{\citenamefont {D'Souza}\ \emph {et~al.}(2020)\citenamefont
  {D'Souza}, \citenamefont {Cao}, \citenamefont {{Querales-Flores}},
  \citenamefont {Fahy},\ and\ \citenamefont {Savi{\'c}}}]{2020DSouzaMobility}%
  \BibitemOpen
  \bibfield  {author} {\bibinfo {author} {\bibfnamefont {Ransell}\ \bibnamefont
  {D'Souza}}, \bibinfo {author} {\bibfnamefont {Jiang}\ \bibnamefont {Cao}},
  \bibinfo {author} {\bibfnamefont {Jos{\'e}~D.}\ \bibnamefont
  {{Querales-Flores}}}, \bibinfo {author} {\bibfnamefont {Stephen}\
  \bibnamefont {Fahy}}, \ and\ \bibinfo {author} {\bibfnamefont {Ivana}\
  \bibnamefont {Savi{\'c}}},\ }\bibfield  {title} {\enquote {\bibinfo {title}
  {Electron-phonon scattering and thermoelectric transport in p -type {{PbTe}}
  from first principles},}\ }\href {\doibase 10.1103/PhysRevB.102.115204}
  {\bibfield  {journal} {\bibinfo  {journal} {Physical Review B}\ }\textbf
  {\bibinfo {volume} {102}},\ \bibinfo {pages} {115204} (\bibinfo {year}
  {2020})}\BibitemShut {NoStop}%
\bibitem [{\citenamefont {Vitale}\ \emph {et~al.}(2020)\citenamefont {Vitale},
  \citenamefont {Pizzi}, \citenamefont {Marrazzo}, \citenamefont {Yates},
  \citenamefont {Marzari},\ and\ \citenamefont {Mostofi}}]{2020VitaleWannier}%
  \BibitemOpen
  \bibfield  {author} {\bibinfo {author} {\bibfnamefont {Valerio}\ \bibnamefont
  {Vitale}}, \bibinfo {author} {\bibfnamefont {Giovanni}\ \bibnamefont
  {Pizzi}}, \bibinfo {author} {\bibfnamefont {Antimo}\ \bibnamefont
  {Marrazzo}}, \bibinfo {author} {\bibfnamefont {Jonathan~R.}\ \bibnamefont
  {Yates}}, \bibinfo {author} {\bibfnamefont {Nicola}\ \bibnamefont {Marzari}},
  \ and\ \bibinfo {author} {\bibfnamefont {Arash~A.}\ \bibnamefont {Mostofi}},\
  }\bibfield  {title} {\enquote {\bibinfo {title} {Automated high-throughput
  {{Wannierisation}}},}\ }\href {\doibase 10.1038/s41524-020-0312-y} {\bibfield
   {journal} {\bibinfo  {journal} {npj Computational Materials}\ }\textbf
  {\bibinfo {volume} {6}},\ \bibinfo {pages} {66} (\bibinfo {year}
  {2020})}\BibitemShut {NoStop}%
\bibitem [{\citenamefont {Bistoni}\ \emph {et~al.}(2019)\citenamefont
  {Bistoni}, \citenamefont {Barone}, \citenamefont {Cappelluti}, \citenamefont
  {Benfatto},\ and\ \citenamefont {Mauri}}]{2019BistoniBorn}%
  \BibitemOpen
  \bibfield  {author} {\bibinfo {author} {\bibfnamefont {O}~\bibnamefont
  {Bistoni}}, \bibinfo {author} {\bibfnamefont {P}~\bibnamefont {Barone}},
  \bibinfo {author} {\bibfnamefont {E}~\bibnamefont {Cappelluti}}, \bibinfo
  {author} {\bibfnamefont {L}~\bibnamefont {Benfatto}}, \ and\ \bibinfo
  {author} {\bibfnamefont {F}~\bibnamefont {Mauri}},\ }\bibfield  {title}
  {\enquote {\bibinfo {title} {Giant effective charges and piezoelectricity in
  gapped graphene},}\ }\href {\doibase 10.1088/2053-1583/ab2ce0} {\bibfield
  {journal} {\bibinfo  {journal} {2D Materials}\ }\textbf {\bibinfo {volume}
  {6}},\ \bibinfo {pages} {045015} (\bibinfo {year} {2019})}\BibitemShut
  {NoStop}%
\bibitem [{\citenamefont {Binci}\ \emph {et~al.}(2021)\citenamefont {Binci},
  \citenamefont {Barone},\ and\ \citenamefont {Mauri}}]{2021BinciBorn}%
  \BibitemOpen
  \bibfield  {author} {\bibinfo {author} {\bibfnamefont {Luca}\ \bibnamefont
  {Binci}}, \bibinfo {author} {\bibfnamefont {Paolo}\ \bibnamefont {Barone}}, \
  and\ \bibinfo {author} {\bibfnamefont {Francesco}\ \bibnamefont {Mauri}},\
  }\bibfield  {title} {\enquote {\bibinfo {title} {First-principles theory of
  infrared vibrational spectroscopy of metals and semimetals: {{Application}}
  to graphite},}\ }\href {\doibase 10.1103/PhysRevB.103.134304} {\bibfield
  {journal} {\bibinfo  {journal} {Physical Review B}\ }\textbf {\bibinfo
  {volume} {103}},\ \bibinfo {pages} {134304} (\bibinfo {year}
  {2021})}\BibitemShut {NoStop}%
\bibitem [{\citenamefont {Dreyer}\ \emph {et~al.}(2021)\citenamefont {Dreyer},
  \citenamefont {Coh},\ and\ \citenamefont {Stengel}}]{2021DreyerBorn}%
  \BibitemOpen
  \bibfield  {author} {\bibinfo {author} {\bibfnamefont {Cyrus~E}\ \bibnamefont
  {Dreyer}}, \bibinfo {author} {\bibfnamefont {Sinisa}\ \bibnamefont {Coh}}, \
  and\ \bibinfo {author} {\bibfnamefont {Massimiliano}\ \bibnamefont
  {Stengel}},\ }\bibfield  {title} {\enquote {\bibinfo {title} {Nonadiabatic
  {{Born}} effective charges in metals and the {{Drude}} weight},}\ }\href@noop
  {} {\bibfield  {journal} {\bibinfo  {journal} {arXiv:2103.04425 [cond-mat]}\
  ,\ \bibinfo {pages} {16}} (\bibinfo {year} {2021})},\ \Eprint
  {http://arxiv.org/abs/2103.04425} {arXiv:2103.04425} \BibitemShut {NoStop}%
\bibitem [{\citenamefont {Eiguren}\ and\ \citenamefont
  {Ambrosch-Draxl}(2008)}]{2008Eiguren}%
  \BibitemOpen
  \bibfield  {author} {\bibinfo {author} {\bibfnamefont {Asier}\ \bibnamefont
  {Eiguren}}\ and\ \bibinfo {author} {\bibfnamefont {Claudia}\ \bibnamefont
  {Ambrosch-Draxl}},\ }\bibfield  {title} {\enquote {\bibinfo {title} {Wannier
  interpolation scheme for phonon-induced potentials: Application to bulk
  {${\text{MgB}}_{2}$}, {W}, and the $(1\ifmmode\times\else\texttimes\fi{}1)$
  {H}-covered {W}(110) surface},}\ }\href {\doibase 10.1103/PhysRevB.78.045124}
  {\bibfield  {journal} {\bibinfo  {journal} {Phys. Rev. B}\ }\textbf {\bibinfo
  {volume} {78}},\ \bibinfo {pages} {045124} (\bibinfo {year}
  {2008})}\BibitemShut {NoStop}%
\bibitem [{\citenamefont {Giustino}\ \emph
  {et~al.}(2010{\natexlab{b}})\citenamefont {Giustino}, \citenamefont {Cohen},\
  and\ \citenamefont {Louie}}]{2010GiustinoGW}%
  \BibitemOpen
  \bibfield  {author} {\bibinfo {author} {\bibfnamefont {Feliciano}\
  \bibnamefont {Giustino}}, \bibinfo {author} {\bibfnamefont {Marvin~L.}\
  \bibnamefont {Cohen}}, \ and\ \bibinfo {author} {\bibfnamefont {Steven~G.}\
  \bibnamefont {Louie}},\ }\bibfield  {title} {\enquote {\bibinfo {title}
  {{$GW$} method with the self-consistent {Sternheimer} equation},}\ }\href
  {\doibase 10.1103/PhysRevB.81.115105} {\bibfield  {journal} {\bibinfo
  {journal} {Phys. Rev. B}\ }\textbf {\bibinfo {volume} {81}},\ \bibinfo
  {pages} {115105} (\bibinfo {year} {2010}{\natexlab{b}})}\BibitemShut
  {NoStop}%
\bibitem [{\citenamefont {Umari}\ \emph {et~al.}(2010)\citenamefont {Umari},
  \citenamefont {Stenuit},\ and\ \citenamefont {Baroni}}]{2010UmariGW}%
  \BibitemOpen
  \bibfield  {author} {\bibinfo {author} {\bibfnamefont {P.}~\bibnamefont
  {Umari}}, \bibinfo {author} {\bibfnamefont {Geoffrey}\ \bibnamefont
  {Stenuit}}, \ and\ \bibinfo {author} {\bibfnamefont {Stefano}\ \bibnamefont
  {Baroni}},\ }\bibfield  {title} {\enquote {\bibinfo {title} {{$GW$}
  quasiparticle spectra from occupied states only},}\ }\href {\doibase
  10.1103/PhysRevB.81.115104} {\bibfield  {journal} {\bibinfo  {journal} {Phys.
  Rev. B}\ }\textbf {\bibinfo {volume} {81}},\ \bibinfo {pages} {115104}
  (\bibinfo {year} {2010})}\BibitemShut {NoStop}%
\bibitem [{\citenamefont {Giannozzi}\ \emph {et~al.}(2009)\citenamefont
  {Giannozzi}, \citenamefont {Baroni}, \citenamefont {Bonini}, \citenamefont
  {Calandra}, \citenamefont {Car}, \citenamefont {Cavazzoni}, \citenamefont
  {Ceresoli}, \citenamefont {Chiarotti}, \citenamefont {Cococcioni},
  \citenamefont {Dabo}, \citenamefont {Dal~Corso}, \citenamefont
  {de~Gironcoli}, \citenamefont {Fabris}, \citenamefont {Fratesi},
  \citenamefont {Gebauer}, \citenamefont {Gerstmann}, \citenamefont
  {Gougoussis}, \citenamefont {Kokalj}, \citenamefont {Lazzeri}, \citenamefont
  {Martin-Samos}, \citenamefont {Marzari}, \citenamefont {Mauri}, \citenamefont
  {Mazzarello}, \citenamefont {Paolini}, \citenamefont {Pasquarello},
  \citenamefont {Paulatto}, \citenamefont {Sbraccia}, \citenamefont {Scandolo},
  \citenamefont {Sclauzero}, \citenamefont {Seitsonen}, \citenamefont
  {Smogunov}, \citenamefont {Umari},\ and\ \citenamefont
  {Wentzcovitch}}]{2009GiannozziQE}%
  \BibitemOpen
  \bibfield  {author} {\bibinfo {author} {\bibfnamefont {P.}~\bibnamefont
  {Giannozzi}}, \bibinfo {author} {\bibfnamefont {S.}~\bibnamefont {Baroni}},
  \bibinfo {author} {\bibfnamefont {N.}~\bibnamefont {Bonini}}, \bibinfo
  {author} {\bibfnamefont {M.}~\bibnamefont {Calandra}}, \bibinfo {author}
  {\bibfnamefont {R.}~\bibnamefont {Car}}, \bibinfo {author} {\bibfnamefont
  {C.}~\bibnamefont {Cavazzoni}}, \bibinfo {author} {\bibfnamefont
  {D.}~\bibnamefont {Ceresoli}}, \bibinfo {author} {\bibfnamefont {G.~L.}\
  \bibnamefont {Chiarotti}}, \bibinfo {author} {\bibfnamefont {M.}~\bibnamefont
  {Cococcioni}}, \bibinfo {author} {\bibfnamefont {I.}~\bibnamefont {Dabo}},
  \bibinfo {author} {\bibfnamefont {A.}~\bibnamefont {Dal~Corso}}, \bibinfo
  {author} {\bibfnamefont {S.}~\bibnamefont {de~Gironcoli}}, \bibinfo {author}
  {\bibfnamefont {S.}~\bibnamefont {Fabris}}, \bibinfo {author} {\bibfnamefont
  {G.}~\bibnamefont {Fratesi}}, \bibinfo {author} {\bibfnamefont
  {R.}~\bibnamefont {Gebauer}}, \bibinfo {author} {\bibfnamefont
  {U.}~\bibnamefont {Gerstmann}}, \bibinfo {author} {\bibfnamefont
  {C.}~\bibnamefont {Gougoussis}}, \bibinfo {author} {\bibfnamefont
  {A.}~\bibnamefont {Kokalj}}, \bibinfo {author} {\bibfnamefont
  {M.}~\bibnamefont {Lazzeri}}, \bibinfo {author} {\bibfnamefont
  {L.}~\bibnamefont {Martin-Samos}}, \bibinfo {author} {\bibfnamefont
  {N.}~\bibnamefont {Marzari}}, \bibinfo {author} {\bibfnamefont
  {F.}~\bibnamefont {Mauri}}, \bibinfo {author} {\bibfnamefont
  {R.}~\bibnamefont {Mazzarello}}, \bibinfo {author} {\bibfnamefont
  {S.}~\bibnamefont {Paolini}}, \bibinfo {author} {\bibfnamefont
  {A.}~\bibnamefont {Pasquarello}}, \bibinfo {author} {\bibfnamefont
  {L.}~\bibnamefont {Paulatto}}, \bibinfo {author} {\bibfnamefont
  {C.}~\bibnamefont {Sbraccia}}, \bibinfo {author} {\bibfnamefont
  {S.}~\bibnamefont {Scandolo}}, \bibinfo {author} {\bibfnamefont
  {G.}~\bibnamefont {Sclauzero}}, \bibinfo {author} {\bibfnamefont {A.~P.}\
  \bibnamefont {Seitsonen}}, \bibinfo {author} {\bibfnamefont {A.}~\bibnamefont
  {Smogunov}}, \bibinfo {author} {\bibfnamefont {P.}~\bibnamefont {Umari}}, \
  and\ \bibinfo {author} {\bibfnamefont {R.M.}\ \bibnamefont {Wentzcovitch}},\
  }\bibfield  {title} {\enquote {\bibinfo {title} {{QUANTUM ESPRESSO: a modular
  and open-source software project for quantum simulations of materials}},}\
  }\href {http://stacks.iop.org/0953-8984/21/i=39/a=395502} {\bibfield
  {journal} {\bibinfo  {journal} {J. Phys.: Condens. Matter}\ }\textbf
  {\bibinfo {volume} {21}},\ \bibinfo {pages} {395502} (\bibinfo {year}
  {2009})}\BibitemShut {NoStop}%
\bibitem [{\citenamefont {Giannozzi}\ \emph {et~al.}(2017)\citenamefont
  {Giannozzi}, \citenamefont {Andreussi}, \citenamefont {Brumme}, \citenamefont
  {Bunau}, \citenamefont {Buongiorno~Nardelli}, \citenamefont {Calandra},
  \citenamefont {Car}, \citenamefont {Cavazzoni}, \citenamefont {Ceresoli},
  \citenamefont {Cococcioni}, \citenamefont {Colonna}, \citenamefont
  {Carnimeo}, \citenamefont {Dal~Corso}, \citenamefont {{de Gironcoli}},
  \citenamefont {Delugas}, \citenamefont {DiStasio}, \citenamefont {Ferretti},
  \citenamefont {Floris}, \citenamefont {Fratesi}, \citenamefont {Fugallo},
  \citenamefont {Gebauer}, \citenamefont {Gerstmann}, \citenamefont {Giustino},
  \citenamefont {Gorni}, \citenamefont {Jia}, \citenamefont {Kawamura},
  \citenamefont {Ko}, \citenamefont {Kokalj}, \citenamefont {K{\"u}{\c
  c}{\"u}kbenli}, \citenamefont {Lazzeri}, \citenamefont {Marsili},
  \citenamefont {Marzari}, \citenamefont {Mauri}, \citenamefont {Nguyen},
  \citenamefont {Nguyen}, \citenamefont {{Otero-de-la-Roza}}, \citenamefont
  {Paulatto}, \citenamefont {Ponc{\'e}}, \citenamefont {Rocca}, \citenamefont
  {Sabatini}, \citenamefont {Santra}, \citenamefont {Schlipf}, \citenamefont
  {Seitsonen}, \citenamefont {Smogunov}, \citenamefont {Timrov}, \citenamefont
  {Thonhauser}, \citenamefont {Umari}, \citenamefont {Vast}, \citenamefont
  {Wu},\ and\ \citenamefont {Baroni}}]{2017GiannozziQE}%
  \BibitemOpen
  \bibfield  {author} {\bibinfo {author} {\bibfnamefont {P}~\bibnamefont
  {Giannozzi}}, \bibinfo {author} {\bibfnamefont {O}~\bibnamefont {Andreussi}},
  \bibinfo {author} {\bibfnamefont {T}~\bibnamefont {Brumme}}, \bibinfo
  {author} {\bibfnamefont {O}~\bibnamefont {Bunau}}, \bibinfo {author}
  {\bibfnamefont {M}~\bibnamefont {Buongiorno~Nardelli}}, \bibinfo {author}
  {\bibfnamefont {M}~\bibnamefont {Calandra}}, \bibinfo {author} {\bibfnamefont
  {R}~\bibnamefont {Car}}, \bibinfo {author} {\bibfnamefont {C}~\bibnamefont
  {Cavazzoni}}, \bibinfo {author} {\bibfnamefont {D}~\bibnamefont {Ceresoli}},
  \bibinfo {author} {\bibfnamefont {M}~\bibnamefont {Cococcioni}}, \bibinfo
  {author} {\bibfnamefont {N}~\bibnamefont {Colonna}}, \bibinfo {author}
  {\bibfnamefont {I}~\bibnamefont {Carnimeo}}, \bibinfo {author} {\bibfnamefont
  {A}~\bibnamefont {Dal~Corso}}, \bibinfo {author} {\bibfnamefont
  {S}~\bibnamefont {{de Gironcoli}}}, \bibinfo {author} {\bibfnamefont
  {P}~\bibnamefont {Delugas}}, \bibinfo {author} {\bibfnamefont {R~A}\
  \bibnamefont {DiStasio}}, \bibinfo {author} {\bibfnamefont {A}~\bibnamefont
  {Ferretti}}, \bibinfo {author} {\bibfnamefont {A}~\bibnamefont {Floris}},
  \bibinfo {author} {\bibfnamefont {G}~\bibnamefont {Fratesi}}, \bibinfo
  {author} {\bibfnamefont {G}~\bibnamefont {Fugallo}}, \bibinfo {author}
  {\bibfnamefont {R}~\bibnamefont {Gebauer}}, \bibinfo {author} {\bibfnamefont
  {U}~\bibnamefont {Gerstmann}}, \bibinfo {author} {\bibfnamefont
  {F}~\bibnamefont {Giustino}}, \bibinfo {author} {\bibfnamefont
  {T}~\bibnamefont {Gorni}}, \bibinfo {author} {\bibfnamefont {J}~\bibnamefont
  {Jia}}, \bibinfo {author} {\bibfnamefont {M}~\bibnamefont {Kawamura}},
  \bibinfo {author} {\bibfnamefont {H-Y}\ \bibnamefont {Ko}}, \bibinfo {author}
  {\bibfnamefont {A}~\bibnamefont {Kokalj}}, \bibinfo {author} {\bibfnamefont
  {E}~\bibnamefont {K{\"u}{\c c}{\"u}kbenli}}, \bibinfo {author} {\bibfnamefont
  {M}~\bibnamefont {Lazzeri}}, \bibinfo {author} {\bibfnamefont
  {M}~\bibnamefont {Marsili}}, \bibinfo {author} {\bibfnamefont
  {N}~\bibnamefont {Marzari}}, \bibinfo {author} {\bibfnamefont
  {F}~\bibnamefont {Mauri}}, \bibinfo {author} {\bibfnamefont {N~L}\
  \bibnamefont {Nguyen}}, \bibinfo {author} {\bibfnamefont {H-V}\ \bibnamefont
  {Nguyen}}, \bibinfo {author} {\bibfnamefont {A}~\bibnamefont
  {{Otero-de-la-Roza}}}, \bibinfo {author} {\bibfnamefont {L}~\bibnamefont
  {Paulatto}}, \bibinfo {author} {\bibfnamefont {S}~\bibnamefont {Ponc{\'e}}},
  \bibinfo {author} {\bibfnamefont {D}~\bibnamefont {Rocca}}, \bibinfo {author}
  {\bibfnamefont {R}~\bibnamefont {Sabatini}}, \bibinfo {author} {\bibfnamefont
  {B}~\bibnamefont {Santra}}, \bibinfo {author} {\bibfnamefont {M}~\bibnamefont
  {Schlipf}}, \bibinfo {author} {\bibfnamefont {A~P}\ \bibnamefont
  {Seitsonen}}, \bibinfo {author} {\bibfnamefont {A}~\bibnamefont {Smogunov}},
  \bibinfo {author} {\bibfnamefont {I}~\bibnamefont {Timrov}}, \bibinfo
  {author} {\bibfnamefont {T}~\bibnamefont {Thonhauser}}, \bibinfo {author}
  {\bibfnamefont {P}~\bibnamefont {Umari}}, \bibinfo {author} {\bibfnamefont
  {N}~\bibnamefont {Vast}}, \bibinfo {author} {\bibfnamefont {X}~\bibnamefont
  {Wu}}, \ and\ \bibinfo {author} {\bibfnamefont {S}~\bibnamefont {Baroni}},\
  }\bibfield  {title} {\enquote {\bibinfo {title} {Advanced capabilities for
  materials modelling with {{Quantum ESPRESSO}}},}\ }\href {\doibase
  10.1088/1361-648X/aa8f79} {\bibfield  {journal} {\bibinfo  {journal} {Journal
  of Physics: Condensed Matter}\ }\textbf {\bibinfo {volume} {29}},\ \bibinfo
  {pages} {465901} (\bibinfo {year} {2017})}\BibitemShut {NoStop}%
\bibitem [{\citenamefont {Hamann}(2013)}]{2013HamannONCVPSP}%
  \BibitemOpen
  \bibfield  {author} {\bibinfo {author} {\bibfnamefont {D.~R.}\ \bibnamefont
  {Hamann}},\ }\bibfield  {title} {\enquote {\bibinfo {title} {Optimized
  norm-conserving {Vanderbilt} pseudopotentials},}\ }\href {\doibase
  10.1103/PhysRevB.88.085117} {\bibfield  {journal} {\bibinfo  {journal}
  {Physical Review B}\ }\textbf {\bibinfo {volume} {88}},\ \bibinfo {pages}
  {085117} (\bibinfo {year} {2013})}\BibitemShut {NoStop}%
\bibitem [{\citenamefont {van Setten}\ \emph {et~al.}(2018)\citenamefont {van
  Setten}, \citenamefont {Giantomassi}, \citenamefont {Bousquet}, \citenamefont
  {Verstraete}, \citenamefont {Hamann}, \citenamefont {Gonze},\ and\
  \citenamefont {Rignanese}}]{2018VanSettenPseudoDojo}%
  \BibitemOpen
  \bibfield  {author} {\bibinfo {author} {\bibfnamefont {M.J.}\ \bibnamefont
  {van Setten}}, \bibinfo {author} {\bibfnamefont {M.}~\bibnamefont
  {Giantomassi}}, \bibinfo {author} {\bibfnamefont {E.}~\bibnamefont
  {Bousquet}}, \bibinfo {author} {\bibfnamefont {M.J.}\ \bibnamefont
  {Verstraete}}, \bibinfo {author} {\bibfnamefont {D.R.}\ \bibnamefont
  {Hamann}}, \bibinfo {author} {\bibfnamefont {X.}~\bibnamefont {Gonze}}, \
  and\ \bibinfo {author} {\bibfnamefont {G.-M.}\ \bibnamefont {Rignanese}},\
  }\bibfield  {title} {\enquote {\bibinfo {title} {The {PseudoDojo}: Training
  and grading a 85 element optimized norm-conserving pseudopotential table},}\
  }\href {\doibase 10.1016/j.cpc.2018.01.012} {\bibfield  {journal} {\bibinfo
  {journal} {Computer Physics Communications}\ }\textbf {\bibinfo {volume}
  {226}},\ \bibinfo {pages} {39--54} (\bibinfo {year} {2018})}\BibitemShut
  {NoStop}%
\bibitem [{\citenamefont {Perdew}\ and\ \citenamefont
  {Wang}(1992)}]{1992PerdewLDA}%
  \BibitemOpen
  \bibfield  {author} {\bibinfo {author} {\bibfnamefont {John~P.}\ \bibnamefont
  {Perdew}}\ and\ \bibinfo {author} {\bibfnamefont {Yue}\ \bibnamefont
  {Wang}},\ }\bibfield  {title} {\enquote {\bibinfo {title} {Accurate and
  simple analytic representation of the electron-gas correlation energy},}\
  }\href {\doibase 10.1103/PhysRevB.45.13244} {\bibfield  {journal} {\bibinfo
  {journal} {Physical Review B}\ }\textbf {\bibinfo {volume} {45}},\ \bibinfo
  {pages} {13244--13249} (\bibinfo {year} {1992})}\BibitemShut {NoStop}%
\bibitem [{\citenamefont {Reeber}\ and\ \citenamefont
  {Wang}(1996)}]{1996ReeberSiliconExperiment}%
  \BibitemOpen
  \bibfield  {author} {\bibinfo {author} {\bibfnamefont {Robert~R.}\
  \bibnamefont {Reeber}}\ and\ \bibinfo {author} {\bibfnamefont {Kai}\
  \bibnamefont {Wang}},\ }\bibfield  {title} {\enquote {\bibinfo {title}
  {Thermal expansion and lattice parameters of group {IV} semiconductors},}\
  }\href {\doibase https://doi.org/10.1016/S0254-0584(96)01808-1} {\bibfield
  {journal} {\bibinfo  {journal} {Materials Chemistry and Physics}\ }\textbf
  {\bibinfo {volume} {46}},\ \bibinfo {pages} {259 -- 264} (\bibinfo {year}
  {1996})}\BibitemShut {NoStop}%
\bibitem [{\citenamefont {Lautenschlager}\ \emph {et~al.}(1985)\citenamefont
  {Lautenschlager}, \citenamefont {Allen},\ and\ \citenamefont
  {Cardona}}]{1985LautenschlagerSilicon}%
  \BibitemOpen
  \bibfield  {author} {\bibinfo {author} {\bibfnamefont {P.}~\bibnamefont
  {Lautenschlager}}, \bibinfo {author} {\bibfnamefont {P.~B.}\ \bibnamefont
  {Allen}}, \ and\ \bibinfo {author} {\bibfnamefont {M.}~\bibnamefont
  {Cardona}},\ }\bibfield  {title} {\enquote {\bibinfo {title} {Temperature
  dependence of band gaps in si and ge},}\ }\href {\doibase
  10.1103/PhysRevB.31.2163} {\bibfield  {journal} {\bibinfo  {journal} {Phys.
  Rev. B}\ }\textbf {\bibinfo {volume} {31}},\ \bibinfo {pages} {2163--2171}
  (\bibinfo {year} {1985})}\BibitemShut {NoStop}%
\bibitem [{\citenamefont {Perdew}\ \emph {et~al.}(1996)\citenamefont {Perdew},
  \citenamefont {Burke},\ and\ \citenamefont {Ernzerhof}}]{1996PerdewPBE}%
  \BibitemOpen
  \bibfield  {author} {\bibinfo {author} {\bibfnamefont {John~P.}\ \bibnamefont
  {Perdew}}, \bibinfo {author} {\bibfnamefont {Kieron}\ \bibnamefont {Burke}},
  \ and\ \bibinfo {author} {\bibfnamefont {Matthias}\ \bibnamefont
  {Ernzerhof}},\ }\bibfield  {title} {\enquote {\bibinfo {title} {Generalized
  gradient approximation made simple},}\ }\href {\doibase
  10.1103/PhysRevLett.77.3865} {\bibfield  {journal} {\bibinfo  {journal}
  {Phys. Rev. Lett.}\ }\textbf {\bibinfo {volume} {77}},\ \bibinfo {pages}
  {3865--3868} (\bibinfo {year} {1996})}\BibitemShut {NoStop}%
\bibitem [{\citenamefont {Pizzi}\ \emph {et~al.}(2020)\citenamefont {Pizzi},
  \citenamefont {Vitale}, \citenamefont {Arita}, \citenamefont {Bl{\"u}gel},
  \citenamefont {Freimuth}, \citenamefont {G{\'e}ranton}, \citenamefont
  {Gibertini}, \citenamefont {Gresch}, \citenamefont {Johnson}, \citenamefont
  {Koretsune}, \citenamefont {{Iba{\~n}ez-Azpiroz}}, \citenamefont {Lee},
  \citenamefont {Lihm}, \citenamefont {Marchand}, \citenamefont {Marrazzo},
  \citenamefont {Mokrousov}, \citenamefont {Mustafa}, \citenamefont {Nohara},
  \citenamefont {Nomura}, \citenamefont {Paulatto}, \citenamefont {Ponc{\'e}},
  \citenamefont {Ponweiser}, \citenamefont {Qiao}, \citenamefont {Th{\"o}le},
  \citenamefont {Tsirkin}, \citenamefont {Wierzbowska}, \citenamefont
  {Marzari}, \citenamefont {Vanderbilt}, \citenamefont {Souza}, \citenamefont
  {Mostofi},\ and\ \citenamefont {Yates}}]{2020PizziWannier90}%
  \BibitemOpen
  \bibfield  {author} {\bibinfo {author} {\bibfnamefont {Giovanni}\
  \bibnamefont {Pizzi}}, \bibinfo {author} {\bibfnamefont {Valerio}\
  \bibnamefont {Vitale}}, \bibinfo {author} {\bibfnamefont {Ryotaro}\
  \bibnamefont {Arita}}, \bibinfo {author} {\bibfnamefont {Stefan}\
  \bibnamefont {Bl{\"u}gel}}, \bibinfo {author} {\bibfnamefont {Frank}\
  \bibnamefont {Freimuth}}, \bibinfo {author} {\bibfnamefont {Guillaume}\
  \bibnamefont {G{\'e}ranton}}, \bibinfo {author} {\bibfnamefont {Marco}\
  \bibnamefont {Gibertini}}, \bibinfo {author} {\bibfnamefont {Dominik}\
  \bibnamefont {Gresch}}, \bibinfo {author} {\bibfnamefont {Charles}\
  \bibnamefont {Johnson}}, \bibinfo {author} {\bibfnamefont {Takashi}\
  \bibnamefont {Koretsune}}, \bibinfo {author} {\bibfnamefont {Julen}\
  \bibnamefont {{Iba{\~n}ez-Azpiroz}}}, \bibinfo {author} {\bibfnamefont
  {Hyungjun}\ \bibnamefont {Lee}}, \bibinfo {author} {\bibfnamefont {Jae-Mo}\
  \bibnamefont {Lihm}}, \bibinfo {author} {\bibfnamefont {Daniel}\ \bibnamefont
  {Marchand}}, \bibinfo {author} {\bibfnamefont {Antimo}\ \bibnamefont
  {Marrazzo}}, \bibinfo {author} {\bibfnamefont {Yuriy}\ \bibnamefont
  {Mokrousov}}, \bibinfo {author} {\bibfnamefont {Jamal~I}\ \bibnamefont
  {Mustafa}}, \bibinfo {author} {\bibfnamefont {Yoshiro}\ \bibnamefont
  {Nohara}}, \bibinfo {author} {\bibfnamefont {Yusuke}\ \bibnamefont {Nomura}},
  \bibinfo {author} {\bibfnamefont {Lorenzo}\ \bibnamefont {Paulatto}},
  \bibinfo {author} {\bibfnamefont {Samuel}\ \bibnamefont {Ponc{\'e}}},
  \bibinfo {author} {\bibfnamefont {Thomas}\ \bibnamefont {Ponweiser}},
  \bibinfo {author} {\bibfnamefont {Junfeng}\ \bibnamefont {Qiao}}, \bibinfo
  {author} {\bibfnamefont {Florian}\ \bibnamefont {Th{\"o}le}}, \bibinfo
  {author} {\bibfnamefont {Stepan~S}\ \bibnamefont {Tsirkin}}, \bibinfo
  {author} {\bibfnamefont {Ma{\l}gorzata}\ \bibnamefont {Wierzbowska}},
  \bibinfo {author} {\bibfnamefont {Nicola}\ \bibnamefont {Marzari}}, \bibinfo
  {author} {\bibfnamefont {David}\ \bibnamefont {Vanderbilt}}, \bibinfo
  {author} {\bibfnamefont {Ivo}\ \bibnamefont {Souza}}, \bibinfo {author}
  {\bibfnamefont {Arash~A}\ \bibnamefont {Mostofi}}, \ and\ \bibinfo {author}
  {\bibfnamefont {Jonathan~R}\ \bibnamefont {Yates}},\ }\bibfield  {title}
  {\enquote {\bibinfo {title} {Wannier90 as a community code: New features and
  applications},}\ }\href {\doibase 10.1088/1361-648X/ab51ff} {\bibfield
  {journal} {\bibinfo  {journal} {Journal of Physics: Condensed Matter}\
  }\textbf {\bibinfo {volume} {32}},\ \bibinfo {pages} {165902} (\bibinfo
  {year} {2020})}\BibitemShut {NoStop}%
\bibitem [{\citenamefont {Deslippe}\ \emph {et~al.}(2012)\citenamefont
  {Deslippe}, \citenamefont {Samsonidze}, \citenamefont {Strubbe},
  \citenamefont {Jain}, \citenamefont {Cohen},\ and\ \citenamefont
  {Louie}}]{2012DeslippeBerkeleyGW}%
  \BibitemOpen
  \bibfield  {author} {\bibinfo {author} {\bibfnamefont {Jack}\ \bibnamefont
  {Deslippe}}, \bibinfo {author} {\bibfnamefont {Georgy}\ \bibnamefont
  {Samsonidze}}, \bibinfo {author} {\bibfnamefont {David~A.}\ \bibnamefont
  {Strubbe}}, \bibinfo {author} {\bibfnamefont {Manish}\ \bibnamefont {Jain}},
  \bibinfo {author} {\bibfnamefont {Marvin~L.}\ \bibnamefont {Cohen}}, \ and\
  \bibinfo {author} {\bibfnamefont {Steven~G.}\ \bibnamefont {Louie}},\
  }\bibfield  {title} {\enquote {\bibinfo {title} {{{BerkeleyGW}}: {{A}}
  massively parallel computer package for the calculation of the quasiparticle
  and optical properties of materials and nanostructures},}\ }\href {\doibase
  10.1016/j.cpc.2011.12.006} {\bibfield  {journal} {\bibinfo  {journal}
  {Computer Physics Communications}\ }\textbf {\bibinfo {volume} {183}},\
  \bibinfo {pages} {1269--1289} (\bibinfo {year} {2012})}\BibitemShut {NoStop}%
\bibitem [{\citenamefont {Deslippe}\ \emph {et~al.}(2013)\citenamefont
  {Deslippe}, \citenamefont {Samsonidze}, \citenamefont {Jain}, \citenamefont
  {Cohen},\ and\ \citenamefont {Louie}}]{2013DeslippeGwRemainder}%
  \BibitemOpen
  \bibfield  {author} {\bibinfo {author} {\bibfnamefont {Jack}\ \bibnamefont
  {Deslippe}}, \bibinfo {author} {\bibfnamefont {Georgy}\ \bibnamefont
  {Samsonidze}}, \bibinfo {author} {\bibfnamefont {Manish}\ \bibnamefont
  {Jain}}, \bibinfo {author} {\bibfnamefont {Marvin~L.}\ \bibnamefont {Cohen}},
  \ and\ \bibinfo {author} {\bibfnamefont {Steven~G.}\ \bibnamefont {Louie}},\
  }\bibfield  {title} {\enquote {\bibinfo {title} {Coulomb-hole summations and
  energies for {$GW$} calculations with limited number of empty orbitals: A
  modified static remainder approach},}\ }\href {\doibase
  10.1103/PhysRevB.87.165124} {\bibfield  {journal} {\bibinfo  {journal} {Phys.
  Rev. B}\ }\textbf {\bibinfo {volume} {87}},\ \bibinfo {pages} {165124}
  (\bibinfo {year} {2013})}\BibitemShut {NoStop}%
\bibitem [{\citenamefont {Ponc{\'e}}\ \emph {et~al.}(2016)\citenamefont
  {Ponc{\'e}}, \citenamefont {Margine}, \citenamefont {Verdi},\ and\
  \citenamefont {Giustino}}]{2016PonceEPW}%
  \BibitemOpen
  \bibfield  {author} {\bibinfo {author} {\bibfnamefont {S.}~\bibnamefont
  {Ponc{\'e}}}, \bibinfo {author} {\bibfnamefont {E.R.}\ \bibnamefont
  {Margine}}, \bibinfo {author} {\bibfnamefont {C.}~\bibnamefont {Verdi}}, \
  and\ \bibinfo {author} {\bibfnamefont {F.}~\bibnamefont {Giustino}},\
  }\bibfield  {title} {\enquote {\bibinfo {title} {{{EPW}}:
  {{Electron}}\textendash phonon coupling, transport and superconducting
  properties using maximally localized {{Wannier}} functions},}\ }\href
  {\doibase 10.1016/j.cpc.2016.07.028} {\bibfield  {journal} {\bibinfo
  {journal} {Computer Physics Communications}\ }\textbf {\bibinfo {volume}
  {209}},\ \bibinfo {pages} {116--133} (\bibinfo {year} {2016})}\BibitemShut
  {NoStop}%
\bibitem [{\citenamefont {Rohlfing}\ and\ \citenamefont
  {Louie}(2000)}]{2000Rohlfing}%
  \BibitemOpen
  \bibfield  {author} {\bibinfo {author} {\bibfnamefont {Michael}\ \bibnamefont
  {Rohlfing}}\ and\ \bibinfo {author} {\bibfnamefont {Steven~G.}\ \bibnamefont
  {Louie}},\ }\bibfield  {title} {\enquote {\bibinfo {title} {Electron-hole
  excitations and optical spectra from first principles},}\ }\href {\doibase
  10.1103/PhysRevB.62.4927} {\bibfield  {journal} {\bibinfo  {journal} {Phys.
  Rev. B}\ }\textbf {\bibinfo {volume} {62}},\ \bibinfo {pages} {4927--4944}
  (\bibinfo {year} {2000})}\BibitemShut {NoStop}%
\bibitem [{\citenamefont {Green}\ and\ \citenamefont
  {Keevers}(1995)}]{1995GreenSiExperiment}%
  \BibitemOpen
  \bibfield  {author} {\bibinfo {author} {\bibfnamefont {Martin~A}\
  \bibnamefont {Green}}\ and\ \bibinfo {author} {\bibfnamefont {Mark~J}\
  \bibnamefont {Keevers}},\ }\bibfield  {title} {\enquote {\bibinfo {title}
  {Optical properties of intrinsic silicon at 300 {K}},}\ }\href@noop {}
  {\bibfield  {journal} {\bibinfo  {journal} {Progress in Photovoltaics:
  Research and Applications}\ }\textbf {\bibinfo {volume} {3}},\ \bibinfo
  {pages} {189--192} (\bibinfo {year} {1995})}\BibitemShut {NoStop}%
\bibitem [{\citenamefont {Nastos}\ and\ \citenamefont
  {Sipe}(2006)}]{2006NastosShift}%
  \BibitemOpen
  \bibfield  {author} {\bibinfo {author} {\bibfnamefont {F.}~\bibnamefont
  {Nastos}}\ and\ \bibinfo {author} {\bibfnamefont {J.~E.}\ \bibnamefont
  {Sipe}},\ }\bibfield  {title} {\enquote {\bibinfo {title} {Optical
  rectification and shift currents in {{GaAs}} and {{GaP}} response: {{Below}}
  and above the band gap},}\ }\href {\doibase 10.1103/PhysRevB.74.035201}
  {\bibfield  {journal} {\bibinfo  {journal} {Physical Review B}\ }\textbf
  {\bibinfo {volume} {74}},\ \bibinfo {pages} {035201} (\bibinfo {year}
  {2006})}\BibitemShut {NoStop}%
\bibitem [{\citenamefont {Tsirkin}(2021)}]{2021TsirkinWBerri}%
  \BibitemOpen
  \bibfield  {author} {\bibinfo {author} {\bibfnamefont {Stepan~S.}\
  \bibnamefont {Tsirkin}},\ }\bibfield  {title} {\enquote {\bibinfo {title}
  {High performance {{Wannier}} interpolation of {{Berry}} curvature and
  related quantities with {{WannierBerri}} code},}\ }\href {\doibase
  10.1038/s41524-021-00498-5} {\bibfield  {journal} {\bibinfo  {journal} {npj
  Computational Materials}\ }\textbf {\bibinfo {volume} {7}},\ \bibinfo {pages}
  {33} (\bibinfo {year} {2021})}\BibitemShut {NoStop}%
\bibitem [{\citenamefont {Lee}\ \emph {et~al.}(2015)\citenamefont {Lee},
  \citenamefont {Silva}, \citenamefont {Calderin}, \citenamefont {Nguyen},
  \citenamefont {Hollander}, \citenamefont {Bersch}, \citenamefont {Mallouk},\
  and\ \citenamefont {Robinson}}]{2015LeeWTe2}%
  \BibitemOpen
  \bibfield  {author} {\bibinfo {author} {\bibfnamefont {Chia-Hui}\
  \bibnamefont {Lee}}, \bibinfo {author} {\bibfnamefont {Eduardo~Cruz}\
  \bibnamefont {Silva}}, \bibinfo {author} {\bibfnamefont {Lazaro}\
  \bibnamefont {Calderin}}, \bibinfo {author} {\bibfnamefont {Minh An~T.}\
  \bibnamefont {Nguyen}}, \bibinfo {author} {\bibfnamefont {Matthew~J.}\
  \bibnamefont {Hollander}}, \bibinfo {author} {\bibfnamefont {Brian}\
  \bibnamefont {Bersch}}, \bibinfo {author} {\bibfnamefont {Thomas~E.}\
  \bibnamefont {Mallouk}}, \ and\ \bibinfo {author} {\bibfnamefont {Joshua~A.}\
  \bibnamefont {Robinson}},\ }\bibfield  {title} {\enquote {\bibinfo {title}
  {Tungsten ditelluride: a layered semimetal},}\ }\href {\doibase
  10.1038/srep10013} {\bibfield  {journal} {\bibinfo  {journal} {Scientific
  Reports}\ }\textbf {\bibinfo {volume} {5}},\ \bibinfo {pages} {10013}
  (\bibinfo {year} {2015})}\BibitemShut {NoStop}%
\bibitem [{\citenamefont {Momma}\ and\ \citenamefont
  {Izumi}(2011)}]{2011MommaVESTA}%
  \BibitemOpen
  \bibfield  {author} {\bibinfo {author} {\bibfnamefont {Koichi}\ \bibnamefont
  {Momma}}\ and\ \bibinfo {author} {\bibfnamefont {Fujio}\ \bibnamefont
  {Izumi}},\ }\bibfield  {title} {\enquote {\bibinfo {title} {{{\it VESTA3} for
  three-dimensional visualization of crystal, volumetric and morphology
  data}},}\ }\href {\doibase 10.1107/S0021889811038970} {\bibfield  {journal}
  {\bibinfo  {journal} {Journal of Applied Crystallography}\ }\textbf {\bibinfo
  {volume} {44}},\ \bibinfo {pages} {1272--1276} (\bibinfo {year}
  {2011})}\BibitemShut {NoStop}%
\bibitem [{\citenamefont {Poncé}\ \emph {et~al.}(2015)\citenamefont {Poncé},
  \citenamefont {Gillet}, \citenamefont {Laflamme~Janssen}, \citenamefont
  {Marini}, \citenamefont {Verstraete},\ and\ \citenamefont
  {Gonze}}]{2015PonceJCP}%
  \BibitemOpen
  \bibfield  {author} {\bibinfo {author} {\bibfnamefont {S.}~\bibnamefont
  {Poncé}}, \bibinfo {author} {\bibfnamefont {Y.}~\bibnamefont {Gillet}},
  \bibinfo {author} {\bibfnamefont {J.}~\bibnamefont {Laflamme~Janssen}},
  \bibinfo {author} {\bibfnamefont {A.}~\bibnamefont {Marini}}, \bibinfo
  {author} {\bibfnamefont {M.}~\bibnamefont {Verstraete}}, \ and\ \bibinfo
  {author} {\bibfnamefont {X.}~\bibnamefont {Gonze}},\ }\bibfield  {title}
  {\enquote {\bibinfo {title} {Temperature dependence of the electronic
  structure of semiconductors and insulators},}\ }\href {\doibase
  10.1063/1.4927081} {\bibfield  {journal} {\bibinfo  {journal} {The Journal of
  Chemical Physics}\ }\textbf {\bibinfo {volume} {143}},\ \bibinfo {pages}
  {102813} (\bibinfo {year} {2015})}\BibitemShut {NoStop}%
\bibitem [{\citenamefont {Querales-Flores}\ \emph {et~al.}(2019)\citenamefont
  {Querales-Flores}, \citenamefont {Cao}, \citenamefont {Fahy},\ and\
  \citenamefont {Savić}}]{2019QueralesFlores}%
  \BibitemOpen
  \bibfield  {author} {\bibinfo {author} {\bibfnamefont {José~D.}\
  \bibnamefont {Querales-Flores}}, \bibinfo {author} {\bibfnamefont {Jiang}\
  \bibnamefont {Cao}}, \bibinfo {author} {\bibfnamefont {Stephen}\ \bibnamefont
  {Fahy}}, \ and\ \bibinfo {author} {\bibfnamefont {Ivana}\ \bibnamefont
  {Savić}},\ }\bibfield  {title} {\enquote {\bibinfo {title} {Temperature
  effects on the electronic band structure of {PbTe} from first principles},}\
  }\href {\doibase 10.1103/PhysRevMaterials.3.055405} {\bibfield  {journal}
  {\bibinfo  {journal} {Physical Review Materials}\ }\textbf {\bibinfo {volume}
  {3}},\ \bibinfo {pages} {055405} (\bibinfo {year} {2019})}\BibitemShut
  {NoStop}%
\bibitem [{\citenamefont {Ponc\'e}\ \emph {et~al.}(2021)\citenamefont
  {Ponc\'e}, \citenamefont {Macheda}, \citenamefont {Margine}, \citenamefont
  {Marzari}, \citenamefont {Bonini},\ and\ \citenamefont
  {Giustino}}]{2021PonceMobility}%
  \BibitemOpen
  \bibfield  {author} {\bibinfo {author} {\bibfnamefont {Samuel}\ \bibnamefont
  {Ponc\'e}}, \bibinfo {author} {\bibfnamefont {Francesco}\ \bibnamefont
  {Macheda}}, \bibinfo {author} {\bibfnamefont {Elena~Roxana}\ \bibnamefont
  {Margine}}, \bibinfo {author} {\bibfnamefont {Nicola}\ \bibnamefont
  {Marzari}}, \bibinfo {author} {\bibfnamefont {Nicola}\ \bibnamefont
  {Bonini}}, \ and\ \bibinfo {author} {\bibfnamefont {Feliciano}\ \bibnamefont
  {Giustino}},\ }\bibfield  {title} {\enquote {\bibinfo {title}
  {First-principles predictions of hall and drift mobilities in
  semiconductors},}\ }\href {\doibase 10.1103/PhysRevResearch.3.043022}
  {\bibfield  {journal} {\bibinfo  {journal} {Phys. Rev. Research}\ }\textbf
  {\bibinfo {volume} {3}},\ \bibinfo {pages} {043022} (\bibinfo {year}
  {2021})}\BibitemShut {NoStop}%
\bibitem [{\citenamefont {Adolph}\ \emph {et~al.}(1996)\citenamefont {Adolph},
  \citenamefont {Gavrilenko}, \citenamefont {Tenelsen}, \citenamefont
  {Bechstedt},\ and\ \citenamefont {Del~Sole}}]{1996AdolphVelocity}%
  \BibitemOpen
  \bibfield  {author} {\bibinfo {author} {\bibfnamefont {B.}~\bibnamefont
  {Adolph}}, \bibinfo {author} {\bibfnamefont {V.~I.}\ \bibnamefont
  {Gavrilenko}}, \bibinfo {author} {\bibfnamefont {K.}~\bibnamefont
  {Tenelsen}}, \bibinfo {author} {\bibfnamefont {F.}~\bibnamefont {Bechstedt}},
  \ and\ \bibinfo {author} {\bibfnamefont {R.}~\bibnamefont {Del~Sole}},\
  }\bibfield  {title} {\enquote {\bibinfo {title} {Nonlocality and many-body
  effects in the optical properties of semiconductors},}\ }\href {\doibase
  10.1103/PhysRevB.53.9797} {\bibfield  {journal} {\bibinfo  {journal} {Phys.
  Rev. B}\ }\textbf {\bibinfo {volume} {53}},\ \bibinfo {pages} {9797--9808}
  (\bibinfo {year} {1996})}\BibitemShut {NoStop}%
\bibitem [{\citenamefont {Lopez}\ \emph {et~al.}(2012)\citenamefont {Lopez},
  \citenamefont {Vanderbilt}, \citenamefont {Thonhauser},\ and\ \citenamefont
  {Souza}}]{2012LopezOrbitalMag}%
  \BibitemOpen
  \bibfield  {author} {\bibinfo {author} {\bibfnamefont {M.~G.}\ \bibnamefont
  {Lopez}}, \bibinfo {author} {\bibfnamefont {David}\ \bibnamefont
  {Vanderbilt}}, \bibinfo {author} {\bibfnamefont {T.}~\bibnamefont
  {Thonhauser}}, \ and\ \bibinfo {author} {\bibfnamefont {Ivo}\ \bibnamefont
  {Souza}},\ }\bibfield  {title} {\enquote {\bibinfo {title} {Wannier-based
  calculation of the orbital magnetization in crystals},}\ }\href {\doibase
  10.1103/PhysRevB.85.014435} {\bibfield  {journal} {\bibinfo  {journal}
  {Physical Review B}\ }\textbf {\bibinfo {volume} {85}},\ \bibinfo {pages}
  {014435} (\bibinfo {year} {2012})}\BibitemShut {NoStop}%
\end{thebibliography}%


\begin{thebibliography}{10}%
\makeatletter
\providecommand \@ifxundefined [1]{%
 \@ifx{#1\undefined}
}%
\providecommand \@ifnum [1]{%
 \ifnum #1\expandafter \@firstoftwo
 \else \expandafter \@secondoftwo
 \fi
}%
\providecommand \@ifx [1]{%
 \ifx #1\expandafter \@firstoftwo
 \else \expandafter \@secondoftwo
 \fi
}%
\providecommand \natexlab [1]{#1}%
\providecommand \enquote  [1]{``#1''}%
\providecommand \bibnamefont  [1]{#1}%
\providecommand \bibfnamefont [1]{#1}%
\providecommand \citenamefont [1]{#1}%
\providecommand \href@noop [0]{\@secondoftwo}%
\providecommand \href [0]{\begingroup \@sanitize@url \@href}%
\providecommand \@href[1]{\@@startlink{#1}\@@href}%
\providecommand \@@href[1]{\endgroup#1\@@endlink}%
\providecommand \@sanitize@url [0]{\catcode `\\12\catcode `\$12\catcode
  `\&12\catcode `\#12\catcode `\^12\catcode `\_12\catcode `\%12\relax}%
\providecommand \@@startlink[1]{}%
\providecommand \@@endlink[0]{}%
\providecommand \url  [0]{\begingroup\@sanitize@url \@url }%
\providecommand \@url [1]{\endgroup\@href {#1}{\urlprefix }}%
\providecommand \urlprefix  [0]{URL }%
\providecommand \Eprint [0]{\href }%
\providecommand \doibase [0]{http://dx.doi.org/}%
\providecommand \selectlanguage [0]{\@gobble}%
\providecommand \bibinfo  [0]{\@secondoftwo}%
\providecommand \bibfield  [0]{\@secondoftwo}%
\providecommand \translation [1]{[#1]}%
\providecommand \BibitemOpen [0]{}%
\providecommand \bibitemStop [0]{}%
\providecommand \bibitemNoStop [0]{.\EOS\space}%
\providecommand \EOS [0]{\spacefactor3000\relax}%
\providecommand \BibitemShut  [1]{\csname bibitem#1\endcsname}%
\let\auto@bib@innerbib\@empty
\bibitem [{\citenamefont {Wang}\ \emph {et~al.}(2006)\citenamefont {Wang},
  \citenamefont {Yates}, \citenamefont {Souza},\ and\ \citenamefont
  {Vanderbilt}}]{2006WangAHC}%
  \BibitemOpen
  \bibfield  {author} {\bibinfo {author} {\bibfnamefont {Xinjie}\ \bibnamefont
  {Wang}}, \bibinfo {author} {\bibfnamefont {Jonathan~R.}\ \bibnamefont
  {Yates}}, \bibinfo {author} {\bibfnamefont {Ivo}\ \bibnamefont {Souza}}, \
  and\ \bibinfo {author} {\bibfnamefont {David}\ \bibnamefont {Vanderbilt}},\
  }\bibfield  {title} {\enquote {\bibinfo {title} {{\emph{Ab Initio}}
  calculation of the anomalous {{Hall}} conductivity by {{Wannier}}
  interpolation},}\ }\href {\doibase 10.1103/PhysRevB.74.195118} {\bibfield
  {journal} {\bibinfo  {journal} {Physical Review B}\ }\textbf {\bibinfo
  {volume} {74}},\ \bibinfo {pages} {195118} (\bibinfo {year}
  {2006})}\BibitemShut {NoStop}%
\bibitem [{\citenamefont {Qiao}\ \emph {et~al.}(2018)\citenamefont {Qiao},
  \citenamefont {Zhou}, \citenamefont {Yuan},\ and\ \citenamefont
  {Zhao}}]{2018QiaoSHC}%
  \BibitemOpen
  \bibfield  {author} {\bibinfo {author} {\bibfnamefont {Junfeng}\ \bibnamefont
  {Qiao}}, \bibinfo {author} {\bibfnamefont {Jiaqi}\ \bibnamefont {Zhou}},
  \bibinfo {author} {\bibfnamefont {Zhe}\ \bibnamefont {Yuan}}, \ and\ \bibinfo
  {author} {\bibfnamefont {Weisheng}\ \bibnamefont {Zhao}},\ }\bibfield
  {title} {\enquote {\bibinfo {title} {Calculation of intrinsic spin {{Hall}}
  conductivity by {{Wannier}} interpolation},}\ }\href {\doibase
  10.1103/PhysRevB.98.214402} {\bibfield  {journal} {\bibinfo  {journal}
  {Physical Review B}\ }\textbf {\bibinfo {volume} {98}},\ \bibinfo {pages}
  {214402} (\bibinfo {year} {2018})}\BibitemShut {NoStop}%
\bibitem [{\citenamefont {Ryoo}\ \emph {et~al.}(2019)\citenamefont {Ryoo},
  \citenamefont {Park},\ and\ \citenamefont {Souza}}]{2019RyooSHC}%
  \BibitemOpen
  \bibfield  {author} {\bibinfo {author} {\bibfnamefont {Ji~Hoon}\ \bibnamefont
  {Ryoo}}, \bibinfo {author} {\bibfnamefont {Cheol-Hwan}\ \bibnamefont {Park}},
  \ and\ \bibinfo {author} {\bibfnamefont {Ivo}\ \bibnamefont {Souza}},\
  }\bibfield  {title} {\enquote {\bibinfo {title} {Computation of intrinsic
  spin {{Hall}} conductivities from first principles using maximally localized
  {{Wannier}} functions},}\ }\href {\doibase 10.1103/PhysRevB.99.235113}
  {\bibfield  {journal} {\bibinfo  {journal} {Physical Review B}\ }\textbf
  {\bibinfo {volume} {99}},\ \bibinfo {pages} {235113} (\bibinfo {year}
  {2019})}\BibitemShut {NoStop}%
\bibitem [{\citenamefont {Marzari}\ and\ \citenamefont
  {Vanderbilt}(1997)}]{1997Marzari}%
  \BibitemOpen
  \bibfield  {author} {\bibinfo {author} {\bibfnamefont {Nicola}\ \bibnamefont
  {Marzari}}\ and\ \bibinfo {author} {\bibfnamefont {David}\ \bibnamefont
  {Vanderbilt}},\ }\bibfield  {title} {\enquote {\bibinfo {title}
  {Maximally-localized generalized {{Wannier}} functions for composite energy
  bands},}\ }\href {\doibase 10.1103/PhysRevB.56.12847} {\bibfield  {journal}
  {\bibinfo  {journal} {Physical Review B}\ }\textbf {\bibinfo {volume} {56}},\
  \bibinfo {pages} {12847--12865} (\bibinfo {year} {1997})}\BibitemShut
  {NoStop}%
\bibitem [{\citenamefont {Ponc\'e}\ \emph {et~al.}(2021)\citenamefont
  {Ponc\'e}, \citenamefont {Macheda}, \citenamefont {Margine}, \citenamefont
  {Marzari}, \citenamefont {Bonini},\ and\ \citenamefont
  {Giustino}}]{2021PonceMobility}%
  \BibitemOpen
  \bibfield  {author} {\bibinfo {author} {\bibfnamefont {Samuel}\ \bibnamefont
  {Ponc\'e}}, \bibinfo {author} {\bibfnamefont {Francesco}\ \bibnamefont
  {Macheda}}, \bibinfo {author} {\bibfnamefont {Elena~Roxana}\ \bibnamefont
  {Margine}}, \bibinfo {author} {\bibfnamefont {Nicola}\ \bibnamefont
  {Marzari}}, \bibinfo {author} {\bibfnamefont {Nicola}\ \bibnamefont
  {Bonini}}, \ and\ \bibinfo {author} {\bibfnamefont {Feliciano}\ \bibnamefont
  {Giustino}},\ }\bibfield  {title} {\enquote {\bibinfo {title}
  {First-principles predictions of hall and drift mobilities in
  semiconductors},}\ }\href {\doibase 10.1103/PhysRevResearch.3.043022}
  {\bibfield  {journal} {\bibinfo  {journal} {Phys. Rev. Research}\ }\textbf
  {\bibinfo {volume} {3}},\ \bibinfo {pages} {043022} (\bibinfo {year}
  {2021})}\BibitemShut {NoStop}%
\bibitem [{\citenamefont {Adolph}\ \emph {et~al.}(1996)\citenamefont {Adolph},
  \citenamefont {Gavrilenko}, \citenamefont {Tenelsen}, \citenamefont
  {Bechstedt},\ and\ \citenamefont {Del~Sole}}]{1996AdolphVelocity}%
  \BibitemOpen
  \bibfield  {author} {\bibinfo {author} {\bibfnamefont {B.}~\bibnamefont
  {Adolph}}, \bibinfo {author} {\bibfnamefont {V.~I.}\ \bibnamefont
  {Gavrilenko}}, \bibinfo {author} {\bibfnamefont {K.}~\bibnamefont
  {Tenelsen}}, \bibinfo {author} {\bibfnamefont {F.}~\bibnamefont {Bechstedt}},
  \ and\ \bibinfo {author} {\bibfnamefont {R.}~\bibnamefont {Del~Sole}},\
  }\bibfield  {title} {\enquote {\bibinfo {title} {Nonlocality and many-body
  effects in the optical properties of semiconductors},}\ }\href {\doibase
  10.1103/PhysRevB.53.9797} {\bibfield  {journal} {\bibinfo  {journal} {Phys.
  Rev. B}\ }\textbf {\bibinfo {volume} {53}},\ \bibinfo {pages} {9797--9808}
  (\bibinfo {year} {1996})}\BibitemShut {NoStop}%
\bibitem [{\citenamefont {Giannozzi}\ \emph {et~al.}(2017)\citenamefont
  {Giannozzi}, \citenamefont {Andreussi}, \citenamefont {Brumme}, \citenamefont
  {Bunau}, \citenamefont {Buongiorno~Nardelli}, \citenamefont {Calandra},
  \citenamefont {Car}, \citenamefont {Cavazzoni}, \citenamefont {Ceresoli},
  \citenamefont {Cococcioni}, \citenamefont {Colonna}, \citenamefont
  {Carnimeo}, \citenamefont {Dal~Corso}, \citenamefont {{de Gironcoli}},
  \citenamefont {Delugas}, \citenamefont {DiStasio}, \citenamefont {Ferretti},
  \citenamefont {Floris}, \citenamefont {Fratesi}, \citenamefont {Fugallo},
  \citenamefont {Gebauer}, \citenamefont {Gerstmann}, \citenamefont {Giustino},
  \citenamefont {Gorni}, \citenamefont {Jia}, \citenamefont {Kawamura},
  \citenamefont {Ko}, \citenamefont {Kokalj}, \citenamefont {K{\"u}{\c
  c}{\"u}kbenli}, \citenamefont {Lazzeri}, \citenamefont {Marsili},
  \citenamefont {Marzari}, \citenamefont {Mauri}, \citenamefont {Nguyen},
  \citenamefont {Nguyen}, \citenamefont {{Otero-de-la-Roza}}, \citenamefont
  {Paulatto}, \citenamefont {Ponc{\'e}}, \citenamefont {Rocca}, \citenamefont
  {Sabatini}, \citenamefont {Santra}, \citenamefont {Schlipf}, \citenamefont
  {Seitsonen}, \citenamefont {Smogunov}, \citenamefont {Timrov}, \citenamefont
  {Thonhauser}, \citenamefont {Umari}, \citenamefont {Vast}, \citenamefont
  {Wu},\ and\ \citenamefont {Baroni}}]{2017GiannozziQE}%
  \BibitemOpen
  \bibfield  {author} {\bibinfo {author} {\bibfnamefont {P}~\bibnamefont
  {Giannozzi}}, \bibinfo {author} {\bibfnamefont {O}~\bibnamefont {Andreussi}},
  \bibinfo {author} {\bibfnamefont {T}~\bibnamefont {Brumme}}, \bibinfo
  {author} {\bibfnamefont {O}~\bibnamefont {Bunau}}, \bibinfo {author}
  {\bibfnamefont {M}~\bibnamefont {Buongiorno~Nardelli}}, \bibinfo {author}
  {\bibfnamefont {M}~\bibnamefont {Calandra}}, \bibinfo {author} {\bibfnamefont
  {R}~\bibnamefont {Car}}, \bibinfo {author} {\bibfnamefont {C}~\bibnamefont
  {Cavazzoni}}, \bibinfo {author} {\bibfnamefont {D}~\bibnamefont {Ceresoli}},
  \bibinfo {author} {\bibfnamefont {M}~\bibnamefont {Cococcioni}}, \bibinfo
  {author} {\bibfnamefont {N}~\bibnamefont {Colonna}}, \bibinfo {author}
  {\bibfnamefont {I}~\bibnamefont {Carnimeo}}, \bibinfo {author} {\bibfnamefont
  {A}~\bibnamefont {Dal~Corso}}, \bibinfo {author} {\bibfnamefont
  {S}~\bibnamefont {{de Gironcoli}}}, \bibinfo {author} {\bibfnamefont
  {P}~\bibnamefont {Delugas}}, \bibinfo {author} {\bibfnamefont {R~A}\
  \bibnamefont {DiStasio}}, \bibinfo {author} {\bibfnamefont {A}~\bibnamefont
  {Ferretti}}, \bibinfo {author} {\bibfnamefont {A}~\bibnamefont {Floris}},
  \bibinfo {author} {\bibfnamefont {G}~\bibnamefont {Fratesi}}, \bibinfo
  {author} {\bibfnamefont {G}~\bibnamefont {Fugallo}}, \bibinfo {author}
  {\bibfnamefont {R}~\bibnamefont {Gebauer}}, \bibinfo {author} {\bibfnamefont
  {U}~\bibnamefont {Gerstmann}}, \bibinfo {author} {\bibfnamefont
  {F}~\bibnamefont {Giustino}}, \bibinfo {author} {\bibfnamefont
  {T}~\bibnamefont {Gorni}}, \bibinfo {author} {\bibfnamefont {J}~\bibnamefont
  {Jia}}, \bibinfo {author} {\bibfnamefont {M}~\bibnamefont {Kawamura}},
  \bibinfo {author} {\bibfnamefont {H-Y}\ \bibnamefont {Ko}}, \bibinfo {author}
  {\bibfnamefont {A}~\bibnamefont {Kokalj}}, \bibinfo {author} {\bibfnamefont
  {E}~\bibnamefont {K{\"u}{\c c}{\"u}kbenli}}, \bibinfo {author} {\bibfnamefont
  {M}~\bibnamefont {Lazzeri}}, \bibinfo {author} {\bibfnamefont
  {M}~\bibnamefont {Marsili}}, \bibinfo {author} {\bibfnamefont
  {N}~\bibnamefont {Marzari}}, \bibinfo {author} {\bibfnamefont
  {F}~\bibnamefont {Mauri}}, \bibinfo {author} {\bibfnamefont {N~L}\
  \bibnamefont {Nguyen}}, \bibinfo {author} {\bibfnamefont {H-V}\ \bibnamefont
  {Nguyen}}, \bibinfo {author} {\bibfnamefont {A}~\bibnamefont
  {{Otero-de-la-Roza}}}, \bibinfo {author} {\bibfnamefont {L}~\bibnamefont
  {Paulatto}}, \bibinfo {author} {\bibfnamefont {S}~\bibnamefont {Ponc{\'e}}},
  \bibinfo {author} {\bibfnamefont {D}~\bibnamefont {Rocca}}, \bibinfo {author}
  {\bibfnamefont {R}~\bibnamefont {Sabatini}}, \bibinfo {author} {\bibfnamefont
  {B}~\bibnamefont {Santra}}, \bibinfo {author} {\bibfnamefont {M}~\bibnamefont
  {Schlipf}}, \bibinfo {author} {\bibfnamefont {A~P}\ \bibnamefont
  {Seitsonen}}, \bibinfo {author} {\bibfnamefont {A}~\bibnamefont {Smogunov}},
  \bibinfo {author} {\bibfnamefont {I}~\bibnamefont {Timrov}}, \bibinfo
  {author} {\bibfnamefont {T}~\bibnamefont {Thonhauser}}, \bibinfo {author}
  {\bibfnamefont {P}~\bibnamefont {Umari}}, \bibinfo {author} {\bibfnamefont
  {N}~\bibnamefont {Vast}}, \bibinfo {author} {\bibfnamefont {X}~\bibnamefont
  {Wu}}, \ and\ \bibinfo {author} {\bibfnamefont {S}~\bibnamefont {Baroni}},\
  }\bibfield  {title} {\enquote {\bibinfo {title} {Advanced capabilities for
  materials modelling with {{Quantum ESPRESSO}}},}\ }\href {\doibase
  10.1088/1361-648X/aa8f79} {\bibfield  {journal} {\bibinfo  {journal} {Journal
  of Physics: Condensed Matter}\ }\textbf {\bibinfo {volume} {29}},\ \bibinfo
  {pages} {465901} (\bibinfo {year} {2017})}\BibitemShut {NoStop}%
\bibitem [{\citenamefont {Poncé}\ \emph {et~al.}(2015)\citenamefont {Poncé},
  \citenamefont {Gillet}, \citenamefont {Laflamme~Janssen}, \citenamefont
  {Marini}, \citenamefont {Verstraete},\ and\ \citenamefont
  {Gonze}}]{2015PonceJCP}%
  \BibitemOpen
  \bibfield  {author} {\bibinfo {author} {\bibfnamefont {S.}~\bibnamefont
  {Poncé}}, \bibinfo {author} {\bibfnamefont {Y.}~\bibnamefont {Gillet}},
  \bibinfo {author} {\bibfnamefont {J.}~\bibnamefont {Laflamme~Janssen}},
  \bibinfo {author} {\bibfnamefont {A.}~\bibnamefont {Marini}}, \bibinfo
  {author} {\bibfnamefont {M.}~\bibnamefont {Verstraete}}, \ and\ \bibinfo
  {author} {\bibfnamefont {X.}~\bibnamefont {Gonze}},\ }\bibfield  {title}
  {\enquote {\bibinfo {title} {Temperature dependence of the electronic
  structure of semiconductors and insulators},}\ }\href {\doibase
  10.1063/1.4927081} {\bibfield  {journal} {\bibinfo  {journal} {The Journal of
  Chemical Physics}\ }\textbf {\bibinfo {volume} {143}},\ \bibinfo {pages}
  {102813} (\bibinfo {year} {2015})}\BibitemShut {NoStop}%
\bibitem [{\citenamefont {Noffsinger}\ \emph {et~al.}(2012)\citenamefont
  {Noffsinger}, \citenamefont {Kioupakis}, \citenamefont {{Van de Walle}},
  \citenamefont {Louie},\ and\ \citenamefont {Cohen}}]{2012NoffsingerIndabs}%
  \BibitemOpen
  \bibfield  {author} {\bibinfo {author} {\bibfnamefont {Jesse}\ \bibnamefont
  {Noffsinger}}, \bibinfo {author} {\bibfnamefont {Emmanouil}\ \bibnamefont
  {Kioupakis}}, \bibinfo {author} {\bibfnamefont {Chris~G.}\ \bibnamefont {{Van
  de Walle}}}, \bibinfo {author} {\bibfnamefont {Steven~G.}\ \bibnamefont
  {Louie}}, \ and\ \bibinfo {author} {\bibfnamefont {Marvin~L.}\ \bibnamefont
  {Cohen}},\ }\bibfield  {title} {\enquote {\bibinfo {title} {Phonon-{{Assisted
  Optical Absorption}} in {{Silicon}} from {{First Principles}}},}\ }\href
  {\doibase 10.1103/PhysRevLett.108.167402} {\bibfield  {journal} {\bibinfo
  {journal} {Physical Review Letters}\ }\textbf {\bibinfo {volume} {108}},\
  \bibinfo {pages} {167402} (\bibinfo {year} {2012})}\BibitemShut {NoStop}%
\bibitem [{\citenamefont {Lopez}\ \emph {et~al.}(2012)\citenamefont {Lopez},
  \citenamefont {Vanderbilt}, \citenamefont {Thonhauser},\ and\ \citenamefont
  {Souza}}]{2012LopezOrbitalMag}%
  \BibitemOpen
  \bibfield  {author} {\bibinfo {author} {\bibfnamefont {M.~G.}\ \bibnamefont
  {Lopez}}, \bibinfo {author} {\bibfnamefont {David}\ \bibnamefont
  {Vanderbilt}}, \bibinfo {author} {\bibfnamefont {T.}~\bibnamefont
  {Thonhauser}}, \ and\ \bibinfo {author} {\bibfnamefont {Ivo}\ \bibnamefont
  {Souza}},\ }\bibfield  {title} {\enquote {\bibinfo {title} {Wannier-based
  calculation of the orbital magnetization in crystals},}\ }\href {\doibase
  10.1103/PhysRevB.85.014435} {\bibfield  {journal} {\bibinfo  {journal}
  {Physical Review B}\ }\textbf {\bibinfo {volume} {85}},\ \bibinfo {pages}
  {014435} (\bibinfo {year} {2012})}\BibitemShut {NoStop}%
\end{thebibliography}%

\end{document}


\title{Supplemental Materials for ``Wannier Function Perturbation Theory:\\ Localized Representation and Interpolation of Wave Function Perturbation''}

\author{Jae-Mo Lihm}
\email{jaemo.lihm@gmail.com}
\author{Cheol-Hwan Park}
\email{cheolhwan@snu.ac.kr}
\affiliation{Center for Correlated Electron Systems, Institute for Basic Science, Seoul 08826, Korea}
\affiliation{Department of Physics and Astronomy, Seoul National University, Seoul 08826, Korea}
\affiliation{Center for Theoretical Physics, Seoul National University, Seoul 08826, Korea}

\date{\today}

\maketitle

\tableofcontents

\section{Calculation of the velocity operators in the plane-wave basis} \label{sec:s_velocity}
In this section, we discuss the calculation of the velocity operator $\hat{v}^{a}$, spin-velocity operator $\hat{j}^{s;a}$, and the second-order spin-velocity operator $\hat{j}^{s;ab}$.
For the Wannier interpolation of these operators, most works use the relation of the velocity and the dipole operators as done in Ref.~\cite{2006WangAHC} for the velocity and in Refs.~\cite{2018QiaoSHC,2019RyooSHC} for the spin-velocity operator.
Since the position matrix elements are calculated by a finite-difference method on the coarse $k$ grid~\cite{1997Marzari}, the resulting matrix element may suffer from finite-difference error.

In this work, we use an alternative strategy of directly Wannier interpolating the velocity and spin-velocity operators.
The direct Wannier interpolation requires additional calculations on the coarse $k$-point grid but is beneficial in that there is no finite-difference error.
We note that the direct interpolation of the velocity matrix elements was recently also done in Ref.~\cite{2021PonceMobility}.

Since the Wannier interpolation part is the same as other quantities, we focus in this section only on the exact calculation of the matrix elements on the coarse $k$ grid using the plane-wave basis.
As we are calculating the velocity and related operators, one should take into account the effect of nonlocal potentials coming from the pseudopotentials.

For the velocity operator, we use the following analytical expression~\cite{1996AdolphVelocity}
\begin{equation} \label{eq:s_velocity}
    \hat{v}^a
    = \frac{\hat{p}^a}{m} + \frac{i}{\hbar} \comm{\hat{V}_\mathrm{NL}}{\hat{r}^a}.
\end{equation}
where $m$ is the electron mass and $\hat{V}_\mathrm{NL}$ the nonlocal potential.
Since the nonlocal potential is a sum of atomic contributions where each term has finite support in real space, its commutator with the position operator is well defined.
Equation~\eqref{eq:s_velocity} is implemented in the \texttt{Quantum ESPRESSO}~\cite{2017GiannozziQE} package.
The spin-velocity matrix elements can easily be calculated using
\begin{equation} \label{eq:s_spin_velocity}
    \mel{\psi_\mk}{\hat{j}^{s;a}}{\psi_\nk}
    = \frac{1}{2} \left(
      \mel{\psi_\mk}{\hat{S}_s \hat{v}^a}{\psi_\nk}
    + \mel{\psi_\mk}{\hat{v}^a \hat{S}_s}{\psi_\nk}
    \right).
\end{equation}
For example, for the first term on the right-hand side, we calculate the $\hat{v}^a$ matrix element between $\hat{S}_s \ket{\psi_\mk}$ and $\ket{\psi_\nk}$ using \myeqref{eq:s_velocity}.

For the second-order velocity operator $\hat{j}^{0;ab}$, we use the finite-difference method:
\begin{equation} \label{eq:s_velocity_2nd}
    \mel{\psi_\mk}{\hat{j}^{0;ab}}{\psi_\nk}
    = \mel{u_\mk}{\frac{\partial^2\opH(\mb{k})}{\partial k^a \partial k^b}}{u_\nk}
    \approx \frac{\hbar}{2\delta k} \left[ \mel{u_\mk}{\hat{v}^a(\mb{k}+\delta k \mb{e}^b)}{u_\nk} - \mel{u_\mk}{\hat{v}^a(\mb{k}-\delta k \mb{e}^b)}{u_\nk} \right]
\end{equation}
Here, $\ket{u_\nk}$ is the periodic part of $\ket{\psi_\nk}$, $\mb{e}^b$ is the unit vector along $b$, and we defined
\begin{equation}
    \hat{v}^a(\mb{k}) = e^{-i\mb{k}\cdot\hat{\mb{r}}} \hat{v}^a e^{i\mb{k}\cdot\hat{\mb{r}}}.
\end{equation}
The first-order velocity matrix elements in \myeqref{eq:s_velocity_2nd} are calculated using \myeqref{eq:s_velocity}.
We used $\delta k = 10^{-4}~\mathrm{bohr}^{-1}$.
In practice, we calculate, e.\,g.\,, $\hat{v}^a(\mb{k}+\delta k \mb{e}^b)\ket{u_\nk}$ first and then calculate the second-order spin-velocity by taking its inner product with $\hat{S}_s \ket{u_\mk}$.

We note that we only considered norm-conserving pseudopotentials.
For ultrasoft pseudopotentials or projector augmented-wave potentials, one will need additional consideration when multiplying the spin operator and the velocity operator.

\section{Convergence study} \label{sec:s_convergence}
\subsection{Temperature-dependent electron band structure} \label{sec:s_conv_selfen}
\begin{figure*}[htbp]
\includegraphics[width=0.8\textwidth]{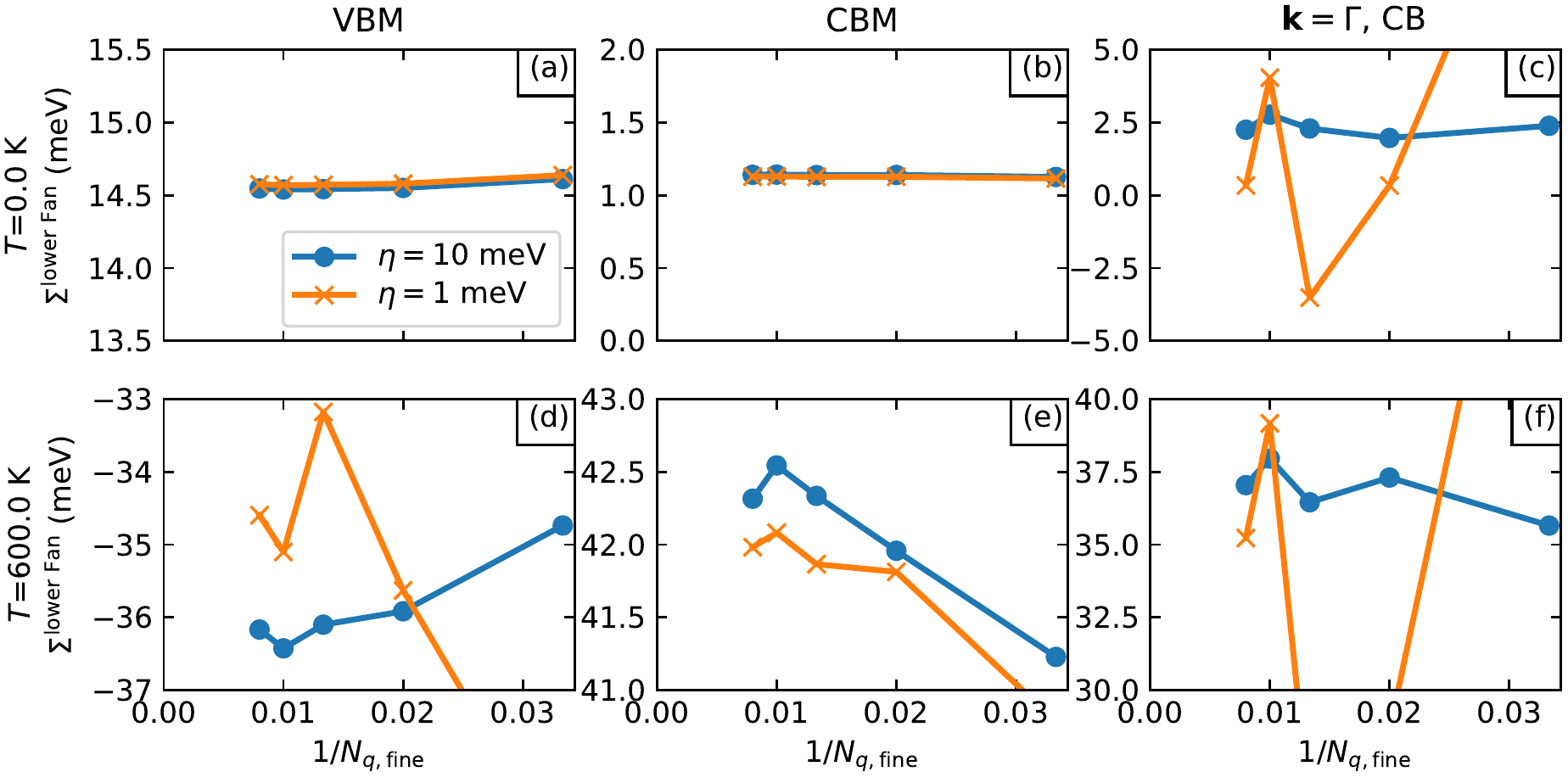}
\caption{
Convergence of the lower Fan self-energy of silicon at (a-c) $T$=0~K and (d-f) $T$=600~K.
}
\label{fig:s_conv_lower_fan}
\end{figure*}

The convergence of the upper Fan and the DW terms are discussed in detail in Appendix \ref{sec:app_doublegrid}.
Here, we discuss the convergence of the lower Fan term with respect to $\eta$ in the denominator of the summand and the fine grid size.

Figure~\ref{fig:s_conv_lower_fan} shows the convergence of the lower Fan self-energy.
At the band extrema, the self-energy converges with error around 2~meV already at $N_{q,\mathrm{fine}} = 50$ and $\eta$=10~meV for both 0~K and 600~K.
Since silicon is an infrared-inactive material with zero Born effective charge, the dependence of the self-energy at the band extrema with respect to $\eta$ and $N_{q,\mathrm{fine}}$ is weak~\cite{2015PonceJCP}.
In contrast, the self-energy converges much slowly for the conduction band at $\mathrm{\Gamma}$ because it is not an extremum point of the band structure.
To make sure that the self-energy is converged for all low-energy states, we used a $125\times125\times125$ fine $q$ grid and $\eta$=10~meV.

\subsection{Indirect optical absorption} \label{sec:s_conv_indabs}

\begin{figure*}[htbp]
\includegraphics[width=0.8\textwidth]{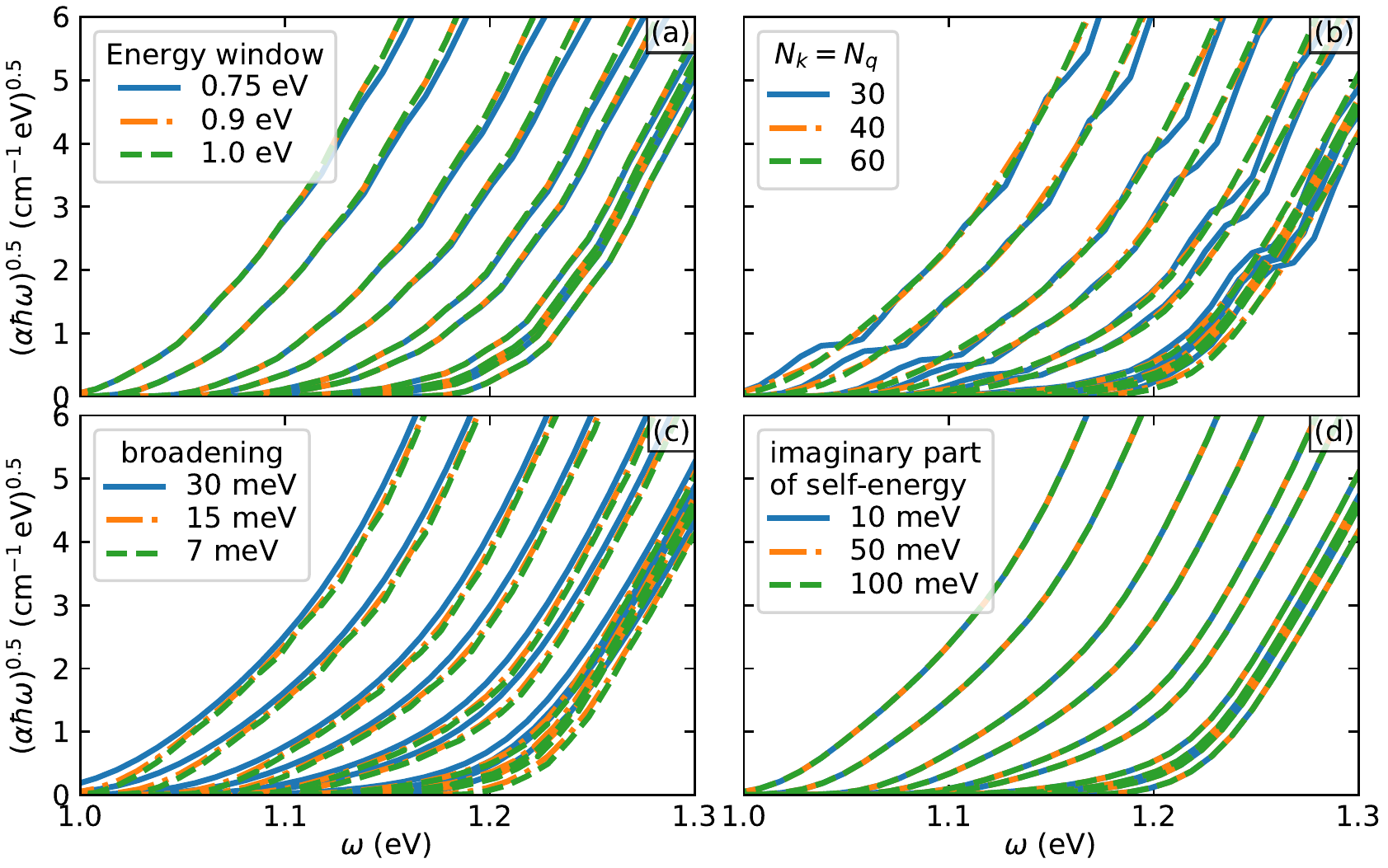}
\caption{
Convergence of the indirect absorption spectra of silicon with respect to (a) the energy window, (b) the $k$- and $q$-point grid, (c) width $\gamma$ for the Gaussian broadening of the delta functions, and (d) the imaginary part of the electron self-energy $\Gamma$.
In all panels, the effect of the phonon-induced renormalization of the electron band structure is not included.
Instead, we rigidly shifted the spectra by 0.10-0.21~eV to roughly match the experimental spectra.
}
\label{fig:s_conv_indabs}
\end{figure*}

Figure~\ref{fig:s_conv_indabs} shows the convergence of the indirect absorption spectra.
In this convergence study, we did not include the effect of the phonon-induced renormalization of the electron band structure.
The energy window in Fig.~\ref{fig:s_conv_indabs}(a) indicates the range where the initial and final states are selected.
For example, with an energy window of 0.75~eV, only the initial and final states inside the $\pm$0.75~eV window around the center of the band gap are included in the calculation.
Note that the intermediate states are not limited by this window.
We find that the energy window of 1.0~eV gives converged spectra up to $\omega=1.3$~eV.
From Fig.~\ref{fig:s_conv_indabs}(b-c), we find that $N_k=60$ and a Gaussian broadening of the delta function in \myeqref{eq:s_indabs_alpha_Sigma} by $\gamma=$15~meV give converged results.
Figure~\ref{fig:s_conv_indabs}(d) shows that the spectra is insensitive to the imaginary part of the self-energy [\myeqref{eq:s_indabs_Gamma}], consistent with the report of Ref.~\cite{2012NoffsingerIndabs}.
We set the imaginary part of the self-energy $\Gamma=$50~meV.

\subsection{Shift charge and spin current} \label{sec:s_conv_shift}
\begin{figure*}[htbp]
\includegraphics[width=1.0\textwidth]{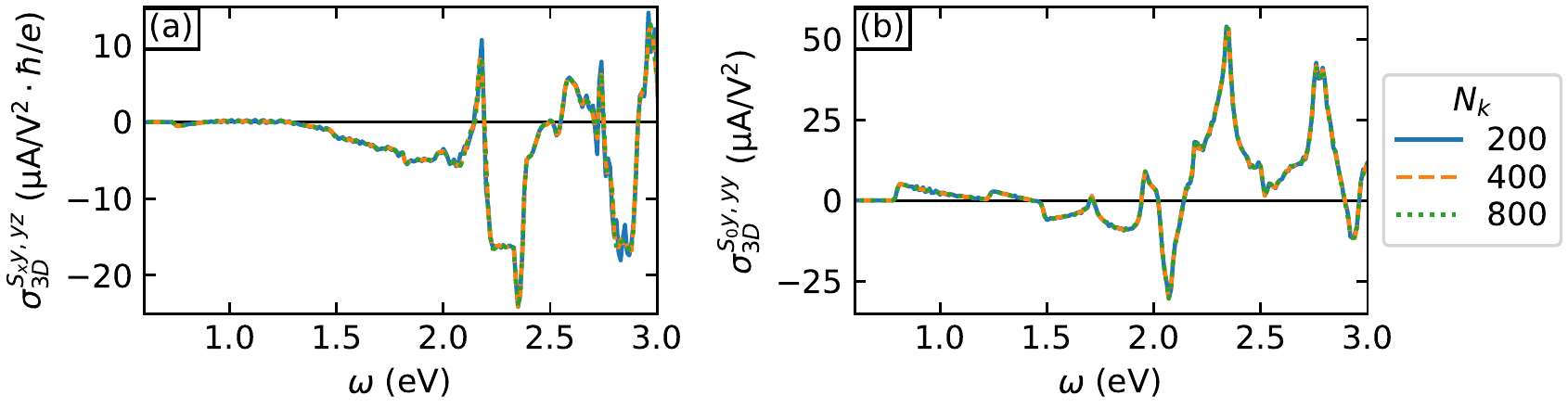}
\caption{
Convergence of the (a) shift spin and (b) shift charge current spectra of monolayer WTe$_2$ with respect to the $N_k\times N_k\times 1$ $k$-point grid. $\gamma=$10~meV is used.
}
\label{fig:s_conv_shift_nk}
\end{figure*}
\begin{figure*}[htbp]
\includegraphics[width=1.0\textwidth]{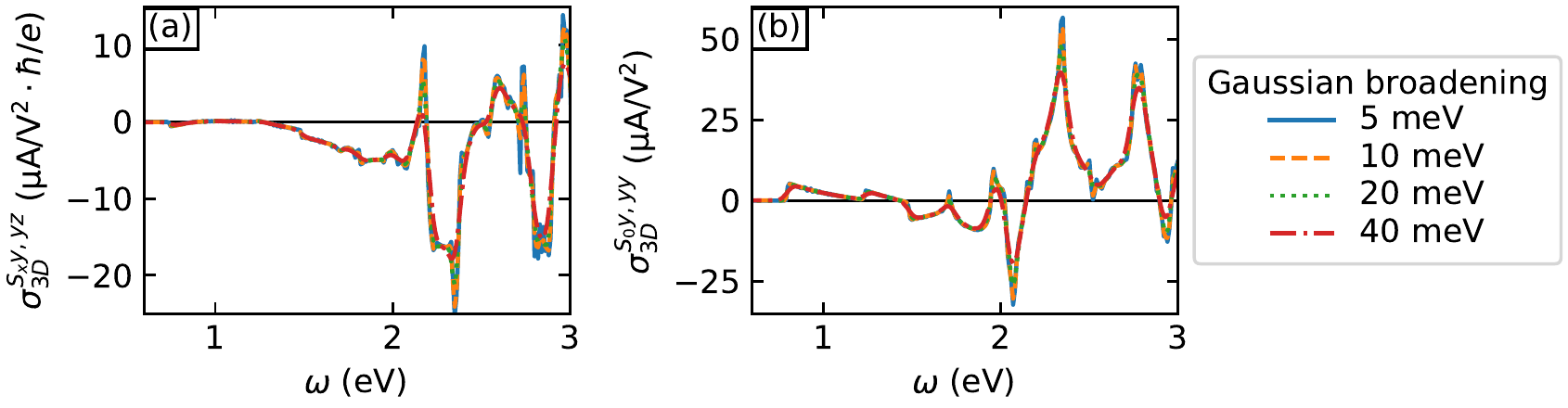}
\caption{
Convergence of the (a) spin and (b) charge shift current spectra of monolayer WTe$_2$ with respect to the broadening parameter $\gamma$. A $400\times400\times1$ $k$-point grid is used.
}
\label{fig:s_conv_shift_smr}
\end{figure*}

Figures~\ref{fig:s_conv_shift_nk} and \ref{fig:s_conv_shift_smr} show the convergence of the shift spin and shift charge current spectra with respect to the size of the $k$-point grid and the parameter $\gamma$ for the Gaussian broadening of the delta function [\myeqref{eq:s_delta_gaussian}].
We find that a $400\times400\times1$ grid and $\gamma=$10~meV give converged results.

\subsection{Spin Hall conductivity} \label{sec:s_conv_shc}
\begin{figure}[htbp]
\includegraphics[width=0.5\columnwidth]{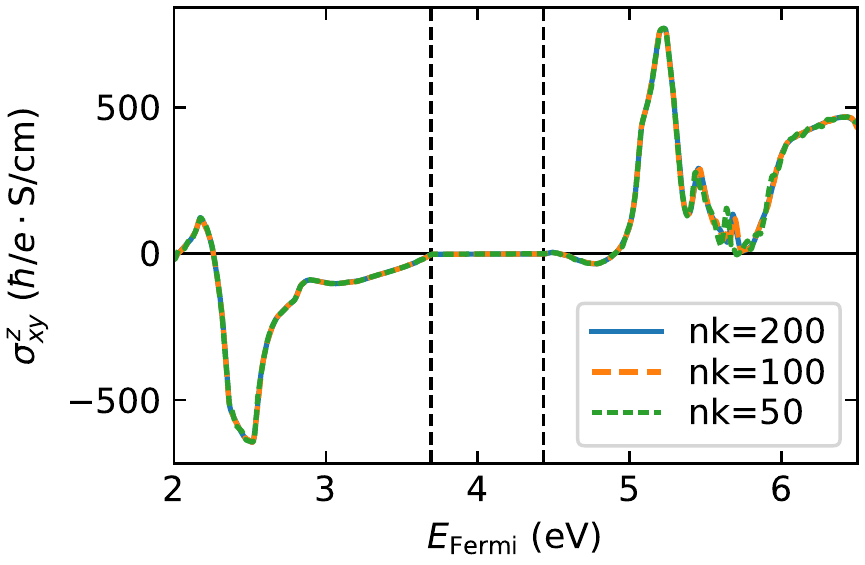}
\caption{
Convergence of the spin Hall conductivity of monolayer WTe$_2$ computed using WFPT with respect to the size of the $k$-point grid. A $N_k\times N_k \times 1$ $k$-point grid and $\eta=10$~meV was used.
The temperature was set to 50~K.
The dashed vertical lines indicate the valence band maximum and the conduction band minimum energies.
}
\label{fig:s_shc_conv_nk}
\end{figure}

\begin{figure}[htbp]
\includegraphics[width=0.5\columnwidth]{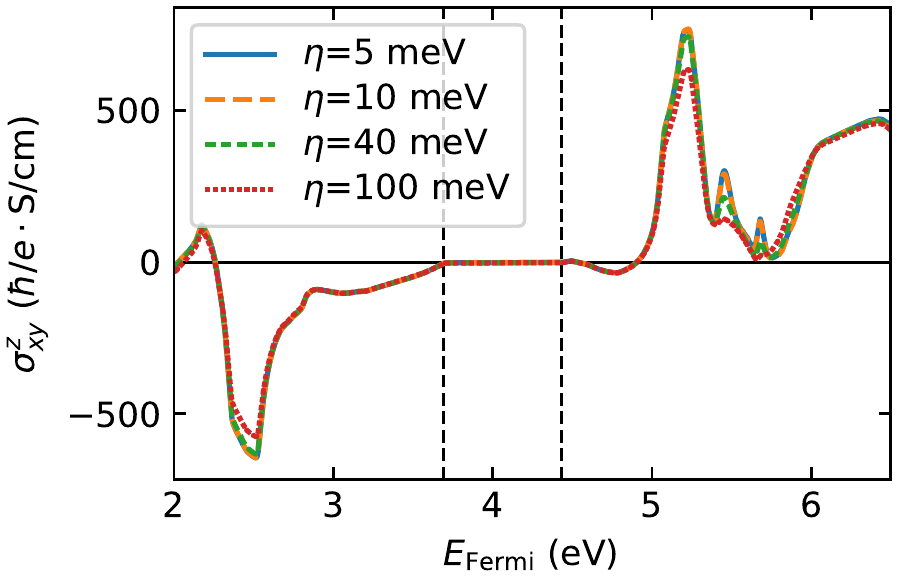}
\caption{
Convergence of the spin Hall conductivity of monolayer WTe$_2$ computed using WFPT with respect to the broadening $\eta$.
A $200\times 200 \times 1$ $k$-point grid was used.
The temperature was set to 50~K.
The dashed vertical lines indicate the valence band maximum and the conduction band minimum energies.
}
\label{fig:s_shc_conv_eta}
\end{figure}

Figures~\ref{fig:s_shc_conv_nk} and \ref{fig:s_shc_conv_eta} show the convergence of the spin Hall conductivity of monolayer WTe$_2$ with respect to the size of the $k$-point grid and the $\eta$ parameter in the definition of spin Berry curvature [\myeqref{eq:shc_berry}].
We find that a $200\times200\times1$ grid and $\eta=$10~meV give converged results.

\begin{figure}[htbp]
\includegraphics[width=0.5\columnwidth]{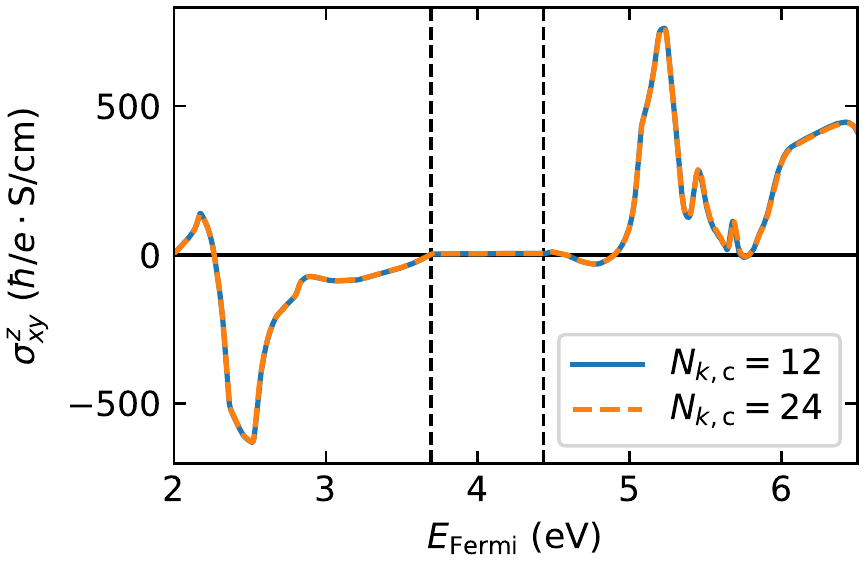}
\caption{
Convergence of the spin Hall conductivity of monolayer WTe$_2$ computed using WFPT with respect to the size of the coarse $k$-point grid, which is set to $N_{k,\mathrm{c}} \times N_{k,\mathrm{c}} \times 1$.
A $200\times 200 \times 1$ fine $k$-point grid and $\eta=10~$meV was used.
The temperature was set to 50~K.
The dashed vertical lines indicate the valence band maximum and the conduction band minimum energies.
}
\label{fig:s_shc_conv_coarse_nk}
\end{figure}

The WF interpolation schemes of Ref.~\cite{2019RyooSHC} and Ref.~\cite{2018QiaoSHC} for the velocity and spin-velocity matrix elements use finite-difference derivatives to compute the real-space matrix elements.
Quantities computed using these methods may converge slower with respect to the size of the coarse $k$ grid than the band structure~\cite{2012LopezOrbitalMag}.
In the WFPT calculation, we exactly evaluated the velocity and spin-velocity matrices on the coarse $k$ grid as discussed in Sec.~\ref{sec:s_velocity}, and hence the results are free from the finite-difference error.

To test the convergence of the WF interpolation with respect to the coarse $k$ grid, we calculated the spin Hall conductivity calculated using the Wannier interpolation method of Ref.~\cite{2019RyooSHC} with $12\times12\times1$ and $24\times24\times1$ coarse $k$ grids.
Figure~\ref{fig:s_shc_conv_coarse_nk} shows that the $12\times12\times1$ grid gives converged results.
Therefore, the small but finite discrepancy between the WFPT and truncated WF calculations in Fig.~\ref{fig:wte2_shc} is indeed the band-truncation error, not the finite-difference error.

\FloatBarrier 
\makeatletter\@input{xx.tex}\makeatother
\bibliography{main}